\documentclass[twocolumn]{aastex631}
\usepackage{amsmath,amsfonts,amssymb,graphicx,chngcntr,multirow, float,booktabs, afterpage}
\usepackage[]{hyperref}
\usepackage{tikz}
\usepackage{graphicx}
\usetikzlibrary{arrows.meta, positioning, shapes.multipart}
\hypersetup{colorlinks=true}

\received{}
\revised{}
\accepted{}

\shorttitle{The sub-parsec density profiles of TDE hosts}

\begin{document}

\title{Resolving the sub-parsec circumnuclear density profiles of quiescent galaxies: \\ Evidence for Bondi accretion flows in tidal disruption event hosts}

\author[0000-0003-3441-8299]{A. J. Goodwin}
\thanks{These authors contributed equally to this work.}
\affiliation{International Centre for Radio Astronomy Research – Curtin University, GPO Box U1987, Perth, WA 6845, Australia}
\email{ajgoodwin.astro@gmail.com}  

\author{A. Mummery} 
\thanks{These authors contributed equally to this work.}
\affiliation{School of Natural Sciences, Institute for Advanced Study, 1 Einstein Drive, Princeton, NJ 08540, USA}
\email{amummery@ias.edu}

\begin{abstract}
\noindent The sub-parsec circumnuclear density profiles of galaxies represent a key element in our understanding of the accretion history and fuel availability of supermassive black holes (SMBHs). 
Observations that directly resolve sub-parsec scales in galaxies require extremely high resolution and generally hot (bright) environments, making this impossible in all but the nearest active galaxies.
Transient accretion events onto previously quiescent SMBHs, such as a tidal disruption event (TDE), offer a new avenue to understand SMBHs and their environments.
Radio-bright outflows from TDEs directly probe the ambient density at $10^{-3}-1$\,pc scales, allowing direct constraints on the circumnuclear density of TDE host galaxies (i.e., quiescent galaxies). 
Here we present, using radio observations of a sample of 11 TDE hosts, a new methodology for fitting observed TDE radio emission to constrain their sub-parsec circumnuclear density profiles. 
Our findings reveal that TDE host galaxies exhibit circumnuclear density profiles remarkably consistent with the expectations of a simple Bondi accretion flow ($n_e\propto R^{-3/2}$). Under the assumption of a Bondi profile, we present a new method to jointly fit the outflow mass and ambient densities, in order to constrain the Bondi accretion rate and temperature. For the TDE host galaxies in our sample, we constrain a sample average Bondi accretion rate Eddington fraction of log\,$_{10}f_{\rm{Edd}} = -3.96^{+0.30}_{-0.38}$ (as well as individual fits to each host). 
This work provides a methodology by which radio observations of TDEs can provide powerful constraints on the sub-parsec density distribution of quiescent SMBHs -- well inside the Bondi sphere. This opens up a new observational avenue to constrain sub-parsec  gas distributions in a broad range of galaxies. 
\end{abstract}


\section{Introduction} \label{sec:intro}
It is well established that most galaxies harbor a supermassive black hole (SMBH) at their center. Feedback from outflows ejected by these SMBHs as they accrete gas seems to be closely linked to galaxy evolution \citep[e.g.][]{Croton2006}. However, the distribution of gas at sub-parsec scales, that ultimately provides the gas reservoir for the accreting material, has proved difficult to observationally constrain. This is primarily due to the extremely high spatial resolution required to resolve these (sub-parsec) scales in all about the most nearby galaxies \citep[e.g.][]{Gillessen2019,Russell2015}. 

A natural theoretical expectation for the background circumnuclear gas distribution in a quiescent galaxy on small radial scales is that of a spherical ``Bondi accretion" profile
\citep{Bondi52}.  Within the region where gas is gravitationally bound to the black hole, i.e. $a_\infty^2/2 - GM_\bullet/r<0$ (where $a_\infty$ is the speed of sound far from the black hole and the other symbols have their usual meaning), gas accelerates rapidly under the influence of the black hole's gravity, settling into a characteristic density profile which will be the focus of much of this paper. 

These Bondi flows may be an important driver of the circumnuclear gas density distribution in quiescent galaxies. Despite neglecting various important physical processes, for example the angular momentum of the accreting gas \citep[e.g. see][for a discussion]{Narayan2011}, magnetic fields, radiative losses and turbulent/viscous dissipation,  the Bondi solution provides a basic framework to describe the distribution of gas around quiescent supermassive black holes. X-ray observations of hot gas in the nuclei of nearby galaxies have been used to provide limits on the Bondi accretion rate \citep[e.g.,][see Figure \ref{fig:gal_comps}]{Allen2006}, although these observations mostly are unable to probe gas density distributions within the Bondi radius and necessarily focus on AGN, as quiescent galaxies by definition do not emit detectable nuclear X-ray emission.  

Tidal disruption events (TDEs) -- when a star is torn apart by tidal forces near a supermassive black hole (SMBH) -- offer a unique opportunity to observe emission from the vicinity of a previously quiescent SMBH.
Over the last decade optical all-sky surveys have discovered some $\sim100$ TDEs, characterised by a $\sim2$ order of magnitude increase in optical luminosity from the nucleus of a previously quiescent galaxy \citep[e.g.,][]{Yao2023}. This significant rise in TDE discoveries has created new opportunities to observe (and understand) galactic nuclei. For example, the switching on of transient ``extreme coronal lines'' \citep[e.g.,][for some recent examples]{Somalwar2023, Yao2023, Newsome2024_22upj, Clark2025} requires a high circumnuclear density $n_e \sim 10^{6}\, {\rm cm}^{-3}$ on small scales $R \sim 10^{16-17}\, {\rm cm}$ \citep[][]{Mummery_et_al25}. These coronal line emitters are rare, however, and the circumnuclear density is hard to measure precisely owing to various parameter degeneracies. TDE radio emission, we shall show in this paper, provides a much more precise probe of the circumnuclear medium. 


Radio emission from the majority of TDEs arises due to synchrotron emission produced by material ejected from the vicinity of the SMBH interacting with the circumnuclear medium \citep[see][for a review]{Alexander2020}. Prompt radio-emitting outflows have been detected in a large fraction of TDEs, with 30--50$\%$ of events showing slow ($\sim10\%\,c$) quasi-spherical outflows at early times ($<50$\,d) post-optical flare \citep[e.g.][]{Alexander16,Cendes_2021_dsg,Goodwin22}. 
Radio observations of these outflows directly probe interactions between the outflow and the medium it propagates through. 

Recently, 
\citet{GoodwinMummery2026} showed that the prompt radio-emitting outflows observed from TDEs are produced by the interaction of a super-Eddington accretion disk wind with the circumnuclear gas. Accretion disk winds are well approximated by a ballistically expanding outflow sweeping up material from the circumnuclear medium, as opposed to a jet-like outflow which may receive energy injection from the base of the jet. In the case of a ballistically expanding outflow, the evolution of the radio emission is only dependent on the total energy in the outflow and the medium it propagates through \citep[e.g.][]{Chevalier1998}. The evolution of the radio emission from TDEs therefore provides direct constraints on the sub-parsec host galaxy environments, including constraining the density of the ambient medium in this region. This makes radio emission of TDEs a powerful probe of the circumnuclear medium around previously quiescent SMBHs.


In this work, we investigate, using a sample of 11 prompt radio-emitting TDEs, the sub-parsec distribution of the circumnuclear medium of each host galaxy, constrained by multi-epoch radio observations. In Section \ref{sec:theory} we derive the theoretical expectations for the ambient density profile assuming a simple Bondi accretion flow, and discuss the implications of this background profile for TDE radio emission in Section \ref{sec:implication}. In Section \ref{sec:results} we describe the radio observations and modeling, and fits to the observed ambient density profiles. In Section \ref{sec:bondi_const} we extend our analysis to constrain the Bondi accretion efficiency and temperature. In Section \ref{sec:discussion} we discuss the implications of our findings, the limitations, and compare to previous studies. Finally, we summarize our results and provide concluding remarks in Section \ref{sec:summary}. 

\section{Expectation for an intra-Bondi density profile}\label{sec:theory}
As much of this paper will involve comparing the resolved circumnuclear density profiles of TDE hosts to the Bondi profile, we provide a full derivation of the classical Bondi result here. The reader familiar with this physics may wish to skip directly to section \ref{sec:implication}, where implications of this profile are discussed in the radio bright TDE context. 

In this section we derive the solution to the classical spherical Bondi accretion problem \citep{Bondi52}, and relate the number density of electrons close to the black hole $n_e(r)$ to two free parameters, the temperature of the circumnuclear gas far from the black hole $T_\infty$ and the Eddington ratio of the background Bondi flow $f_{\rm Edd}$. 

We start with the mass conservation (continuity) equation 
\begin{equation}
    {\partial \rho \over \partial t} + \vec \nabla \cdot (\rho \vec u) = 0, 
\end{equation}
and the Euler (momentum conservation) equation for the gas 
\begin{equation}
    \rho \left[{\partial \vec u \over \partial t} + (\vec u \cdot \vec \nabla)\vec u\right] = -\vec \nabla P -\rho \vec \nabla \Phi .
\end{equation}
Here we have been general, $\rho$ is the density of the gas, $P$ is the gas pressure, $\vec u$ is its velocity and the gas is in some (Newtonian) gravitational potential defined by $\Phi$.

Let us further look for a spherically symmetric steady-state solution (i.e., profiles of the general form $\rho = \rho(r)$, $P=P(r)$ and $\vec u = U(r) \hat r$), in a central point source Newtonian potential $\Phi = -GM_\bullet/r$ with an adiabatic gas equation of state $P\rho^{-\gamma} = K = {\rm constant}$. The constant $K$ here is a measure of the entropy of the gas, and $1<\gamma\leq5/3$ is the adiabatic index. Then mass conservation (in spherical coordinates) becomes 
\begin{equation}
    {1\over r^2}{\partial \over \partial r} \left(r^2 U \rho \right) = 0,
\end{equation}
and (radial) momentum conservation becomes 
\begin{equation}
    \rho U{\partial U\over \partial r} = -{\partial P \over \partial r} - \rho {GM_\bullet \over r^2} .
\end{equation}
Mass conservation implies $r^2 U \rho = {\rm constant}$, or more conventionally 
\begin{equation}
    \dot M_b = -4\pi r^2 \rho U = {\rm constant},
\end{equation}
which defines the (constant) accretion rate $\dot M_b>0$, sourced at large radii from the black hole. Note the minus sign as an accretion flow has a negative radial velocity $U < 0$ (i.e., it flows inwards). 
For the momentum equation it is more convenient to work with the sound speed of the gas $a$, defined via 
\begin{equation}
    a^2 \equiv {{\rm d}P \over {\rm d}\rho } = \gamma K \rho^{\gamma-1} ,
\end{equation}
or 
\begin{equation}
    a^2 = a_\infty^2 \left({\rho\over \rho_\infty}\right)^{\gamma-1} ,
\end{equation}
where we have introduced the sound speed $(a_\infty)$ and density $(\rho_\infty)$ of the gas far from the black hole. Substituting this into the radial momentum equation we find 
\begin{equation}
     U{\partial U\over \partial r} = -{a^2\over \rho}{\partial  \rho \over \partial r} -  {GM_\bullet \over r^2} ,
\end{equation}
or 
\begin{equation}
     U{\partial U\over \partial r} = -{2a\over \gamma-1} {\partial a \over \partial r} -  {GM_\bullet \over r^2} ,
\end{equation}
which can be integrated once to leave the Bernoulli constraint 
\begin{equation}
    {\cal B} = {1\over 2}U^2 + {a^2 \over \gamma -1} - {GM_\bullet\over r} = {\rm constant}. 
\end{equation}
We have two integration constants $\dot M_b$ and ${\cal B}$ which we wish to relate to the two quantities of interest (the temperature of the circumnuclear gas far from the black hole $T_\infty$ and the Eddington ratio of the background Bondi flow $f_{\rm Edd}$). If we assume that the gas is at rest at infinity ($U^2\to0$ as $r\to \infty$) then the Bernoulli constant is 
\begin{equation}
    {\cal B} = {a_\infty^2 \over \gamma-1} = {k_B T_\infty \over \mu m_p(\gamma-1)} .
\end{equation}
Here $\mu\approx 0.6$ is the mean molecular weight of the gas. It will be convenient to relate the Bondi accretion rate to the density and sound speed at infinity. This can be done by noting that there is an additional constraint on the solution, namely the continuity of all variables at the sonic point. Expanding the radial derivative in the expression of conservation of mass, we find
\begin{equation}
    U {\partial \rho \over \partial r}+ \rho {\partial U \over \partial r} + {2U\rho \over r}  = 0,
\end{equation}
which can be substituted into radial momentum conservation to give 
\begin{equation}
    U\left[1 - {a^2 \over U^2}\right]{\partial U \over \partial r} = -{GM_\bullet\over r^2 } \left[1 - {2a^2r \over GM_\bullet}\right]. 
\end{equation}
One sees that at $r_s = GM/2a_s^2$, one requires $U_s^2=a_s^2$. Imposing this requirement on the Bernoulli integral one finds 
\begin{equation}
    U_s^2 = a_s^2 = {2a_\infty^2 \over 5-3\gamma}, 
\end{equation}
which when substituted back into mass conservation gives 
\begin{equation}
    \dot M_b = {\alpha \pi G^2 M_\bullet^2 \rho_\infty \over a_\infty^3}, 
\end{equation}
where 
\begin{equation}
    \alpha \equiv \left({2\over 5-3\gamma}\right)^{(5-3\gamma)/(2-2\gamma)} .
\end{equation}
Note that $\alpha =1$ for $\gamma = 5/3$. We now have the full solution to our problem in terms of the parameters we wish to specify. Taking $\gamma = 5/3$, and $\dot M_b = f_{\rm Edd} \dot M_{\rm Edd}$ we have
\begin{equation}
     {\pi G^2 M_\bullet^2 \rho_\infty \over a_\infty^3} =  f_{\rm Edd} \dot M_{\rm Edd} =f_{\rm Edd} {4\pi G M_\bullet \over \kappa \eta c}, 
\end{equation}
where $\eta = 0.1$ is a fiducial efficiency (we stress that this value is a convention, and is only applicable as a {\it radiative} efficiency for a thin disk, which a Bondi flow  certainly is not $\eta_{\rm bondi} \ll \eta$), and $\kappa$ is the electron scattering opacity. Equivalently 
\begin{equation}
     {\rho_\infty} =  {4 f_{\rm Edd}\over GM_\bullet \kappa \eta c}\left({k_B T_\infty \over \mu m_p}\right)^{3/2} . 
\end{equation}
Defining $n_e \equiv \rho/m_p$ as the electron number density, and using mass conservation to remove $U$, we are left with a formal solution of our Bondi problem:
\begin{multline}\label{eq:bondi_full}
    {f_{\rm Edd}^2 G^2 M_\bullet^2 m_p^2 \over 2 \kappa^2 \eta^2 c^2} {1 \over n_e^2 r^4} + {3\over 2} \left({  GM_\bullet m_p\kappa \eta c \over 4f_{\rm Edd} }\right)^{2/3}  n_e^{2/3} \\ - {GM_\bullet\over r} = {3k_B T_\infty \over 2\mu m_p} ,
\end{multline}
which can be solved numerically for $n_e$ at each radius. 

\section{Implications of a background Bondi profile for TDE radio emission}\label{sec:implication}
Equation \ref{eq:bondi_full} shows that as $r\to \infty$ one has $n_e \to n_\infty$, but that for $r\to 0$ one finds 
\begin{equation}\label{eq:ner_bondi}
    n_e(r) \approx {\xi f_{\rm Edd} c^2 \over  GM_\bullet m_p\kappa\eta }\, \left({rc^2 \over 2GM_\bullet}\right)^{-3/2} \, ,
\end{equation}
where $\xi = (4/3\sqrt{3}) \approx 0.77$ for $\gamma = 5/3$. While the above derivation assumed a given equation of state $\gamma = 5/3$, the same profile $(n_e \propto r^{-3/2})$ is found for all $1< \gamma \leq 5/3$, simply with slightly different numerical prefactors $\xi$ (for $\gamma < 5/3$ the flow becomes super-sonic at sufficiently small $r$, whereafter $U^2/2 - GM_\bullet/r \approx 0$; when combined with $\rho = {\rm constant}/r^2 U$ one always finds the characteristic $n_e \sim r^{-3/2}$ scaling). Indeed, the following expression provides an excellent approximation to $n_e(r)$ at all relevant radii (for $\gamma = 5/3$)
\begin{equation}
    n_e(r) \approx  {4 f_{\rm Edd}  \over  GM_\bullet m_p\kappa\eta c } \left[{k_B T_\infty \over \mu m_p} + {2GM_\bullet \over 3 r}\right]^{3/2}. 
\end{equation}

We see that at fixed (physical) radius close to the black hole, $n_e \propto f_{\rm Edd} M_\bullet^{1/2}$. The radius at which the electron number density begins to flatten out (and deviate from its $n_e \propto r^{-3/2}$ profile) is the classical Bondi radius
\begin{multline}
    r_b \approx {2 GM_\bullet \over a_\infty^2} \approx {2\mu GM_\bullet  m_p\over k_B T_\infty} \\ \approx 2\times 10^{20 } \, {\rm cm}\, \left({M_\bullet  \over 10^7 M_\odot}\right)\, \left({T_{\rm \infty}\over 10^5 \,{\rm K}}\right)^{-1}. 
\end{multline}

\begin{figure}
    \centering
    \includegraphics[width=0.95\linewidth]{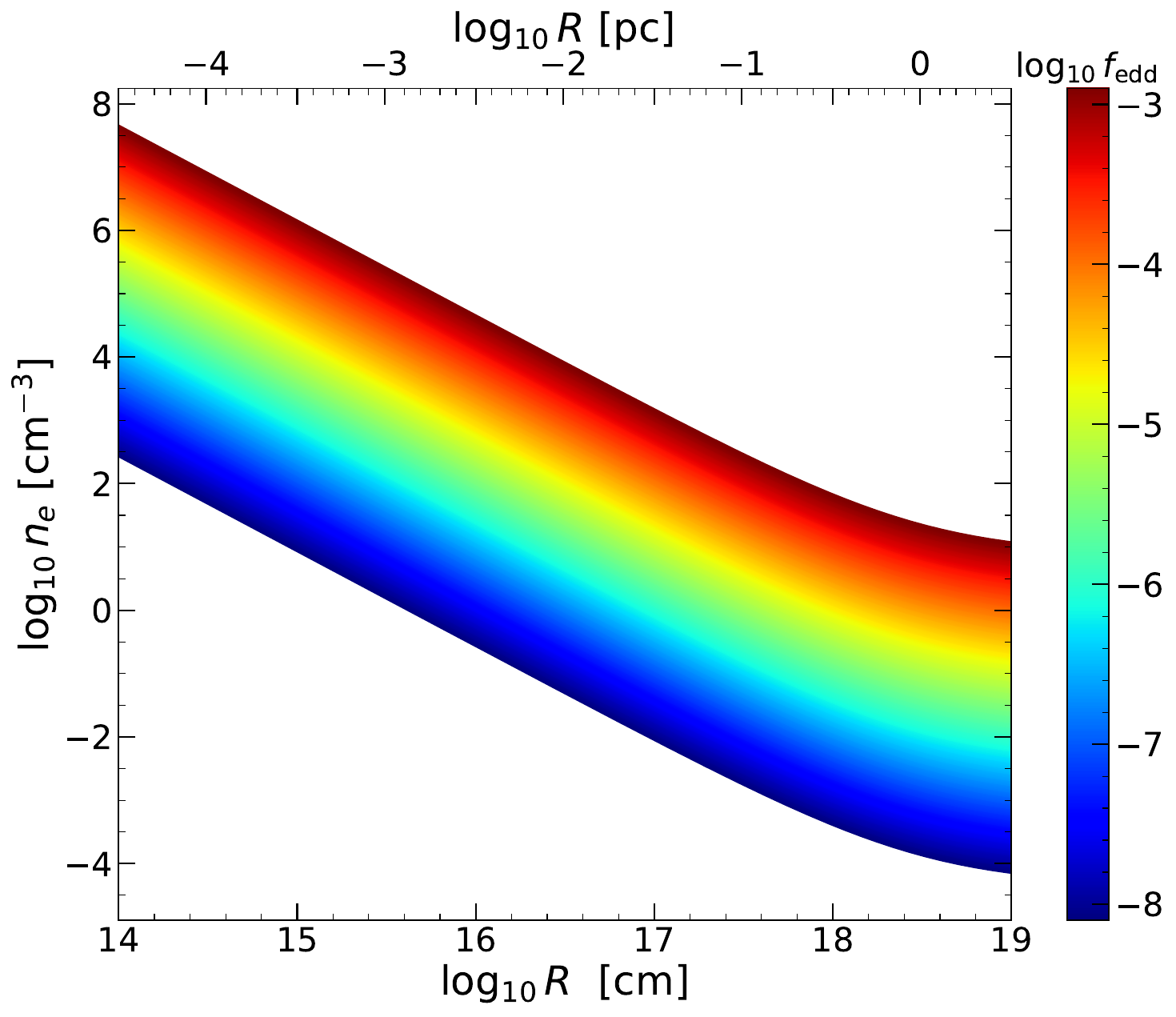}
    \includegraphics[width=0.95\linewidth]{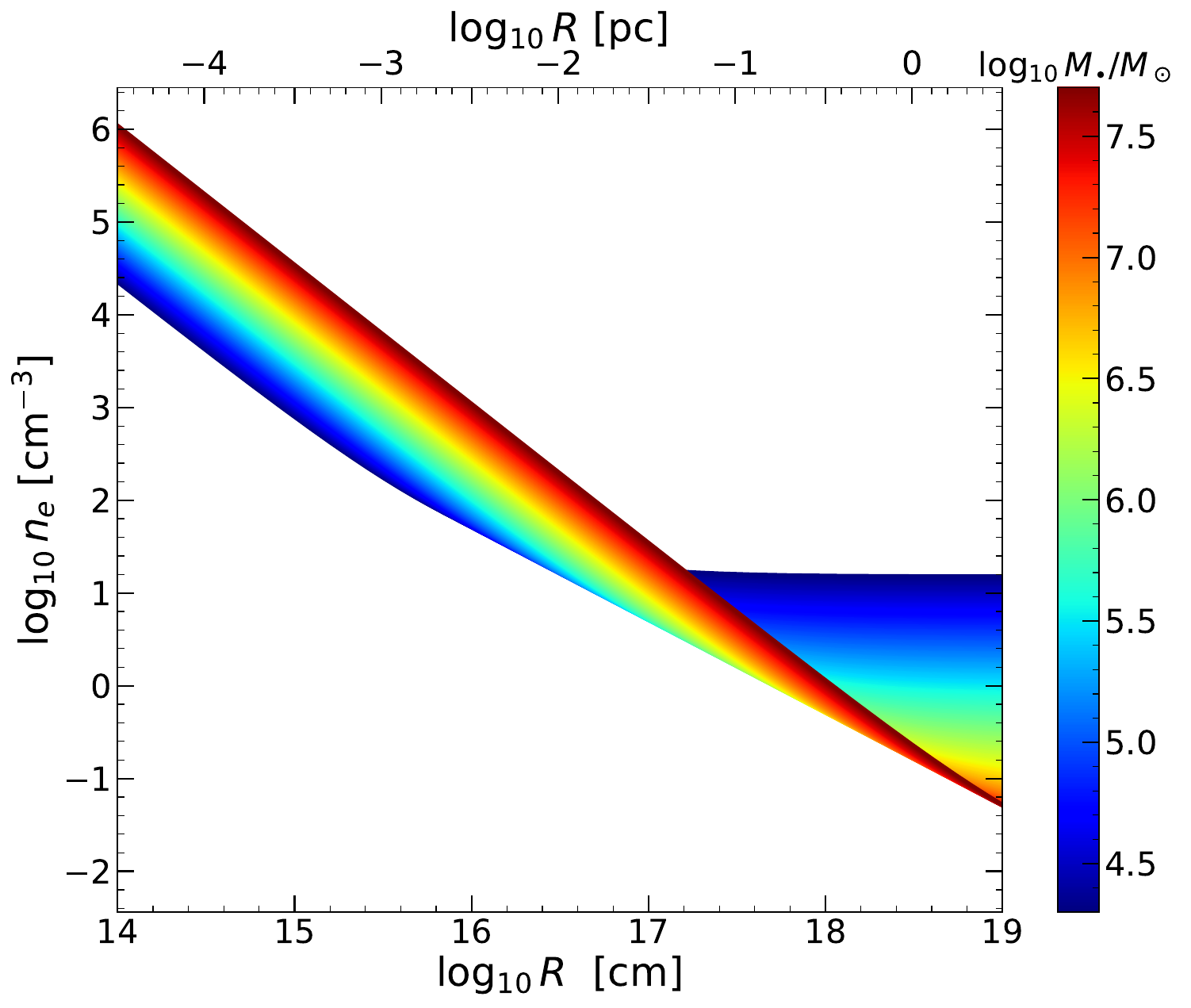}
    \includegraphics[width=0.95\linewidth]{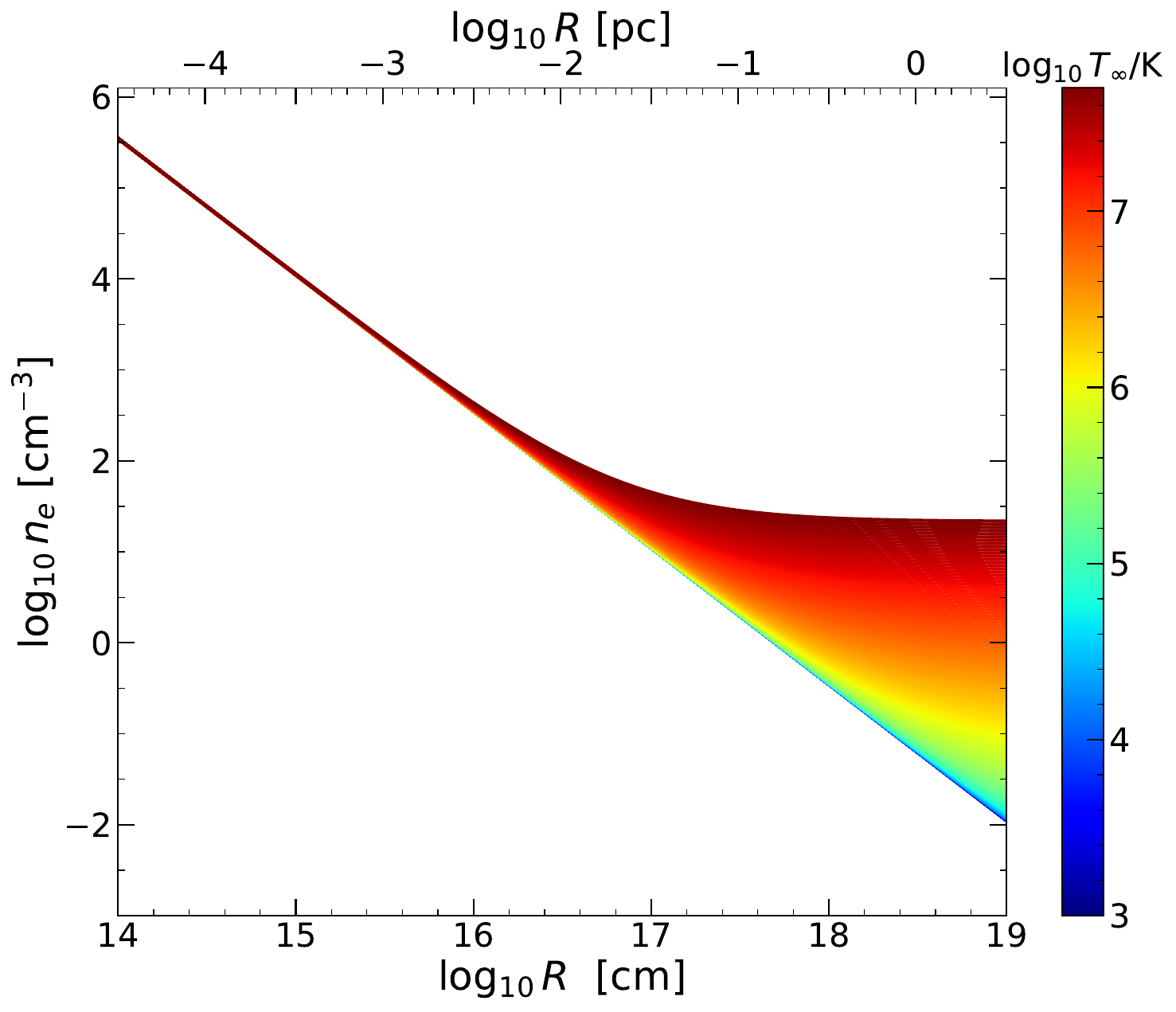}
    \caption{The dependence of the electron number densities of Bondi-accretion profiles on the three parameters of interest (i) the background Bondi accretion rate in units of Eddington (top), (ii) the black hole mass in the TDE (middle) and (iii) the temperature of the gas far from the black hole (bottom). When not varied, the parameters take the  values $M_\bullet=10^{6.5}M_\odot$, $f_{\rm Edd} = 10^{-5}$ and $T_\infty = 10^6$ K.  }
    \label{fig:theory_bondi}
\end{figure}

Figure \ref{fig:theory_bondi} shows the numerical solutions of the above Bondi equations, with the three input parameters varied. When not specified, the parameters take the  values $M_\bullet=10^{6.5}M_\odot$, $f_{\rm Edd} = 10^{-5}$ and $T_\infty = 10^6$ K. We see that the amplitude of the electron density is primarily controlled by the background Eddington ratio $f_{\rm Edd}$, although the black hole mass does impact the profile amplitude in a more subtle manner. Given the relatively small dynamic range in black hole masses in our TDE sample ($\sim 1$ dex), the amplitude of $n_e$ will be primarily controlled by $f_{\rm Edd}$. 

Both the black hole mass and the background gas temperature $T_\infty$ impact the radius at which the density profile flattens, as is clearly visible.

\subsection{Can a super-Eddington  wind from a TDE escape the Bondi sphere?}\label{sec:windsec}
The number density of electrons close to the black hole is  well described by a
\begin{equation}
    n_e(r) = {\xi f_{\rm Edd} c^2 \over GM_\bullet m_p\kappa\eta }\, \left({rc^2 \over 2GM_\bullet }\right)^{-3/2},
\end{equation}
profile. What this means is that the amount of matter swept up by a super-Eddington wind with opening solid angle $\Omega$ is 
\begin{equation}
    M_{\rm swept}(r) = \Omega \int_0^r m_p n_e(r') r'^2\, {\rm d}r', 
\end{equation}
or (a more general expression is derived later)
\begin{equation}
    M_{\rm swept}(r) = {16\Omega\over 3} \, {\xi f_{\rm Edd} G^2 M_\bullet^2  \over \kappa\eta c^4 }   \, \left({c^2r\over 2GM_\bullet}\right)^{3/2} .
\end{equation}
A super-Eddington wind from a TDE will launch an outflow with mass  $M_{\rm out} = f_{\rm out} M_\star$, where $0<f_{\rm out}<1/2$ is the fraction of the incoming stellar mass $M_\star$ launched in the wind (only one half of the stellar debris remains bound to the black hole and has a chance of entering the disk, and therefore wind). The wind launched from a TDE disk will propagate through the Bondi sphere until it has swept up a mass $M_{\rm swept}(r_{\rm dec}) \simeq M_{\rm out}$, after which point it will be decelerated by the ambient medium. This means it can reach a maximum radius (before the outflow begins to decelerate) of
\begin{equation}
    r_{\rm dec} \approx {2GM_\bullet \over c^2}\, \left({3f_{\rm out}M_\star  c^4 \kappa \eta \over 16\Omega G^2 M_\bullet^2 \xi f_{\rm Edd}}\right)^{2/3} ,
\end{equation}
or in more convenient units
\begin{multline}
    r_{\rm dec} \approx 4 \times 10^{17}\, {\rm cm} \, \,\,\left({M_\star\over M_\odot}\right)^{2/3} \, \left({ \Omega \over 4\pi }\right)^{-2/3} \\ \, 
    \left({M_\bullet \over 10^7 M_\odot}\right)^{-1/3}
    \left({f_{\rm out} \over 10^{-2}}\right)^{2/3} \left({f_{\rm Edd} \over 10^{-4}}\right)^{-2/3}   .
\end{multline}
It appears to us unlikely that a super-Eddington wind launched from a typical TDE will contain  sufficient mass to propagate beyond the galactic center Bondi radius, as  
\begin{multline}
    {r_{\rm dec} \over r_b} \approx 2 \times 10^{-3}\,\,\ \left({M_\star\over M_\odot}\right)^{2/3} \, \left({ \Omega \over 4\pi }\right)^{-2/3} \left({M_\bullet \over 10^7 M_\odot}\right)^{-4/3} \\ \, 
    \left({f_{\rm out} \over 10^{-2}}\right)^{2/3} \left({f_{\rm Edd} \over 10^{-4}}\right)^{-2/3}\left({T_{\rm \infty}\over 10^5 \,{\rm K}}\right)   .
\end{multline}
In other words, the Bondi radius  is typically  $\sim 1-3$  orders of magnitude larger than the typical scale at which the outflow begins to decelerate and the radio emission begins to fade. One should therefore expect to see a clear signature of a characteristic $n_e\sim r^{-3/2}$ profile in TDE radio data, if the density of SMBH environments are shaped by Bondi flows. 

\begin{figure}
    \centering
    \includegraphics[width=0.95\linewidth]{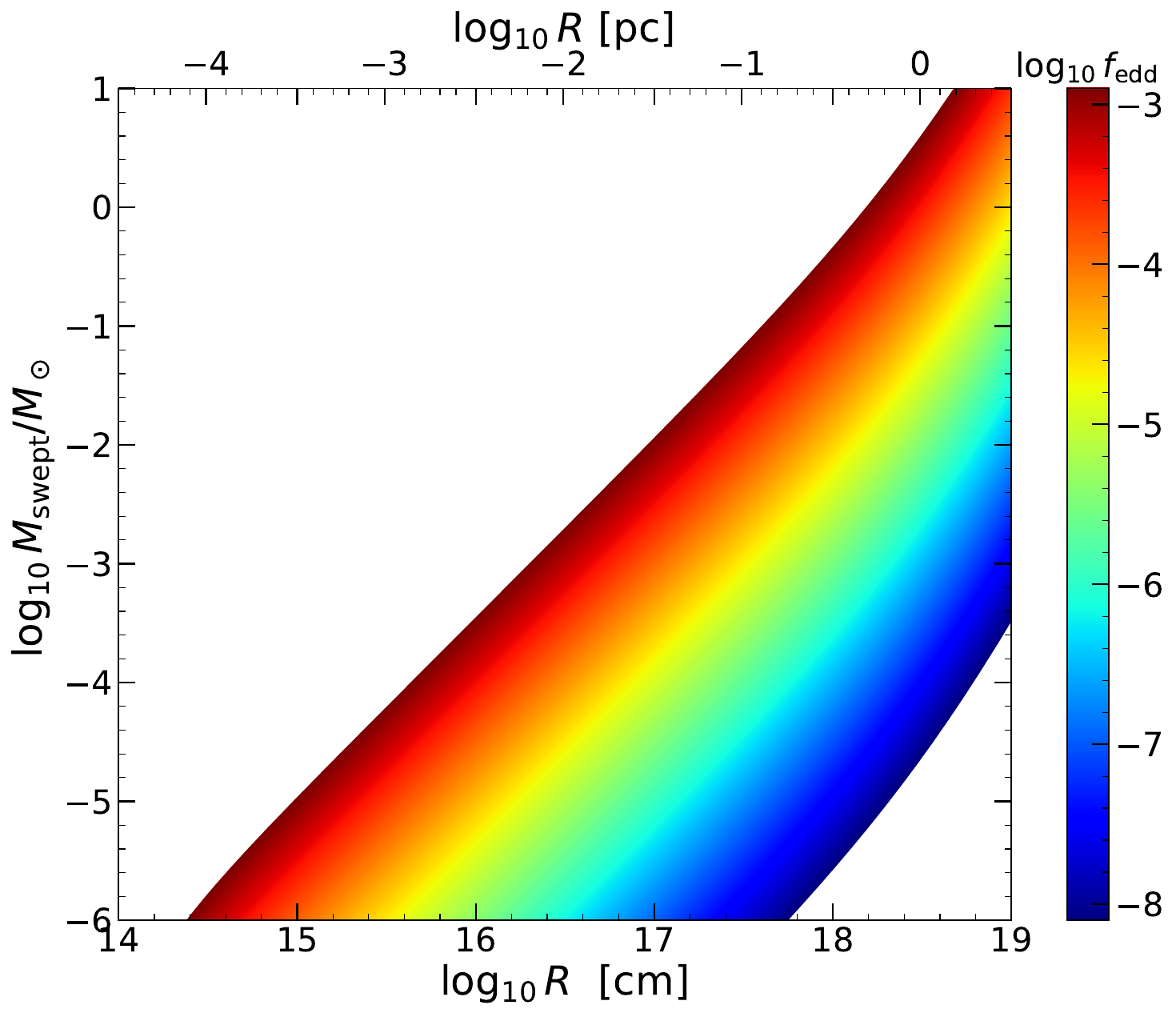}
    \includegraphics[width=0.95\linewidth]{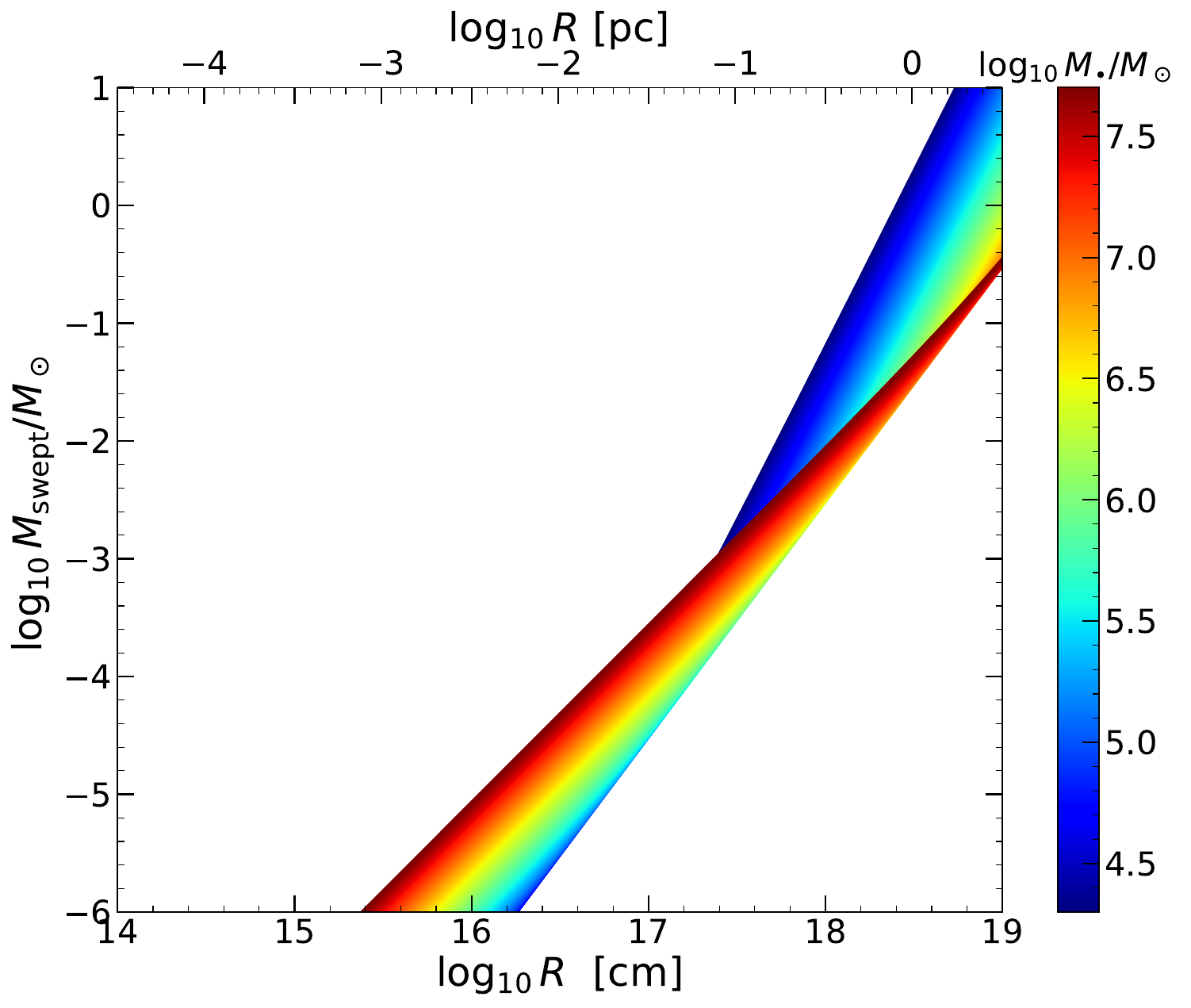}
    \includegraphics[width=0.95\linewidth]{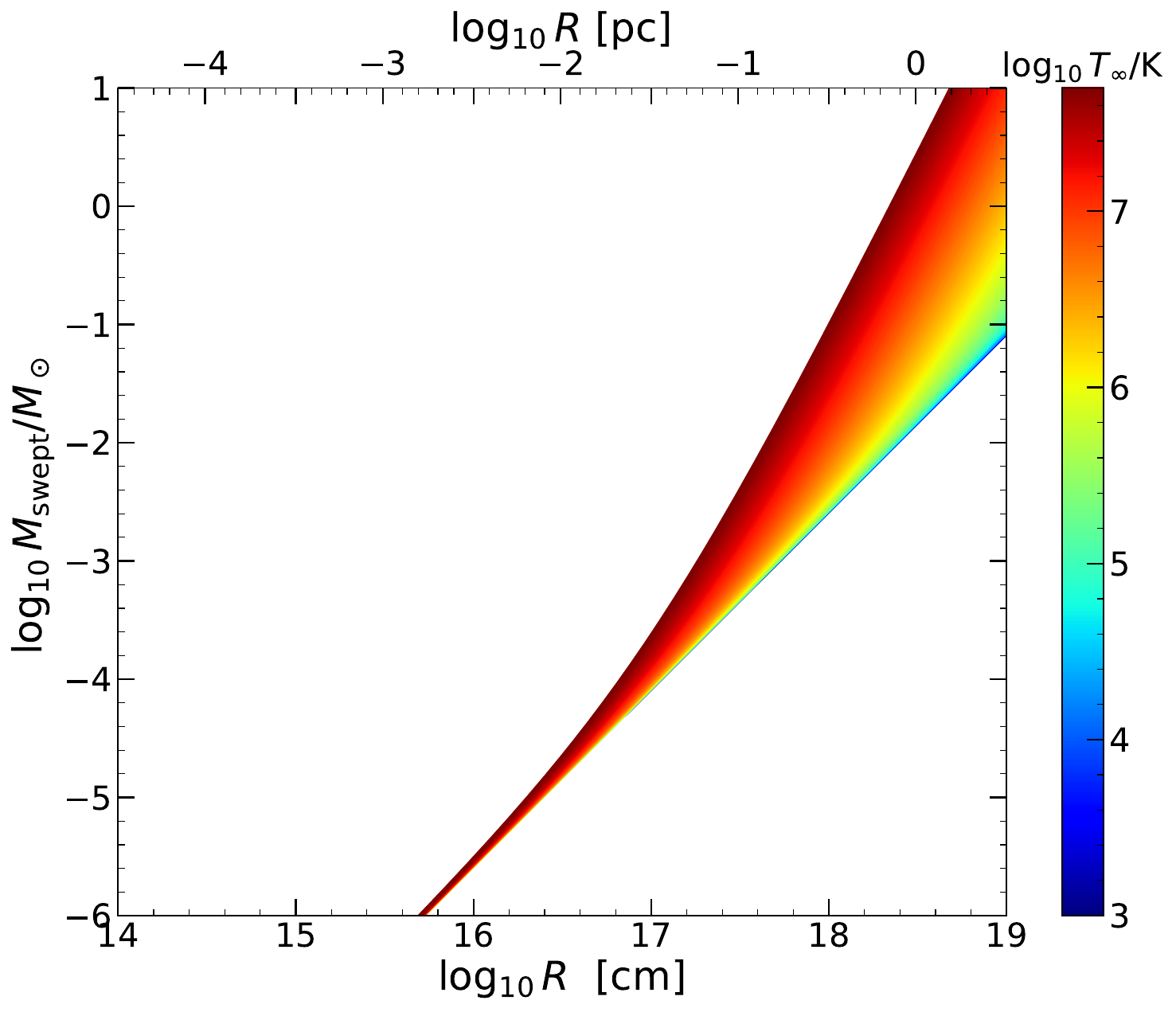}
    \caption{The mass swept up by spherical winds launched into Bondi profiles, as a function of the three parameters of interest (i) the background Bondi accretion rate in units of the Eddington  (top), (ii) the black hole mass in the TDE (middle) and (iii) the temperature of the gas far from the black hole (bottom). When not varied, the parameters take the  values $M_\bullet=10^{6.5}M_\odot$, $f_{\rm Edd} = 10^{-5}$ and $T_\infty = 10^6$ K.  }
    \label{fig:theory_bondi_mass}
\end{figure}

The time taken to sweep up this much matter then bounds the lifetime of the rising portion of a TDE radio light curve (with the radio light curve changing qualitatively during the decelerating phase),  by $t_{\rm rise} \approx r_{\rm dec} / \beta c$, or 
\begin{multline}\label{eq:trise}
    t_{\rm rise} \approx 4 \, {\rm yr} \, \,\, \, \left({\beta \over 0.1}\right)^{-1} \left({M_\star\over M_\odot}\right)^{2/3} \, \left({ \Omega \over 4\pi }\right)^{-2/3} \\ \, 
    \left({M_\bullet \over 10^7 M_\odot}\right)^{-1/3}
    \left({f_{\rm out} \over 10^{-2}}\right)^{2/3} \left({f_{\rm Edd} \over 10^{-4}}\right)^{-2/3}  .
\end{multline}
We see that, in principle, prompt (super-Eddington wind) radio light curves launched from TDEs could be extremely long lived. This is entirely in keeping with the observed population of TDEs, in which rising radio emission has been observed for hundreds of days in some events \citep[e.g.][]{Stein21,Goodwin22,Goodwin24}. 

In Figure \ref{fig:theory_bondi_mass} we show the theoretical radial profiles of the swept up mass, again as a function of the three free parameters of the theory.  When not specified, the parameters take the  values $M_\bullet=10^{6.5}M_\odot$, $f_{\rm Edd} = 10^{-5}$ and $T_\infty = 10^6$ K. Again, the background Eddington ratio $f_{\rm Edd}$ is the primary driver of changes in $M_{\rm swept}$. Breaks in the simple power-law growth of $M_{\rm swept}$ can be identified with Bondi profiles transitioning through their Bondi radius. We see that for $M_{\rm out} \sim 10^{-2} M_\star$, TDE radio outflows are unlikely to propagate much beyond $\sim 10^{18}$ cm before decelerating (although this value is sensitive to the opening angle of the outflow $\Omega$, taken to be $4\pi$ here, a perfectly reasonable value for a wind launched at super-Eddington rates). 

\section{Deriving observational constraints on  SMBH ambient density profiles}\label{sec:results}

Radio observations of TDEs with sufficient spectral coverage to resolve the evolution of the radio spectral peak over time provide strong constraints on the outflow radius and ambient density of the medium through which the outflow propagates \citep[e.g.][]{Alexander16,Goodwin22}. In this section we compile a sample of TDEs with constraining radio spectral observations and fit the  profiles to infer the observed distribution of the ambient circumnuclear electron density. 

\newpage
\subsection{Radio observations and modelling}
The radio spectrum of a TDE outflow and its subsequent evolution over time result from  ambient electrons being accelerated by a passing blastwave into a non-thermal distribution, $N(\gamma)\propto \gamma^{-p}$, where $\gamma$ is the electron Lorentz factor ($\gamma\geq \gamma_{\rm m}$, where $\gamma_{\rm m}$ is the minimum Lorentz factor) and $p$ is the synchrotron electron energy index. The resulting observed spectrum has characteristic break frequencies corresponding to the minimum frequency ($\nu_m$, corresponding to the synchrotron frequency of an electron with $\gamma=\gamma_m$), the self-absorption frequency ($\nu_a$), and the cooling frequency ($\nu_c$) \citep[e.g.][]{Rybicki79}. TDEs with well-constrained radio spectral observations are commonly observed to be in the regime $\nu_m < \nu_a < \nu_c$, where the radio spectral peak is associated with synchrotron self-absorption \citep[e.g.,][]{Goodwin23}. In this limit, the frequency at which the spectrum peaks $(\nu_a)$ and the flux density at that frequency $(F_{\nu_a})$ strongly constrain the physical source attributes, such as the emitting region size, magnetic field, and the density of the ambient medium which the shock propagates through. By measuring the peak of the synchrotron spectrum at multiple epochs, the outflow and ambient medium properties at different radii from the black hole can be directly constrained. 

Using the TDE sample and radio observations modelled in \citet{GoodwinMummery2026}, we select the 11 sources with prompt radio-emitting outflows that are consistent with being launched by a super-Eddington wind. Our sample of TDEs is listed in Table \ref{tab:TDEs}, including the number of radio spectral epochs available for each source, and properties of their host galaxies. In total, the dataset contains 90 unique measurements of ambient density, radius, and emitting-region mass. The radio spectral modelling is described in \citet{GoodwinMummery2026}, which provides constraints on the synchrotron spectrum peak flux density, peak frequency, and spectral index, from which $p$ (note $p>2$ on physical grounds) is inferred. We use the synchrotron spectral properties to infer the host galaxy ambient density, outflow radius, and mass in the outflow emitting region for each radio spectrum of each source.

\begin{table*}
    \centering
    \begin{tabular}{cccccc}
    Tidal disruption event & Redshift & log$_{10}M_{\rm{BH}}$\,(M$_{\odot}$) & $N_{\rm{epoch}}$ & Host Type & Ref. \\
    \hline
    \hline
ASASSN-14li &0.0206   & $6.24\pm0.14$ & 8 &  PSB &[1] \\
ASASSN-15oi & 0.0484 & $6.23\pm0.20$ & 5 & QBS & [2,3] \\
AT2019azh &0.02 & $6.51\pm0.17$ & 18 & PSB & [4] \\
AT2019dsg & 0.051 & $6.74\pm0.17$ & 7 & SF & [5] \\
eJ2344 &0.1 & $7.60\pm0.18$ & 10 & AGN ? & [6] \\
AT2019ahk & 0.0262& $6.78\pm0.15$ & 5 & PSB & [7] \\
AT2019qiz & 0.015& $6.22\pm0.17$ &6 & SF & [8]\\
AT2020opy &0.16& $6.69\pm0.42$ &3 & ? & [9]\\
AT2020vwl & 0.033 & $6.57\pm0.33$ & 6 & Q & [10,11] \\
AT2020zso &0.061& $6.43\pm0.32$& 10 & Seyfert II AGN & [12] \\
AT2021sdu &0.059& $6.75\pm0.37$& 11 & SF? &[12]\\
\hline
    \end{tabular}
    \caption{The 11 TDEs used in this work with prompt radio-emitting outflows launched by Super-Eddington accretion disk winds, as shown in paper 1. 
    $N_{\rm{epoch}}$ is the number of radio spectral epochs modelled. References for radio data: [1] \cite{Alexander16}, [2] \cite{Horesh21}, [3] \cite{Hajela2025}, [4] \cite{Goodwin22}, [5] \cite{Cendes_2021_dsg}, [6] \cite{Goodwin24}, [7] \cite{Christy2024}, [8] \cite{Alexander2025}, [9] \cite{Goodwin23b}, [10] \cite{Goodwin23}, [11] \cite{Goodwin25}, [12] \cite{Christy2025}. For AT2020vwl and ASASSN-15oi, a radio re-flaring episode was observed \citep{Goodwin25,Horesh21}, thought to correspond to a delayed radio jet switching on. For these two sources, we only include epochs that correspond to the first radio flare in our ambient density analysis, i.e. when the radio emission was dominated by a super-Eddington wind, although 11 epochs of each source are available.  Host galaxy types are Post-Starburst (PB), Quiescent Balmer Strong (QBS), Quiescent (Q), and Star-forming (SF) as described in \citet{French2016}. A ``?'' indicates the galaxy has not been reliably classified in the literature.}
    \label{tab:TDEs}
\end{table*}

The radius of the outflow is estimated from the observed radio spectrum peak flux density ($F_{\mathrm{\nu,a}}$), and the frequency at which the spectrum peaks ($\nu_{\mathrm{a}}$), following the derivation in \citet{BarniolDuran2013}. Explicitly, the radius is given by
\begin{equation}\label{eq:R_eq}
\begin{aligned}
\begin{split}
    R_{\mathrm{eq}} = 1\times10^{17} \,\,(21.8 (525^{(p-1)})^{\frac{1}{13+2p}}
    \chi_{\rm e}^{\frac{2-p}{13+2p}}
    F_{\mathrm{\nu,a}}^{\frac{6+p}{13+2p}}\\ \left(\frac{d}{10^{28}\,\rm{cm}}\right)^{\frac{2(p+6)}{13+2p}}\left(\frac{\nu_{\mathrm{a}}}{10\,\rm{GHz}}\right)^{-1}
    (1+z)^{-\frac{19+3p}{13+2p}}
    f_{\mathrm{A}}^{-\frac{5+p}{13+2p}}\\ f_{\mathrm{V}}^{-\frac{1}{13+2p}} 4^{\frac{1}{13+2p}} \xi^{\frac{1}{13+2p}}\quad\rm{cm}. 
\end{split}
\end{aligned}
\end{equation}
where $d$ is the luminosity distance to the source, $z$ is the redshift, $\chi_{\rm e} = \left({p-2)}/{(p-1}\right) \epsilon_{\rm e} {(m_{\rm p}}/{m_{\rm e}})$ where $m_{\rm e}$ is the electron mass and $m_{\rm p}$ is the proton mass, and $\xi = 1 + {1}/{\epsilon_{\rm e}}$. The factor $\epsilon_e$ is  the fraction of the thermal energy in the post-shocked medium which accelerates the electrons, and will be constrained as part of this work. 
This equipartition radius is appropriate for a source in which $\nu_m<\nu_a$, as is appropriate for non-relativistic TDE outflows, and consistent with the spectral shape of our sources. The factor $4^{\frac{1}{13+2p}}$ arises due to a correction to the isotropic number of radiating electrons in the non-relativistic limit. In the above expression $f_A$ and $f_V$ are geometric factors, given by $f_{\mathrm{A}}=A/(\pi R^2/\Gamma^2)$ and $f_{\mathrm{V}} = V/(\pi R^3/\Gamma^4)$, for an outflow with area, $A$, volume, $V$, and distance from the origin of the outflow, $R$. For a spherical outflow, $f_{\mathrm{A}}=1$ and $f_{\mathrm{V}}=4/3$, and for a conical outflow (mildly collimated with half-opening angle $30$\,degrees), $f_{\mathrm{A}}=0.13$ and $f_{\mathrm{V}}=1.15$.

The isotropic equivalent number of electrons is given by \citep{BarniolDuran2013}
\begin{equation}\label{eq:Ne}
\begin{aligned}
\begin{split}
    N_{\mathrm{e}} \approx 4\times10^{54}\, 
    F_{\mathrm{\nu,a}}^3 
    \left(\frac{d}{10^{28}\,\rm{cm}}\right)^{6} \left(\frac{\nu_{\mathrm{a}}}{10\,\rm{GHz}}\right)^{-5} (1+z)^{-8}\\
    f_{\mathrm{A}}^{-2} 
    \left(\frac{R}{10^{17}\,\rm{cm}}\right)^{-4} \left(\frac{\gamma_{\rm m}}{\gamma_{\rm a}}\right)^{1-p}\quad\rm{electrons},
\end{split}
\end{aligned}
\end{equation}
and $\gamma_m = 2$ is the minimum Lorentz factor of the accelerated electrons, while $\gamma_{\rm a}$ is given by
\begin{multline}
    \gamma_{\rm a} = 525\,  F_{\mathrm{\nu,a}} 
    \left(\frac{d}{10^{28}\,\rm{cm}}\right)^2 \left(\frac{\nu_{\mathrm{a}}}{10\,\rm{GHz}}\right)^{-2}  \\ (1+z)^{-3} f_A^{-1} \left(\frac{R}{10^{17}\,\rm{cm}}\right)^{-2} . 
\end{multline}
The ambient electron density is then inferred via 
\begin{equation}\label{eq:n}
n_{\mathrm{e}} = N_{\mathrm{e}}/V.
\end{equation}
An important result to note is that the inferred density {\it profile} is generally only weakly dependent on geometric factors, as 
\begin{align}
    \log(R) &\propto -\left({5+p\over 13+2p}\right) \log f_A -\left({1\over 13+2p}\right)\log f_V, \\
    \log(n_e) &\propto \left({10p - 3 + 2p^2\over 13+2p}\right)\log f_A - \left({11\over 13+2p}\right) \log f_V. 
\end{align}
While not immediately obvious, what occurs for typical $p \approx 2.5$ is that the inferred density drops for more collimated outflows, but the inferred radius increases, such that the inferred trend of density with radius is almost unchanged (see Fig. \ref{fig:geo_epse_comps} for an explicit example).

We additionally infer the equipartition energy from the observed peak flux density, peak frequency, and $p$ of the radio spectra using the following relation from \citet{BarniolDuran2013}. In the Newtonian regime, the equipartition energy, corresponding to the minimum total energy in the observed region when $\nu_{\rm a} > \nu_{\rm m}$, is given by 
\begin{equation}\label{eq:E_eq}
\begin{aligned}
\begin{split}
    E_{\mathrm{eq}} = 1.3\times10^{48}\, 21.8^{-\frac{2(p+1)}{13+2p}}
    (525^{(p-1)} \chi_{\rm e}^{(2-p)})^{\frac{11}{13+2p}}\\
    F_{\mathrm{\nu,a}}^{\frac{14+3p}{13+2p}} \left(\frac{d}{10^{28}\,\rm{cm}}\right)^{\frac{2(3p+14)}{13+2p}} \left(\frac{\nu_{\mathrm{a}}}{10\,\rm{GHz}}\right)^{-1} (1+z)^{\frac{-27+5p}{13+2p}}\\
    f_{\mathrm{A}}^{-\frac{3(p+1)}{13+2p}}
    f_{\mathrm{V}}^{\frac{2(p+1)}{13+2p}} 4^{\frac{11}{13+2p}} \xi^{\frac{11}{13+2p}}\quad\rm{erg}. 
    \end{split}
\end{aligned}
\end{equation}
From the equipartition energy, we infer the  mass in the emitting region of the outflow via
\begin{equation}\label{eq:Mej}
\begin{aligned}
    M_{\mathrm{\rm{out}}} = \frac{2E}{\beta^2 c^2},
\end{aligned}
\end{equation}
where the velocity, $\beta$, is inferred via
\begin{equation}\label{eq:vej}
    t = \frac{R_{\rm eq} (1-\beta)(1+z)}{\beta c},
\end{equation}
where $t$ is time since outflow launch in the observer frame. 

\begin{figure}
    \centering
    \includegraphics[width=0.99\linewidth]{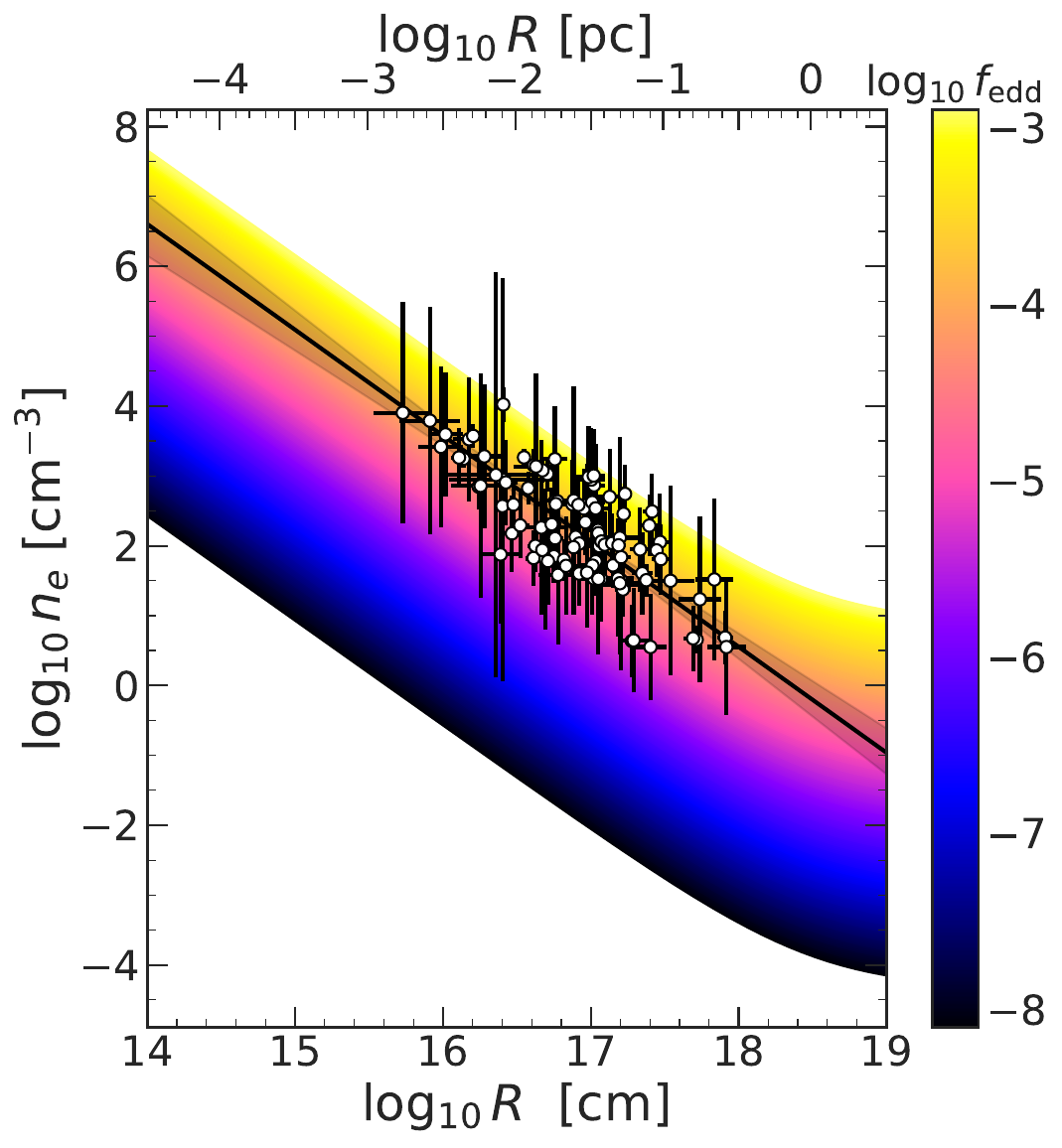}
    \includegraphics[width=0.99\linewidth]{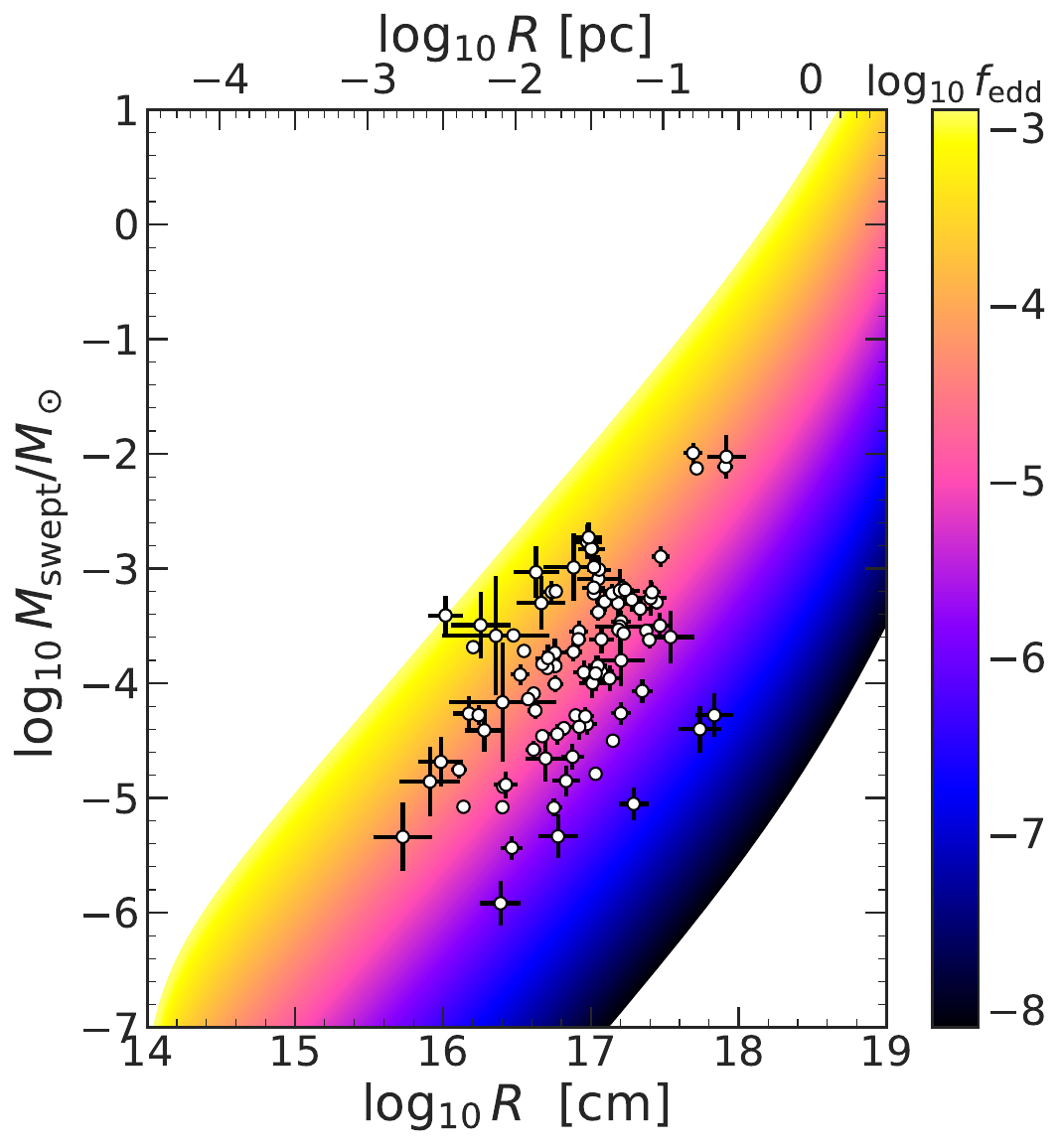}
    \caption{\textit{Top:} The ambient density with radius.  \textit{Bottom:} The mass swept up by the outflow with radius. Both profiles assume $\epsilon_e = 0.1$ and a spherical geometry, assumptions which will be improved upon later in this work. These profiles are shown only to highlight the broad scale of the parameters. In both panels the observed TDE population is plotted (white points with black error bars) over the theoretical density profile for Bondi accretion with $T_\infty = 10^6$ K, $\log_{10}M_\bullet/M_\odot=6.5$. The colour shading indicates different $f_{\rm{Edd}}$, the fraction of the Eddington accretion rate of the Bondi accretion rate.}
    \label{fig:theory+data}
\end{figure}

As an initial view of the TDE inferred circumnuclear density profiles, in Figure \ref{fig:theory+data} we plot the ambient density at radii $R$ (top panel) as well as the  mass in the outflow (bottom panel) for the combined TDE sample ($N=90$). These profiles make the standard assumption that $\epsilon_e = 0.1$, and that the outflow is spherical. We shall improve upon these assumptions in later sections.  We immediately note that the density profiles appear remarkably similar to the expectations of a Bondi accretion profile, warranting further detaield analysis.

\subsection{Naive profile fitting }\label{sec:ne_fits}
We shall show in later sections that by performing a combined fit of the swept up mass profile and the density profile, one can constrain both the  background density profile {\it and} the shock microphysics parameter $\epsilon_e$. The reason this second fact is so important can be highlighted  by fitting {\it only} to the inferred density profiles (for which one cannot uniquely constrain $\epsilon_e$) and examining the impact of different governing assumptions of the geometry and shock microphysics. 

Therefore, as an initial exercise, we fit the observed density profile of all TDEs combined with a simple power-law density profile $n_e \propto R^{k}$, where $k$ the slope of the density profile. The equation we fit takes the form 
\begin{equation}
    \log_{10} n_e = k \log_{10}\frac{R}{R_0} + A.
\end{equation}
We fix $R_0$ at $10^{16}$\,cm, and then fit for the values of $k$ and $A = \log_{10}n_0$ (where $n_0$ is the value of $n_e$ at $R =R_0$). 

We use a Python implementation of a Monte Carlo Markov Chain  (MCMC) algorithm, \texttt{emcee} \citep{EMCEE} to fit the data assuming flat priors on each parameter. We define a likelihood function such that
\begin{multline}
\log {\cal L} = - {1\over 2}\sum_{i=1}^N {(k \log_{10}(R_i/R_0) + A - \log_{10}(n_{e, i}))^2\over \delta \log_{10}n_{e,i}^2 + \sigma^2}\\ + \log(\delta \log_{10}n_{e,i}^2 + \sigma^2),
\end{multline}
where $\sigma$ is an intrinsic scatter (in log space), which we let vary freely, and $\delta \log_{10} n_{e, i}$ is the propagated measurement uncertainty. We assume uniform priors, allowing the fit parameters to vary in the range $0 < A < 10 $,  $-3 < k < 0$, and $ 0 < \sigma < 1$.
We run the chain for 1000 steps with 100 walkers. We extract the 50th percentile and report $1\sigma$ errors as the 16th and 84th percentiles of the posterior distributions. The best-fit results for the two different geometries and two different equipartition fractions examined in this work are listed in Table \ref{tab:fit_results}.

\begin{table}[]
    \centering
    \begin{tabular}{ll|lll}
         Geometry & $\epsilon_e$ & $k$ & $A$ & $\sigma$\\ 
         \hline

        Spherical & 0.1 & $-1.52^{+0.15}_{-0.14}$ & $3.58^{+0.14}_{-0.14}$ & $0.31^{+0.06}_{-0.05}$ \\
        Conical & 0.1& $-1.49^{+0.15}_{-0.15}$ & $3.24^{+0.19}_{-0.20}$ & $0.34^{+0.06}_{-0.06}$ \\
        Spherical & $5\times10^{-4}$& $-1.74^{+0.11}_{-0.11}$ & $1.73^{+0.12}_{-0.13}$ & $0.13^{+0.06}_{-0.06}$\\
        \hline
    \end{tabular}
    \caption{Naive MCMC fit results for the simple profile $\log_{10} n_e = k \log_{10}R/R_0 + A$ to the combined TDE sample, where $R_0=10^{16}$\,cm. Note that $A$ has dimensions of $\log_{10}\, {\rm cm}^{-3}$. Clearly one cannot constrain the background density profile from fits to $n_e$ alone, as the unknown shock microphysics parameter $\epsilon_e$ has a strong impact (while outflow geometry is unimportant).  }
    \label{tab:fit_results}
\end{table}

Table \ref{tab:fit_results} highlights a key point, namely that  one cannot constrain the background density profile from fits to $n_e$ alone, as the unknown shock microphysics parameter $\epsilon_e$ has a strong impact on the inferred density amplitude (while outflow geometry is unimportant). The density {\it index} $k$ is broadly independent of both shock microphysics and geometry (they are consistent at $1\sigma$ in the above).

\begin{figure}
    \centering
    \includegraphics[width=\linewidth]{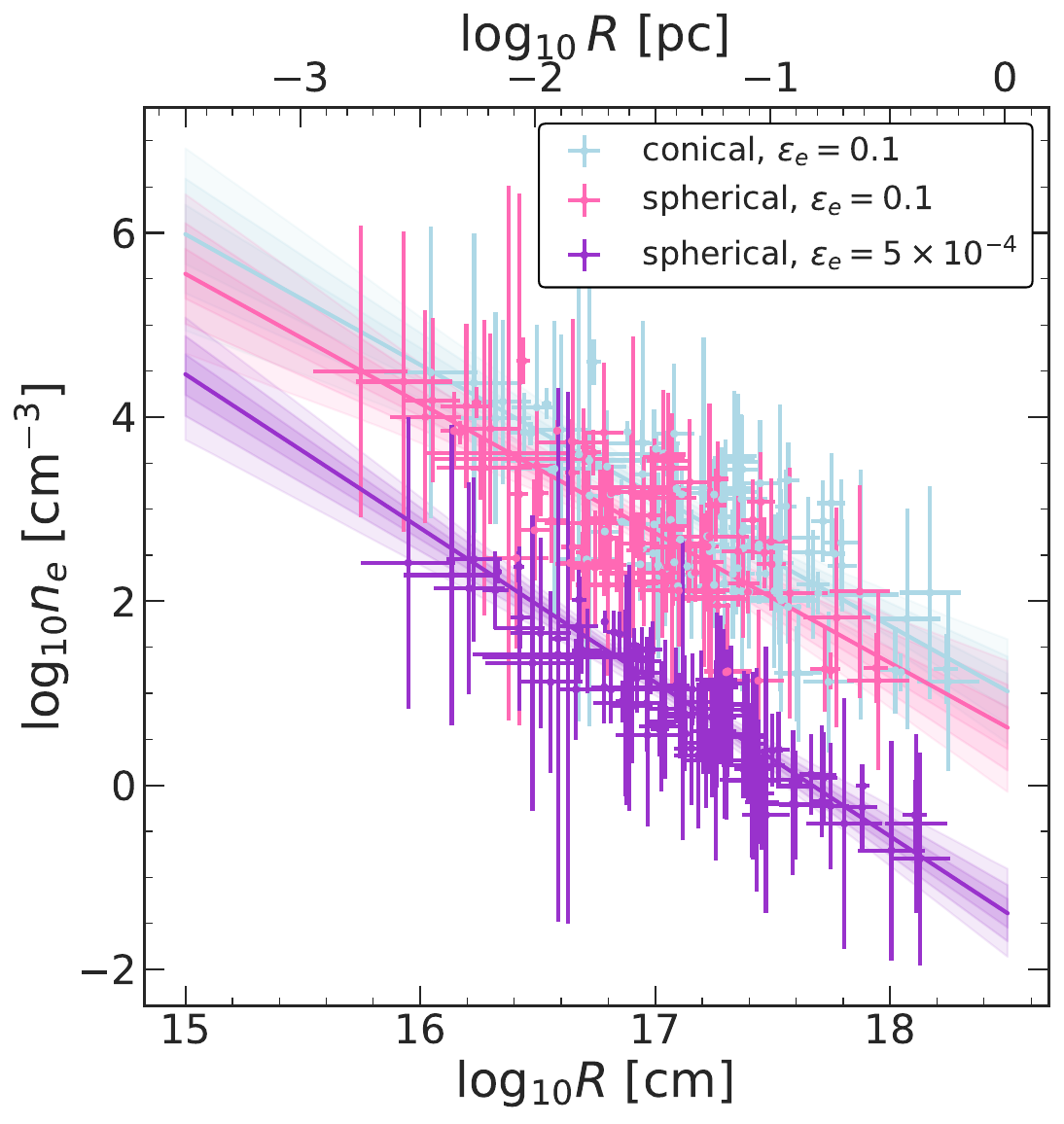}
    \caption{The different fits for $n_e(R)$ under different geometry and equipartition fraction assumptions. We compare conical geometry (blue), spherical geometry (pink), and reduced equipartition fraction of $\epsilon_e=5\times10^{-4}$ (purple). Evidently, changes in geometry assumptions do not affect the measured density profile, but changes in the relative equipartition fraction may. This motivates developing a framework where $\epsilon_e$ is constrained in parallel with the density profile. }
    \label{fig:geo_epse_comps}
\end{figure}

To see this visually, in Figure \ref{fig:geo_epse_comps} we plot a comparison between the equipartition radius and ambient density for a conical or spherical outflow and for a $\epsilon_e=5\times10^{-4}$ compared to the typically assumed $\epsilon_e=0.1$. As discussed, the changes in geometry have no statistically significant affect on the profiles, with collimated geometries shifting points to larger $R$ and smaller $n_e$ on  broadly the same profile. However, the assumed electron energy fraction has a greater effect and shifts the inferred amplitude of the background density (although the fit value of $k$ agrees within 1$\sigma$ errors). Therefore, the relative equipartition fraction within the shock would alter the inferred Bondi accretion efficiency, if it was let to freely vary. 

In the absence of further constraints, we could measure a density amplitude $n_0$ (i.e., an Eddington fraction $f_{\rm Edd}$ in the Bondi case) up to a reasonably strong function of the shock microphysics ($\epsilon_e$). This dependence can be traced through, and we find that in general one would find (for a given choice of $\epsilon_e$) the constraints 
\begin{equation}
    n_0( \epsilon_e) = n_{0, -1} \times  \Psi(\epsilon_e)/\Psi_{-1},
\end{equation}
where 
\begin{equation}
    \Psi( \epsilon_e) = \epsilon_e^{(5+2p)(p-2)/(13+2p)} (1 + 1/\epsilon_e)^{-(5+2p)/(13+2p)},
\end{equation}
and $n_{0, -1}, \Psi_{-1}$ are the respective values for $\epsilon_e = 0.1$. 

Clearly, it is essential to also constrain the parameter $\epsilon_e$ from the data. 

\subsection{Joint mass-density fits}\label{jointmassdens}
We do, however, have more information in TDE radio spectra than we have so far utilized. In addition to the number density of electrons, we also know the energy (or mass swept up) which is in each of the outflows, and a consistent solution satisfies {\it both} the electron number density and the energy constraints. 

The mass swept up by an outflow with opening solid angle $\Omega$ moving through a phenomenological density profile $n_e(R) \propto R^k$
\begin{equation}
    M_{\rm swept}(r) =  \Omega \int_0^r m_p n_0 r'^2 \left({r'\over r_0}\right)^k\, {\rm d}r' .
\end{equation}
This integral is trivial 
\begin{equation}
    M_{\rm swept}(r) = {\Omega \over 3+k} m_p n_0 r_0^3 \left({r\over r_0}\right)^{3+k},
\end{equation}
where $k > -3$ to converge at the lower limit. As we have already demonstrated that geometry has minimal impact on the inferred density profile, we shall henceforth fix $\Omega = 4\pi$. 

Then, if one specifies the parameters 
\begin{equation}
    \mathbf{p} = (A, k, \epsilon_e),
\end{equation}
one can jointly fit to both $n_{e, i}(R_i)$ and $M_{{\rm out}, i}(R_i)$ by maximizing the likelihood 
\begin{equation}
    \log {\cal L} = \log {\cal L}_n + \log {\cal L}_M, 
\end{equation}
where 
\begin{multline}
\log {\cal L}_n = - {1\over 2}\sum_{i=1}^N {(\log_{10}n_e(R_i | \mathbf{p})- \log_{10}(n_{e, i}))^2\over \delta \log_{10}n_{e,i}^2 + \sigma_n^2}\\ + \log(\delta \log_{10}n_{e,i}^2 + \sigma_n^2),
\end{multline}
and 
\begin{multline}
\log {\cal L}_M = - {1\over 2}\sum_{i=1}^N {(\log_{10}M_{\rm swept}(R_i | \mathbf{p})- \log_{10}(M_{{\rm out}, i}))^2\over \delta \log_{10}M_{{\rm out},i}^2 + \sigma_M^2}\\ + \log(\delta \log_{10}M_{{\rm out},i}^2 + \sigma_M^2).
\end{multline}
This allows us to simultaneously constrain $\epsilon_e$ from the data, breaking the degeneracy observed in the naive fits performed earlier. We provide distinct intrinsic scatters in our likelihoods ($\sigma_n$ and $\sigma_M$), which we allow to vary as part of the fitting process, to account for the different measurement systematics that can impact the inference of $n_e$ and $M_{\rm out}$ from radio data.

\subsection{Individual fits}
Now that we are also seeking to constrain $\epsilon_e$ from the data, we move to an individual source framework, as $\epsilon_e$ will likely vary source-by-source \citep[or at least could in principle, e.g.][]{Park2015}. Previous analyses of individual events have also found varying ambient density profiles, with some TDEs, such as AT2019azh, showing evidence for flatter density profiles \citep[$k\approx-0.5$][]{Burn2025}, and others showing evidence for steeper density profiles such as AT2019dsg \citep[$k\approx-2$][]{Cendes_2021_dsg}. In this section we jointly fit the $n_e(R)$ and $M_{\rm swept}(R)$ profile for each individual TDE in order to search for any correlation between $k$ and other system properties such as black hole mass. 

\begin{figure*}
    \centering
    \includegraphics[width=0.31\linewidth]{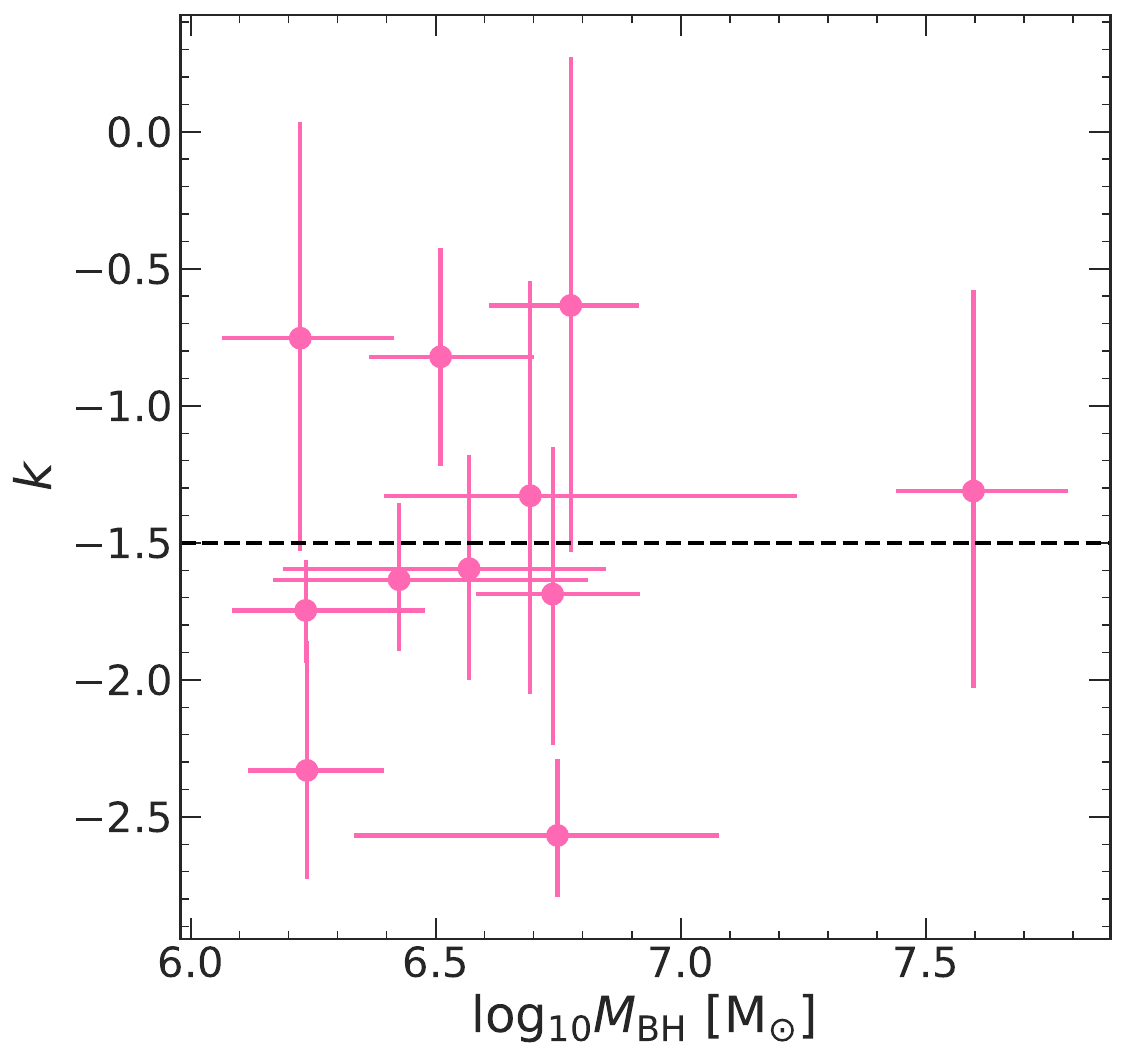}
    \includegraphics[width=0.31\linewidth]{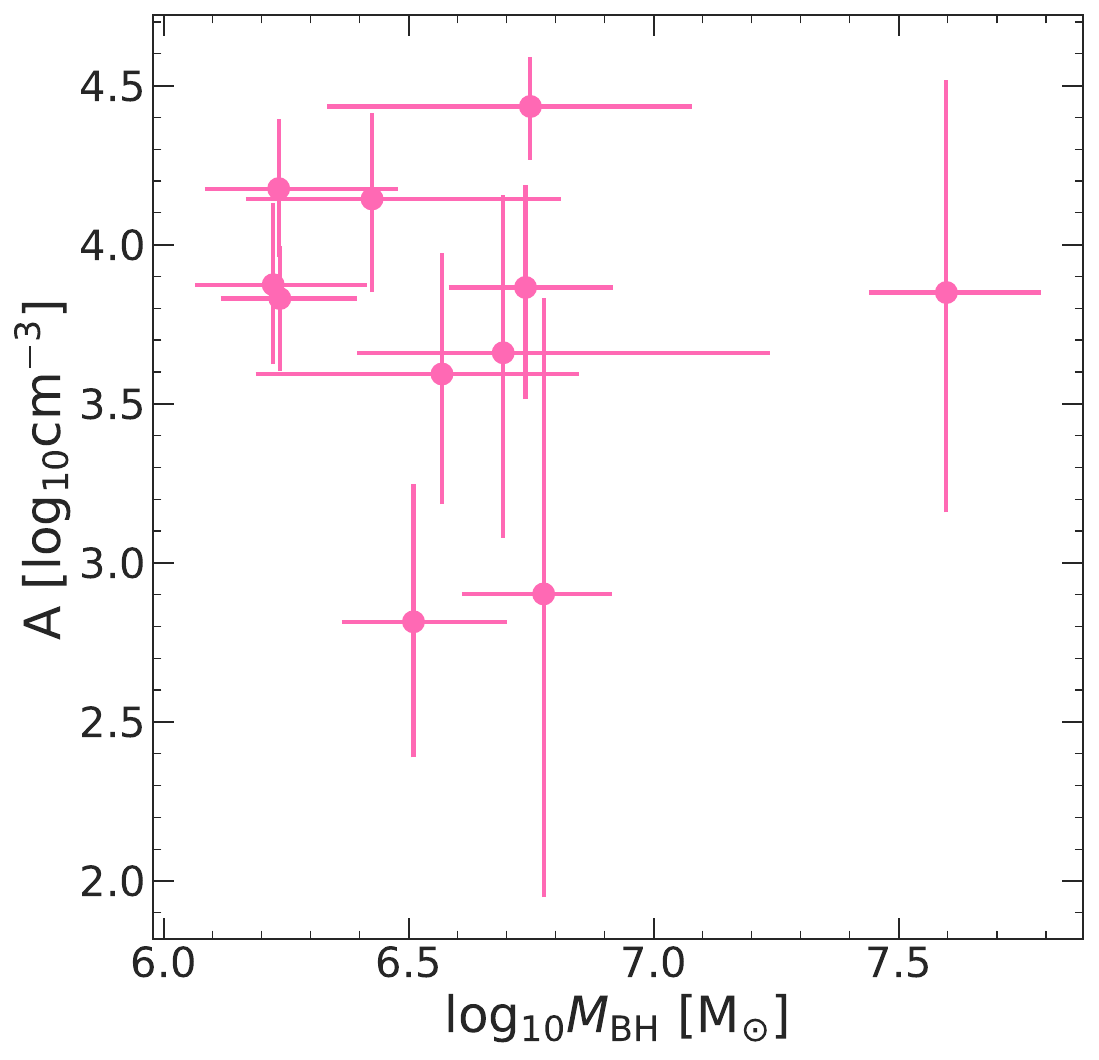}
    \includegraphics[width=0.33\linewidth]{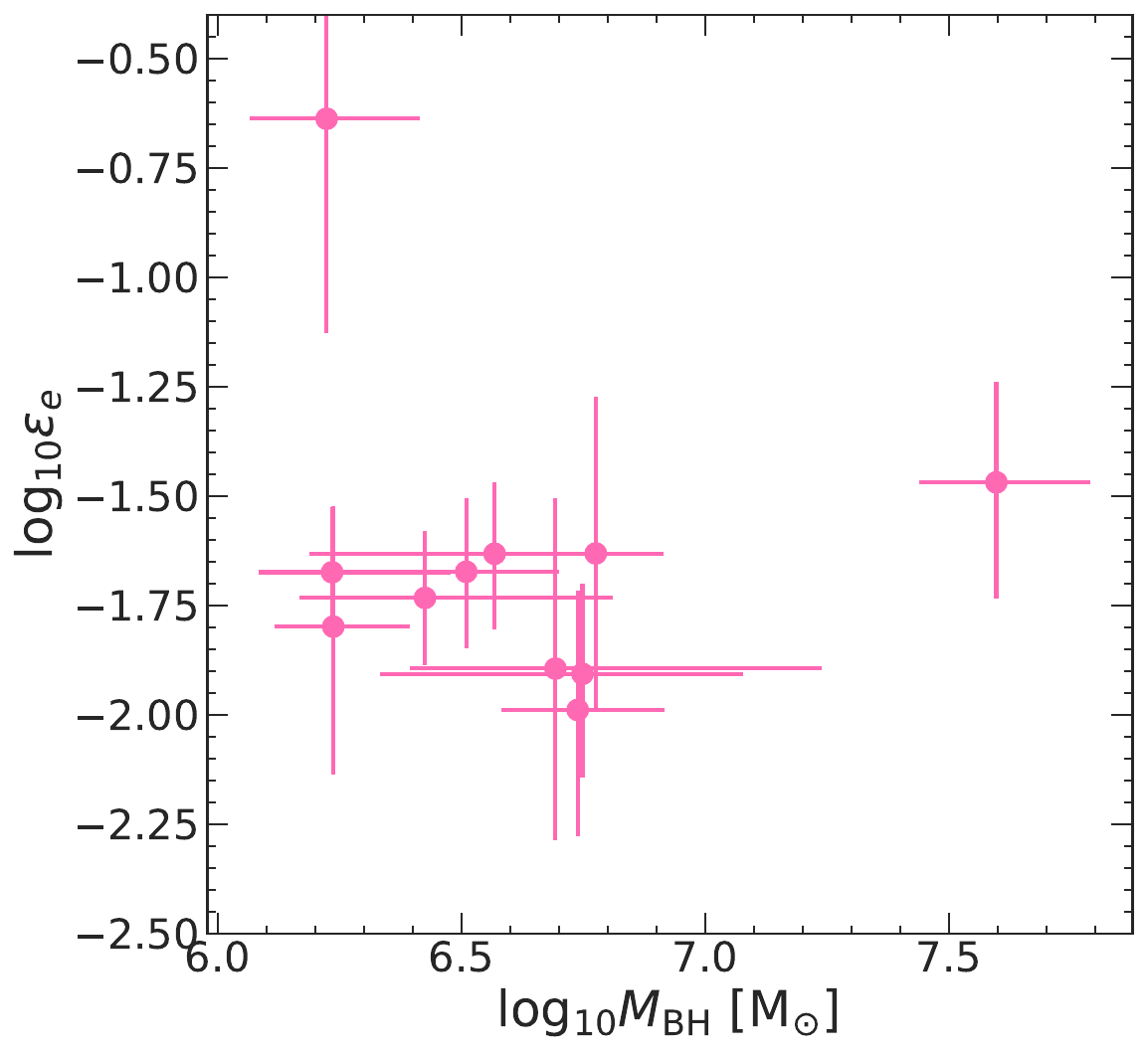}
    \includegraphics[width=0.3\linewidth]{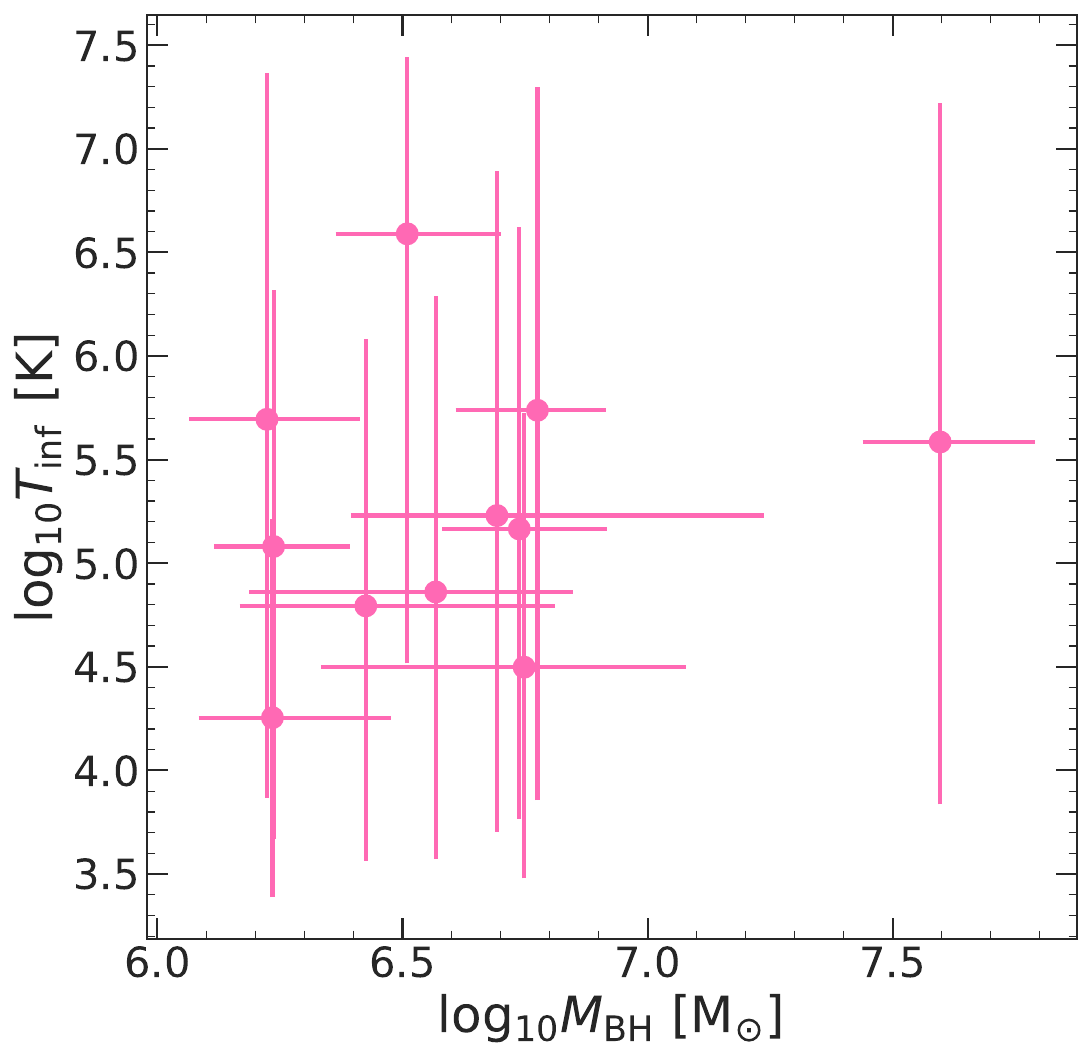}
    \includegraphics[width=0.32\linewidth]{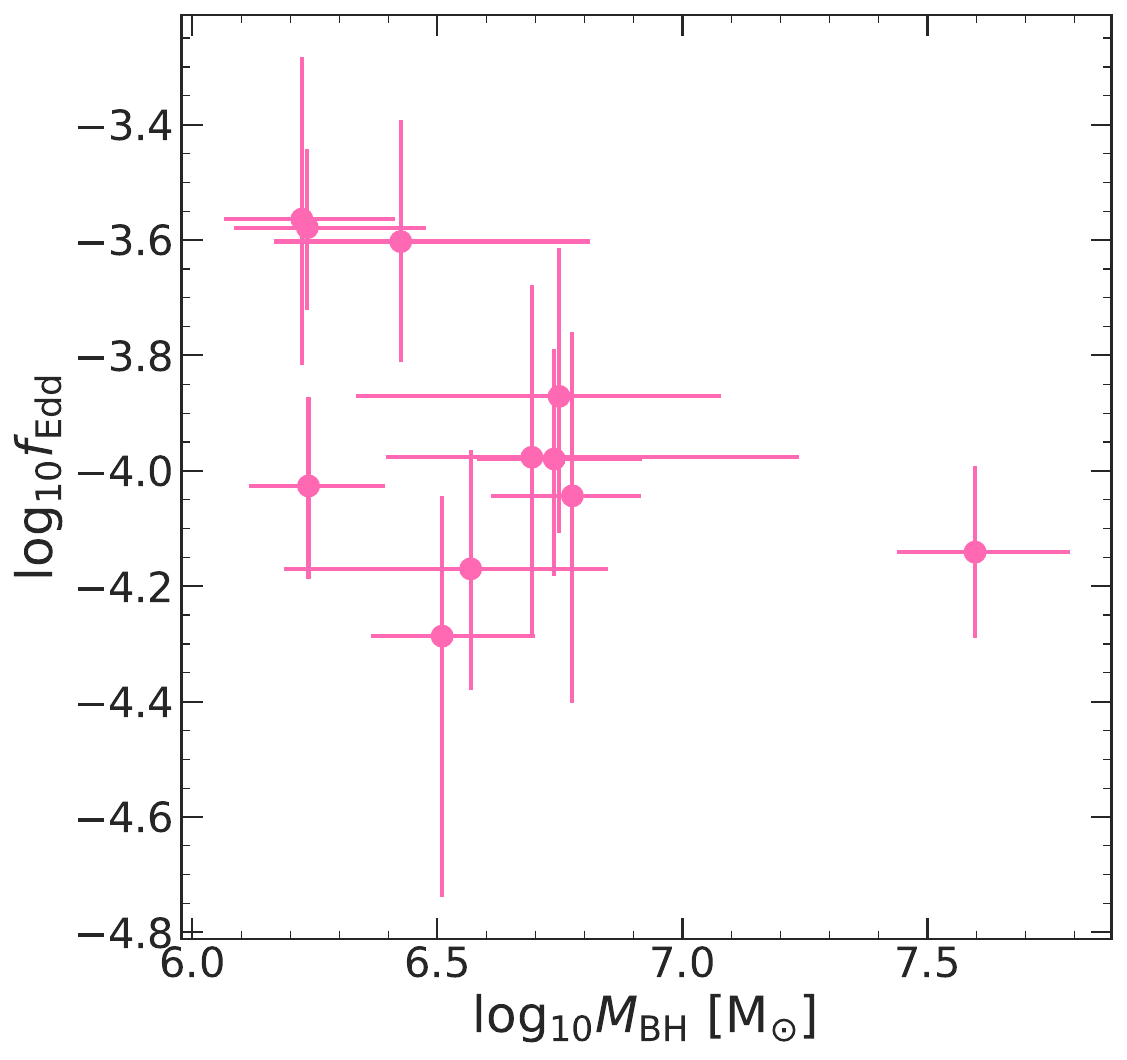}
    \includegraphics[width=0.32\linewidth]{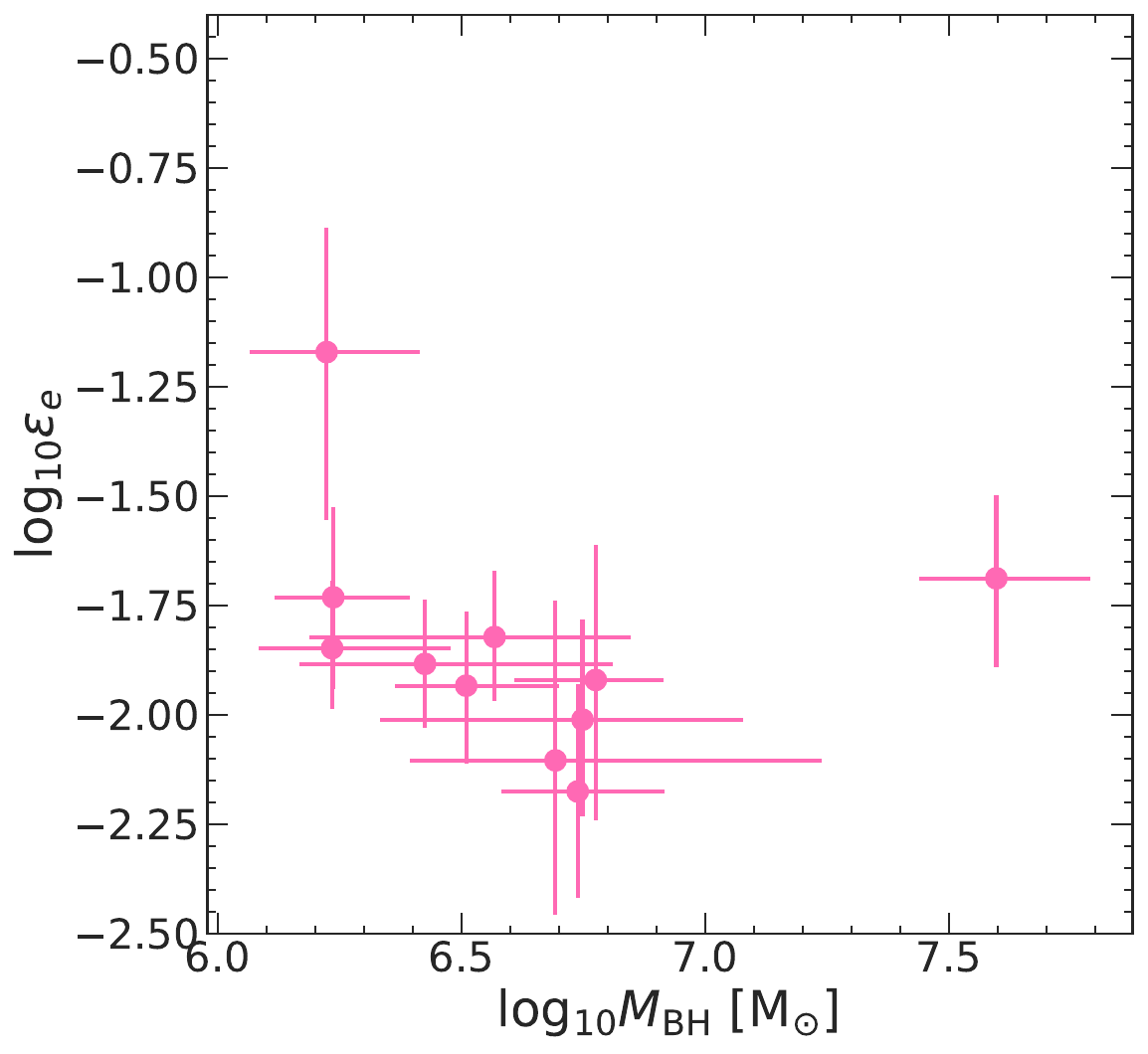}
    \caption{\textit{Top row:} The individual joint mass and density TDE fits for free $k$, $A$, and $\epsilon_e$ plotted against black hole mass, where $\log_{10} n_e = k \log_{10}R/R_0 + A$ and $R_0=10^{16}$\,cm. \textit{Bottom row:} The individual TDE Bondi profile fits for $f_{\rm{Edd}}$, $T_{\inf}$, and $\epsilon_e$ plotted against black hole mass. We find no apparent correlation between black hole mass and either of the fit parameters, although there are large uncertainties for many of the individual TDE fits due to low numbers of epochs of observations.}
    \label{fig:ksmass}
\end{figure*}

We follow the joint fitting procedure introduced above. The individual TDE fits for $k$, $A$, and $\epsilon_e$ for the spherical geometry are listed in Table \ref{tab:ind_fit_results}. In Figure \ref{fig:ksmass} we plot the measured $k$, $A$ and $\log_{10}\epsilon_e$ for each TDE against black hole mass. The uncertainty in $k$ is large when each event is fitted individually, but we find no indication of a dependence of any of the fit parameters on black hole mass. 

\section{Constraining the Bondi accretion rate}\label{sec:bondi_const}
In the previous section we demonstrated that the TDE population analysed in this work all show remarkably similar ambient density distributions around the SMBH, with the majority closely following the canonical $n_e\propto R^{-3/2}$ as expected for a Bondi accretion flow. 

If we make the assumption that the circumnuclear density distribution is indeed a Bondi profile, then we can constrain its governing parameters ($f_{\rm Edd}, T_\infty$) by fitting instead  the solutions of Equation \ref{eq:bondi_full} to the data, rather than a purely phenomenological power law profile. 

This fit in principle is underconstrained, as both the black hole mass and $f_{\rm Edd}$ impact the density normalisation. However, for these events the black hole masses are already constrained by the disk modelling described in \citet{GoodwinMummery2026}, which we take as a  prior probability distribution for the purpose of fitting the density profiles of each source. The assumed black hole masses are listed in Table \ref{tab:TDEs}. Even if we were to assume simple ``TDE priors'' of $M_\bullet \sim 10^6-10^{7.5} M_\odot$ (i.e., the broad and uninformative range expected for these transients assuming no additional multiwavelength information) our results would not change significantly.

To use our  joint fitting analysis, we model the background density field by the Bondi profiles derived above, and compute 
\begin{equation}
    M_{\rm swept}(r) = m_p \Omega \int_0^r n_e(r') \, r'^2\, {\rm d}r',
\end{equation}
or for our Bondi profiles
\begin{equation}
    M_{\rm swept}(r) = {f_{\rm Edd}\Omega r_b^3\over 162GM_\bullet\kappa\eta c}  \left({k_BT_\infty \over \mu m_p}\right)^{3/2}  F\left({3r\over r_b}\right), 
\end{equation}
where the function
\begin{equation}
    F(x) = \sqrt{x(x+1)} (2x+3)(4x+1) - 3\tanh^{-1}\left(\sqrt{x\over 1+x}\,\right),
\end{equation}
comes from performing the integral of the Bondi profile over radius. Note that for small $x\ll1$ we have the approximate result $F(x) \approx 16 x^{3/2}$, and the above reverts to the approximation used in the earlier section. Given the lack of sensitivity to changes in geometry of the function $n_e(R)$ (Figure \ref{fig:geo_epse_comps}), we assume spherical geometry $\Omega = 4\pi$\footnote{We also performed fits allowing the geometry to be a conical outflow with general opening half-angle $\phi$, but found consistent results for $f_{\rm Edd}$ and generally unconstrained angles which wanted to be large $\phi \gtrsim 60^\circ$, and do not report these results here.}. 

Then, if one specifies the parameters 
\begin{equation}
    \mathbf{p} = (f_{\rm Edd}, T_\infty, M_\bullet, \epsilon_e),
\end{equation}
one can jointly fit to both $n_{e, i}(R_i)$ and $M_{{\rm out}, i}(R_i)$ 
by following the same fitting approach outlined in Section \ref{jointmassdens}, treating each TDE in the sample individually. We implement log uniform priors and allow the parameters to vary in the range $-5<\log_{10}\epsilon_e<-0.7$, $-10<$log\,$_{10}f_{\rm{Edd}}<0$, and $3 <$log\,$_{10}T_\infty<8$. For the black hole mass we use the posterior distribution obtained in \citet{GoodwinMummery2026} for each source as a prior. 

\begin{figure}[htbp]
\centering

\begin{tikzpicture}[
    font=\small,
    flowstep/.style={
        rectangle split,
        rectangle split parts=2,
        draw,
        rounded corners,
        thick,
        align=center,
        text width=0.78\linewidth,
        inner sep=6pt
    },
    flowarrow/.style={-Latex, thick}
]

\node[flowstep] (s1) {
    \textbf{1. Fit observed radio spectra at time $t$}\\[3pt]
    Constrain: $F_p$, $\nu_p$, $p$
\nodepart{second}
    \includegraphics[width=\linewidth]{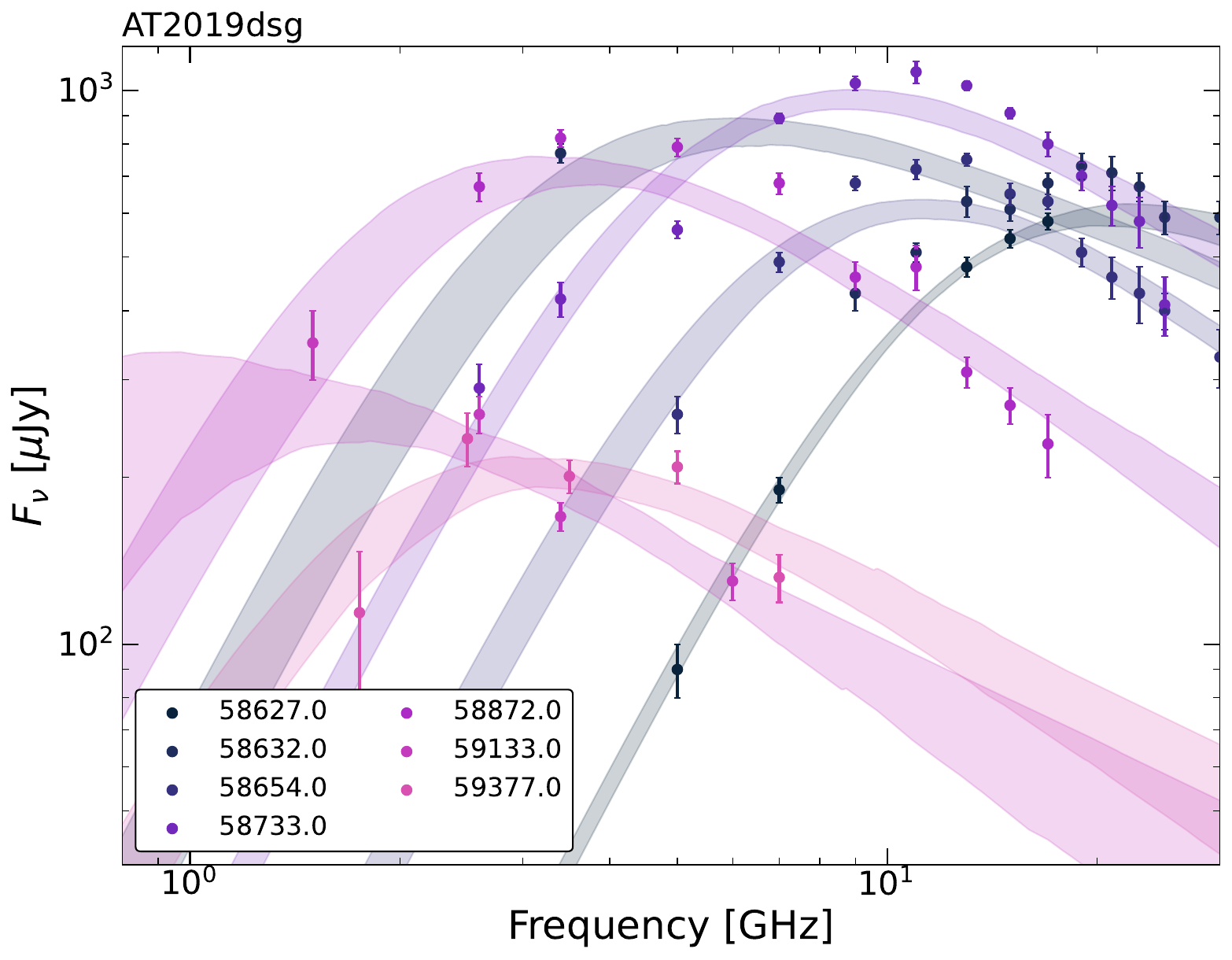}
};

\node[flowstep, below=1.cm of s1] (s2) {
    \textbf{2. Fit $n_e(r)$ and $M_{\rm swept}(r)$ simultaneously}\\[3pt]
    Output:  best-fit profiles and uncertainties
\nodepart{second}
    \includegraphics[width=\linewidth]{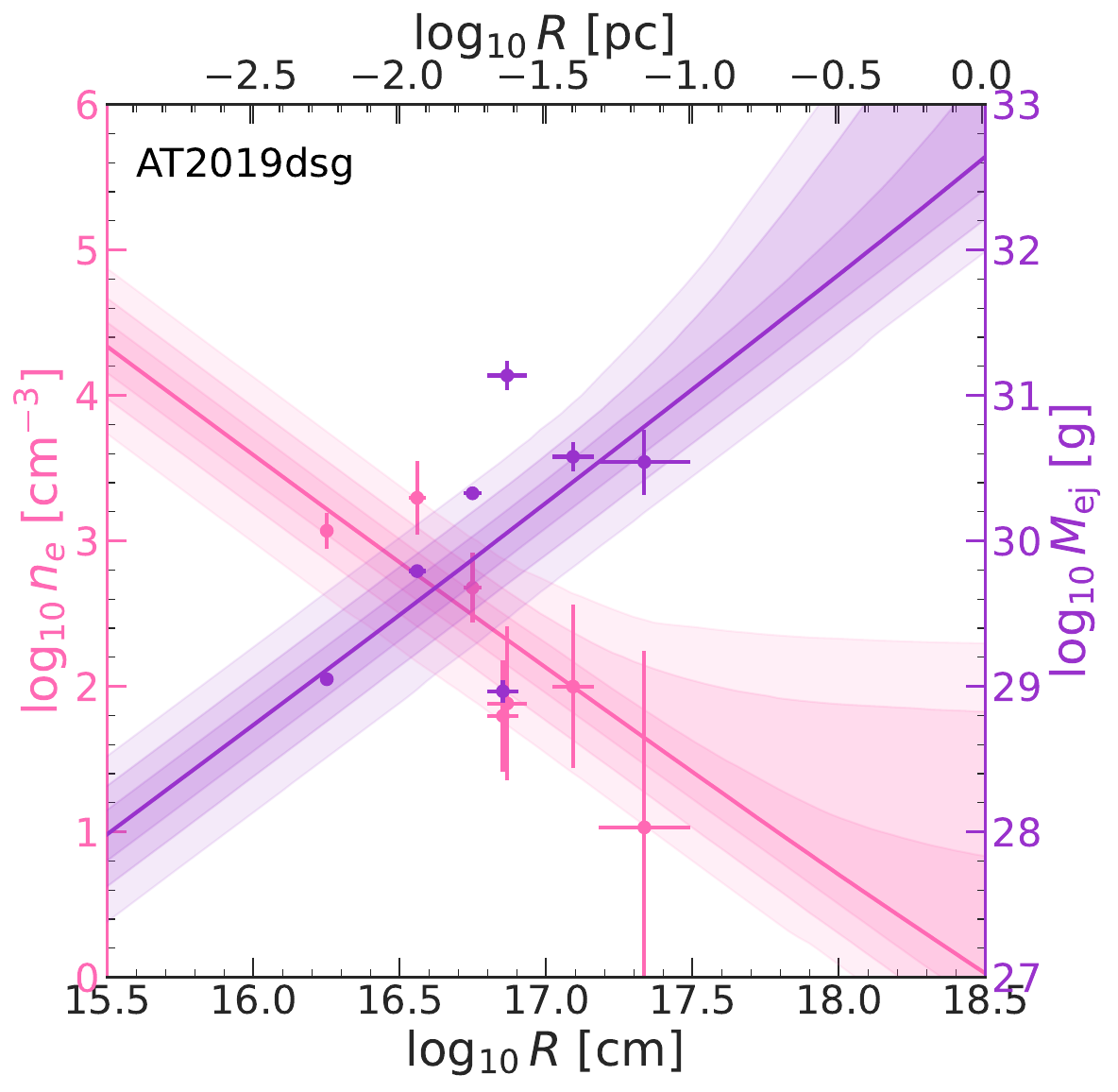}
};

\node[flowstep, below=1.cm of s2] (s3) {
    \textbf{3. Constrain: $f_{\rm Edd}$, $T_\infty$, $\epsilon_e$}\\[3pt]
    
\nodepart{second}
    \includegraphics[width=0.7\linewidth]{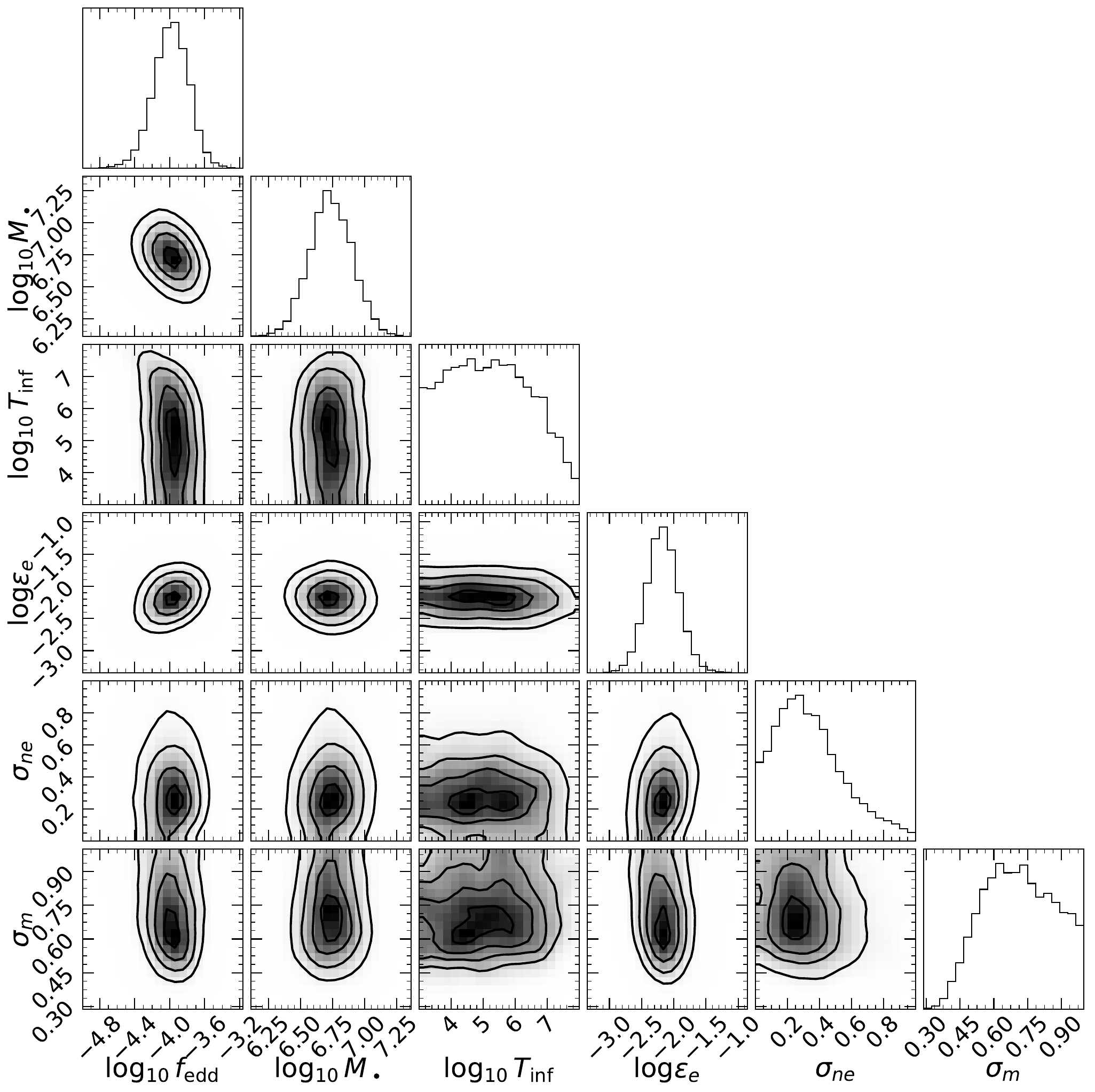}
};

\draw[flowarrow] (s1.south) -- (s2.north);
\draw[flowarrow] (s2.south) -- (s3.north);

\end{tikzpicture}

\caption{Flowchart of the modeling procedure: radio spectral observations constrain $(F_p,\nu_p,p)$ at time $t$, which can be related to $n_e(r)$ and $M_{\rm swept}(r)$ profiles, which ultimately constrain $(f_{\rm Edd},T_\infty,\epsilon_e)$.}
\label{fig:radio_model_flowchart}
\end{figure}

As an example of the methodology used to fit the parameters describing $n_e(r)$, in Figure \ref{fig:radio_model_flowchart} we show a flowchart of the fitting procedure. The upper panel shows the evolving multi-band radio spectra, from which constraints on the peak spectral flux $F_p$ and the peak frequency $\nu_p$ can be made at different times $t$. These spectral parameters then imply different evolving equipartition energy and radius profiles $E_{\rm eq}(t)$ and $R_{\rm eq}(t)$, which can be related to the ambient density of the flow and shock micro-physical parameter $\epsilon_e$. By jointly fitting these two profiles (central panel), constraints can be placed on the governing parameters of the background density profile (lower panel). Fits to every TDE in our sample are shown in Figure \ref{fig:all_ne_fits} in the Appendix.

\begin{figure*}
    \centering
    \includegraphics[width=0.3\linewidth]{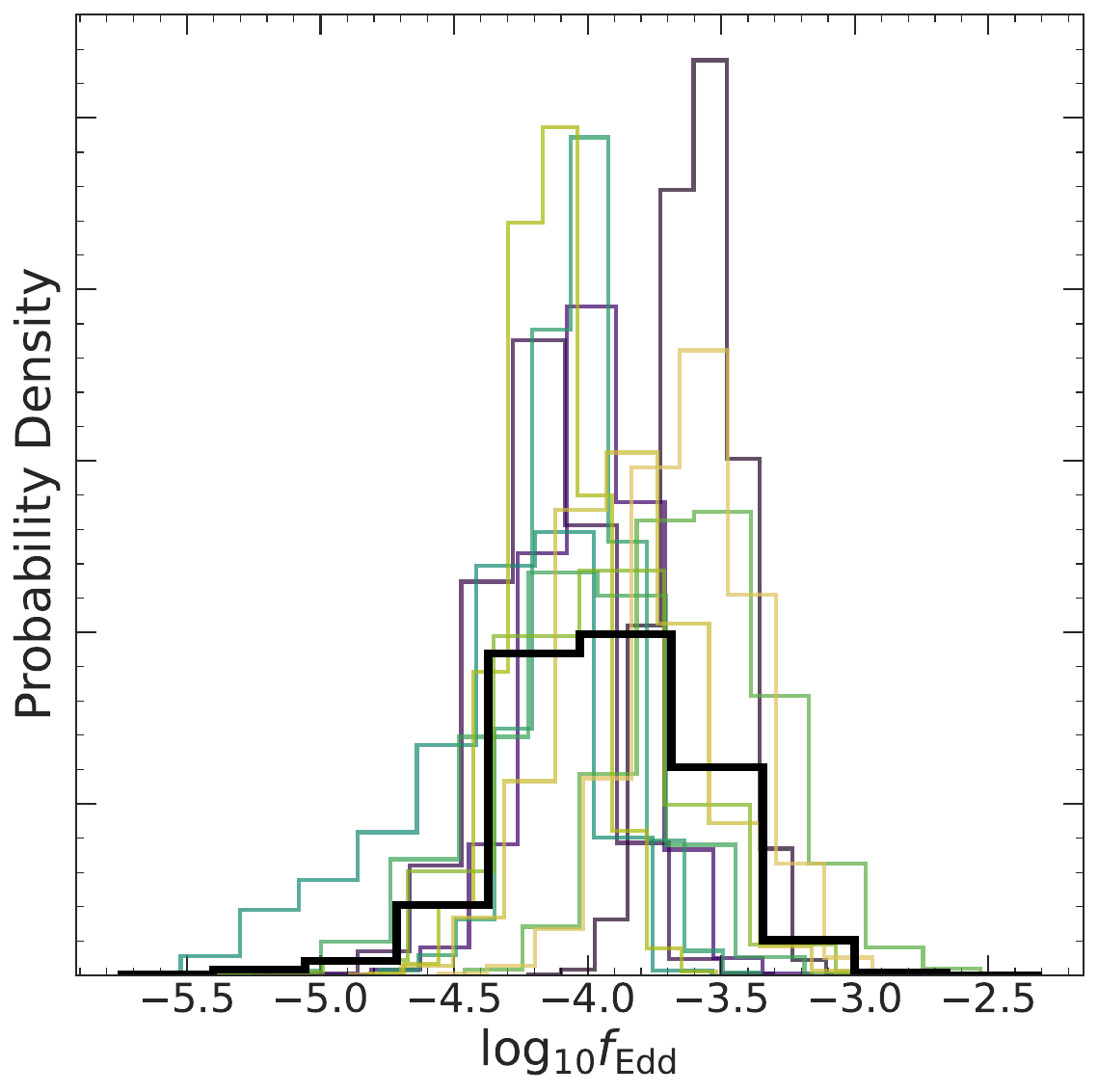}
    \includegraphics[width=0.3\linewidth]{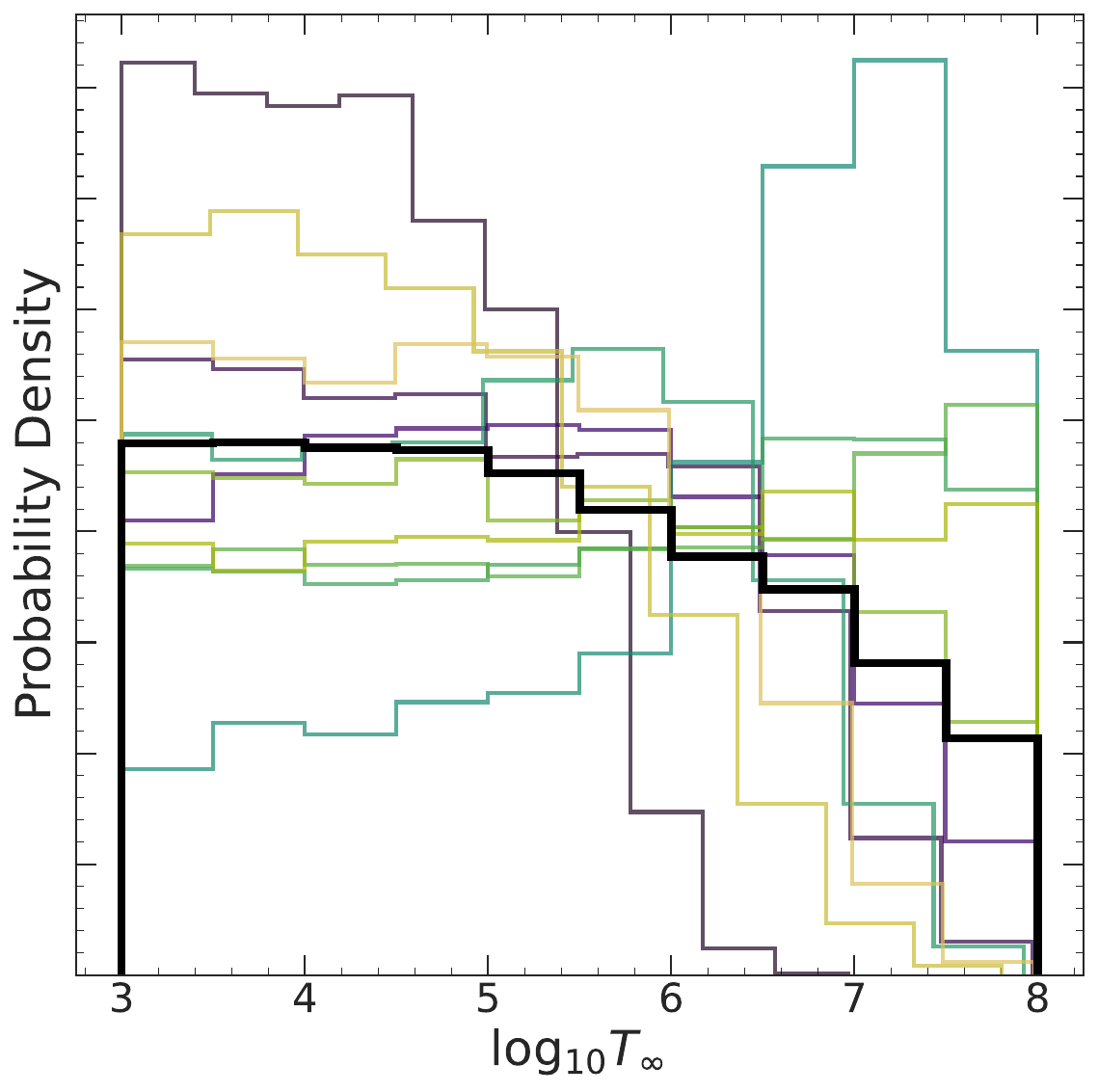}
    \includegraphics[width=0.3\linewidth]{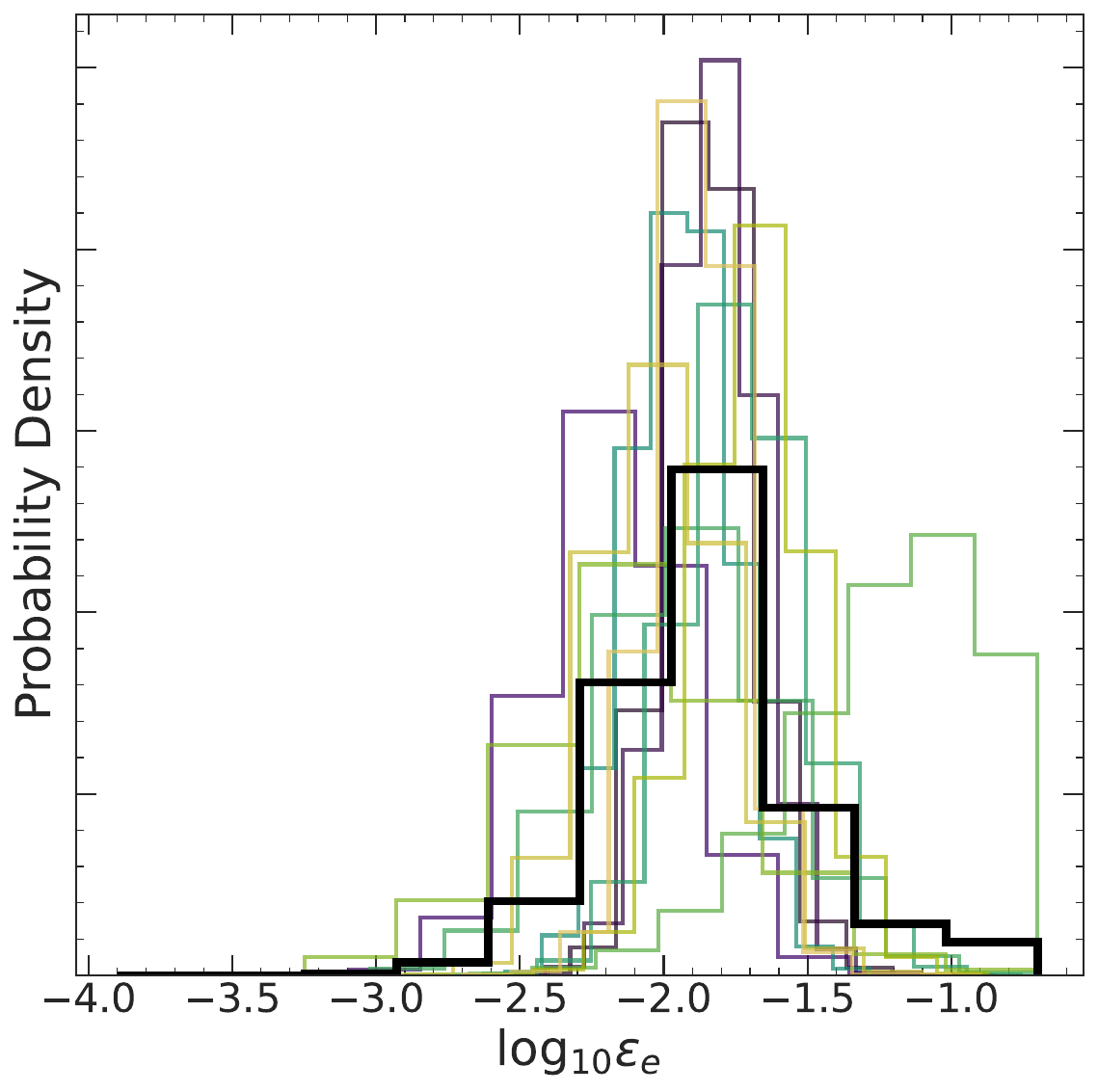}
    \caption{The posterior distribution of $f_{\rm{Edd}}$ (left), $T_{\infty}$ (middle), and $\epsilon_e$ (right) for the joint mass and density fits (Section \ref{sec:jointfits}) for each TDE modelled (coloured lines) and the sum total of all of the posterior distributions for the 11 TDEs (thick black line). We assume spherical geometry. $f_{\rm{Edd}}$ is well-constrained by the fitting but $T_{\infty}$ is largely unconstrained.}
    \label{fig:epse_fits}
\end{figure*}

The resulting posterior distributions for $\log_{10}\epsilon_e$, log\,$_{10}f_{\rm{Edd}}$, and log\,$_{10}T_\infty$ are shown in Figure \ref{fig:epse_fits}. The fitted parameter results for each TDE are listed in Table \ref{tab:ind_fit_results}.
Combining the posterior distributions of each TDE, we find the following parameter constraints at a population level: $\epsilon_e = 0.0138^{+0.02}_{-0.005}$, log\,$_{10}f_{\rm{Edd}} = -3.96^{+0.30}_{-0.38}$, and log\,$_{10}T_\infty$ remains largely unconstrained. 

We are generally only able to place (broadly uninteresting) upper limits on $T_\infty$, which originates from the lack of curvature in the $n_e-R$ profiles found in the data. Low frequency radio observations which track the radio spectral peak to low frequencies ($<1$\,GHz), such as in the era of SKA-low, would provide stronger constraints on $T_{\infty}$. 

\section{Discussion}\label{sec:discussion}
This work provides strong observational evidence that the sub-parsec distribution of gas in quiescent (TDE) host galaxies is well-described by a simple background Bondi accretion flow. 

\begin{figure}
    \centering
    \includegraphics[width=\linewidth]{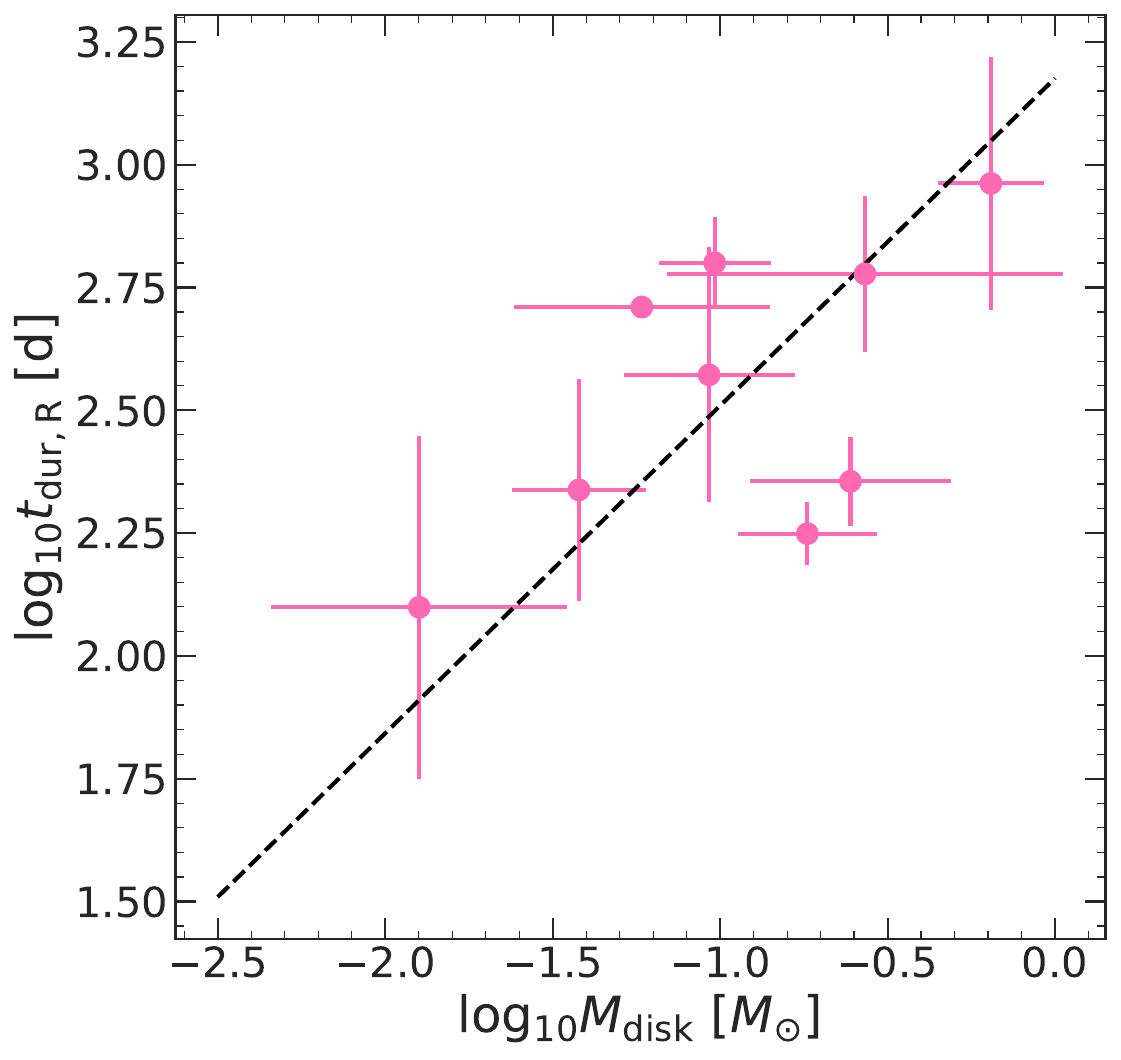}
    \caption{The observed radio flare duration for the TDEs studied in this work and measured accretion disk masses, which is a proxy for stellar mass. The black dashed line shows $t_{\rm{dur,R}}\propto M_{\rm{disk}}^{2/3}$, as expected if $M_{\rm out} \propto M_{\rm{disk}} \propto M_{*}$, and if $t_{\rm{dur,R}}\propto M_{\rm out}^{2/3}$ as derived in Equation \ref{eq:trise}.}
    \label{fig:tdurmdisk}
\end{figure}

The key assumptions that underpin this analysis are: (a) that the circumnuclear density gas distributions are well described by spherically symmetric profiles $n_e = n_e(R)$, and (b) that the observed prompt radio-emitting outflows in TDEs are well-described by a ballistically expanding shock front from material ejected by an accretion disk wind. The gas density distribution is of course unlikely to be as uniform as  we assume on small scales, but on the length scales we are probing such small scale inhomogeneities are likely not important. For the outflow mechanism, numerous recent studies have deduced that accretion disk winds are a likely explanation for the prompt radio-emitting outflows observed in a number of events \citep[e.g.][]{Alexander16,Goodwin23b,Christy2024}. Furthermore, \citet{GoodwinMummery2026} carried out detailed accretion disk modelling for all of the TDEs in our sample, finding that when such constraints could be placed the accretion rate at the time at which the outflow was launched was always consistent with being super-Eddington. 
In the case of an accretion disk wind, both the mass ejected in the wind and the mass of the accretion disk should be proportional to the stellar mass of the disrupted star. We show in Section \ref{sec:windsec} that the duration of the radio flare for a super-Eddington wind is proportional to $M_{\star}^{2/3}$, which implies $t_{\rm{dur}}\propto M_{\rm{disk}}^{2/3}$ if $M_{\rm{disk}}\propto fM_{\star}$ where $f$ is the fraction of the disrupted star that is circularised into the accretion disk. Indeed, as shown in Figure \ref{fig:tdurmdisk}, the radio flare durations of our sample are entirely consistent with $t_{\rm{dur,R}}\propto M_{\rm{disk}}^{2/3}$. Although with a relatively small sample of 11 events this relation  could of course be more statistically robust, and enlarging the sample size we have here is of real interest, the clear correlation provides further evidence that the radio emission used in this work is consistent with the theoretical expectations of an accretion disk wind. 

\subsection{Impact of equipartition assumption}
One further assumption we have made in this work is that the shocked circumnuclear gas reaches an equipartition between the accelerated electron population and the enhanced magnetic energy. Practically this means that we do not also include a magnetic energy density factor $\epsilon_B$, as in equipartition $\epsilon_B = (6/11)\epsilon_e$ \citep[e.g.][]{BarniolDuran2013}.

This is a particularly well motivated assumption for TDEs, as the energy budget required to power a given observed synchrotron spectrum is a strong function of the non-equipartition parameter $\epsilon \equiv (11\epsilon_B/6\epsilon_e)$, namely \cite{BarniolDuran2013}
\begin{equation}
    E = E_{\rm eq} \left[(11/17)\epsilon^{-6/17}+ (6/17)\epsilon^{
11/17}\right], 
\end{equation}
and a TDE outflow has a relatively limited mass (and therefore energy) budget $M_{\rm out} = f_{\rm out}M_\star$. 

While the energy could in principle be larger than the equipartition energy reported here (but, we stress, not by much owing to the limited mass budget), the equipartition radius is only weakly sensitive to deviation from equipartition, as $R=R_{\rm eq} \epsilon^{1/17}$, an entirely negligible change. With the same radius and a higher energy requirement, formally our Eddington fractions therefore represent {\it lower limits} owing to this equipartition assumption. We do not expect a non-equipartition analysis to quantitatively modify our results by any significant degree, owing to the small mass budget available in a TDE, but this assumption should be kept in mind. 

\newpage
\subsection{Comparison to theoretical expectations}
The properties of Bondi accretion flows in galactic nuclei have been extensively analyzed from a theoretical standpoint in the literature. From a practical perspective, the Bondi model we have used here assumes a non-rotating gas and neglects the impacts of turbulent motions, magnetic fields and radiative losses. Previous studies have suggested that the influence of even a small amount of angular momentum contained in the accreting gas may well be important on the smallest scales \citep[e.g.,][a not unsurprising result given the rotational barrier induced by any angular momentum at the smallest radial scales]{Narayan2011,proga_numerical_2003}, and of course a realistic galactic nucleus will not be spherically symmetric. 

Perhaps most relevant is this final point. Recent GRMHD simulations of a resolved Bondi sphere \citep{Guo25} found a $\rho \propto r^{-1}$ profile, rather than the canonical $r^{-3/2}$. The reason for this flattening appears to us to be the highly non-spherical and turbulent nature of the flow found by \cite{Guo25}, with mass outflow along the poles and mass inflow along more equatorial regions (as well as turbulent eddies at all scales). Indeed, while the average of $\dot M \equiv \dot M_{\rm in} - \dot M_{\rm out}$ was flat with radius, both inflow and outflow mass rates grew quite strongly with radius, with $\dot M_{\rm in} \sim r^{1/2}$ a reasonable fit. The net Bondi accretion rate $\dot M$ was indeed only of order a percent of $\dot M_{\rm in}$, with $\dot M_{\rm out}$ carrying a large (order 1) mass flux. This growing $\dot M_{\rm in}$ would be sufficient to explain the flattening of the \cite{Guo25} density profiles. It is not immediately obvious how a super-Eddington wind would interact with such a turbulent non-spherical profile, or how ambient mass outflow would impact the evolution of a spherical blast wave. We leave this discrepancy between recent GRMHD simulations and our observational probes as a somewhat open question from our work, while noting that a handful of our sources are consistent with a shallower $k = -1$  profile (although some very much favor more negative $k$). 

The fact that we are able to successfully reproduce multiple epochs of ambient density inferences with the simplest Bondi profiles suggests either that the impacts of such effects are physically small, or that TDE radio emission (even with tens of epochs) are not sensitive enough to identify such effects. It is an exciting development that we now have a way to systematically probe the sub-Bondi radius density profiles of quiescent galaxies, and future detailed comparison to theoretical models is clearly warranted with larger samples of TDE density constraints. 

\subsection{Comparison to previous studies of Bondi accretion in galactic nuclei}
From an observational perspective, high resolution X-ray observations of a handful of nearby galaxies that constrain the nuclear gas temperature and density have found gas density profiles consistent with the simplest Bondi accretion profiles \citep[e.g.][]{Allen2006,Russell2015,Gillessen2019}. 
The extremely high resolution required for such X-ray observations severely limits the number of galaxies where these measurements have been able to be obtained.  

In this study, we have presented, and demonstrated the utility of, a new method of constraining extragalactic circumnuclear gas density profiles. Our approach means that a much larger sample of galaxies may be studied, by leveraging the influence of the environment on the evolution of radio-bright TDE outflows. 

\begin{figure}
    \centering
    \includegraphics[width=\linewidth]{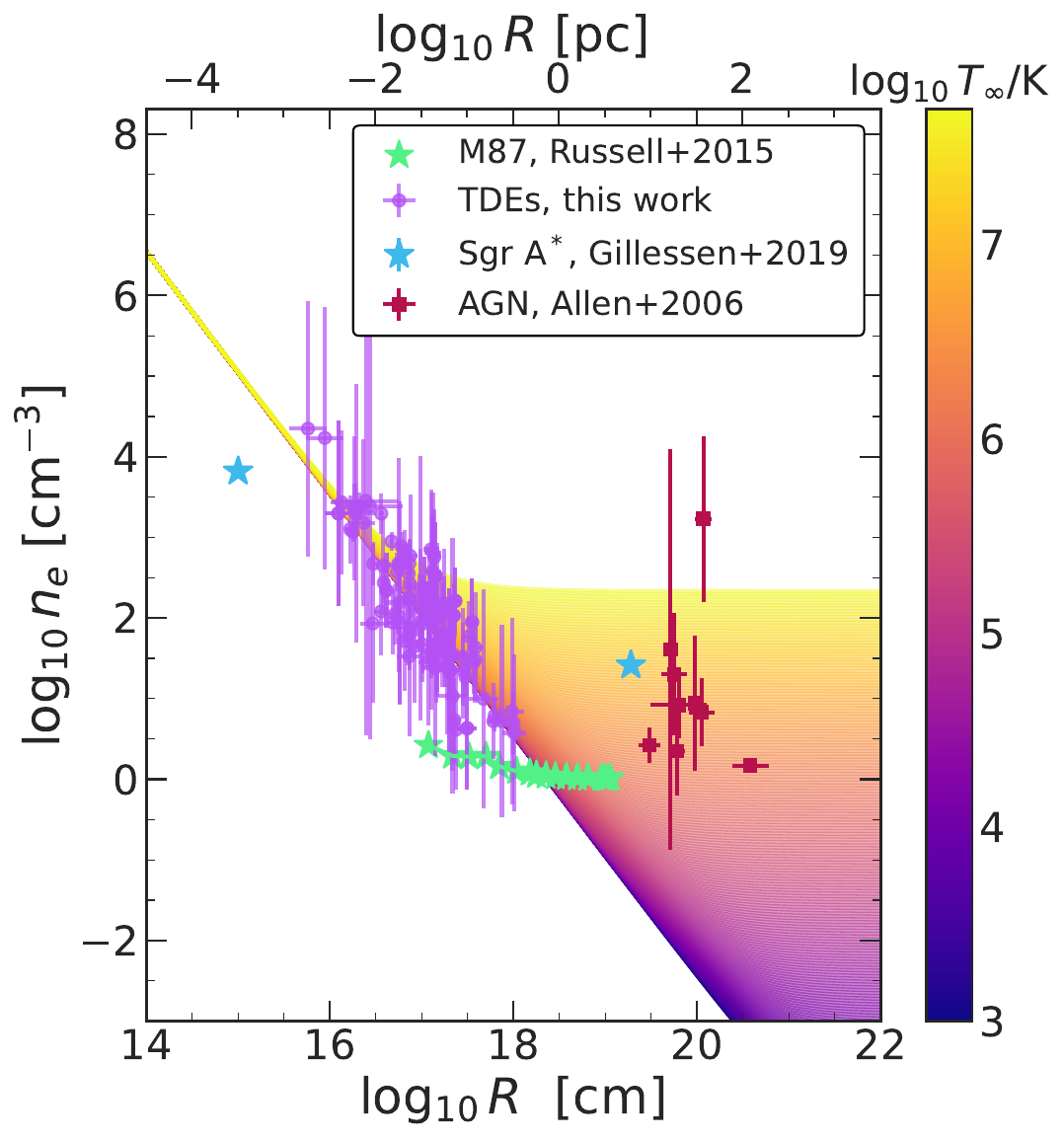}
    \caption{Electron density measurements in the nuclei of galaxies. Purple points indicate the quiescent TDE hosts analysed in this work (we plot the $n_e$ and $R$ values from Section \ref{sec:bondi_const}, corresponding to our best-fit measurements assuming spherical outflow geometry), blue is Sgr A$^*$ from \citet{Gillessen2019}, green is M87 from \citet{Russell2015}, and red are AGN in elliptical galaxies from \citet{Allen2006}. The shaded area indicates the Bondi profile for $f_{\rm{Edd}}=10^{-4}$, BH mass $10^{6.5}$\,M$_{\odot}$ (appropriate for the TDE sample), and varying temperatures. TDEs are evidently a powerful probe of the intra-Bondi region in galaxies, which is extremely difficult to observe otherwise. The density and radius values plotted here for the TDE hosts are tabulated in Table \ref{tab:nes_all}. }
    \label{fig:gal_comps}
\end{figure}

In Figure \ref{fig:gal_comps} we plot observational constraints available in the literature of the density distribution in the nuclei of various galaxies, to which we add the density constraints obtained for the TDE sample in this work. Only Sgr A$^*$ and M87 have measurements within the Bondi radius, at radii that we are able to probe with radio observations of TDE outflows, and only a handful of nearby AGN have constraints close to the Bondi radii. 

Interestingly, the density profiles for our TDE sample appear to lie between the two Sgr A$^*$ points, which is a black hole with mass comparable to our sample. Although the SMBH masses in the AGN sample (occurring in large elliptical galaxies) are almost certainly larger than the TDE sample, the Bondi density at large radii scales as $f_{\rm Edd}/M_\bullet$, a relatively weak dependence. The large radii density profiles of these AGN would be consistent with our results if they have higher $f_{\rm Edd}$, a result which is entirely expected for those galaxies which are currently actively accreting (something we now examine).

\subsection{The observed Bondi accretion rate}
\begin{figure}
    \centering
    \includegraphics[width=\linewidth]{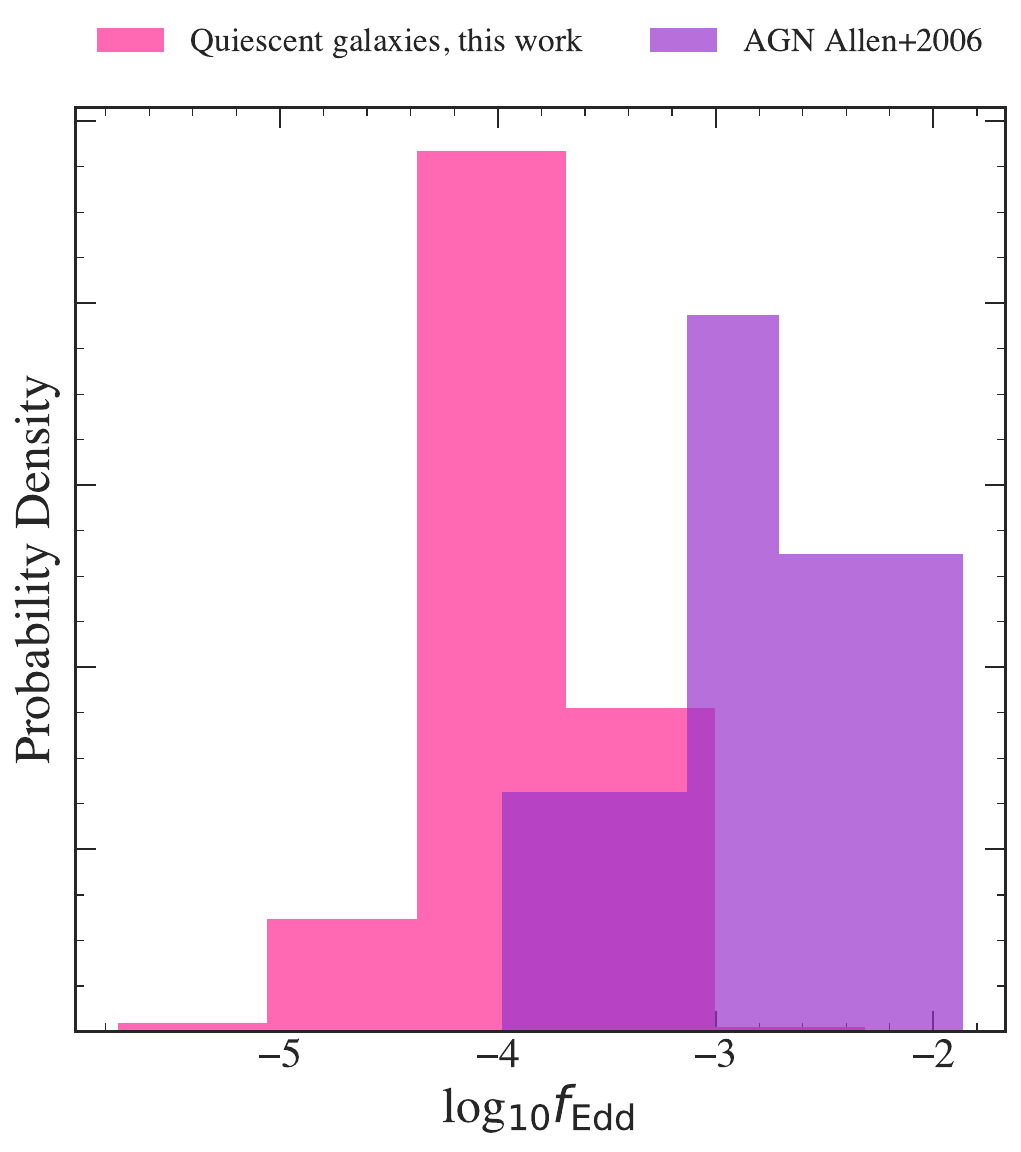}
    \includegraphics[width=\linewidth]{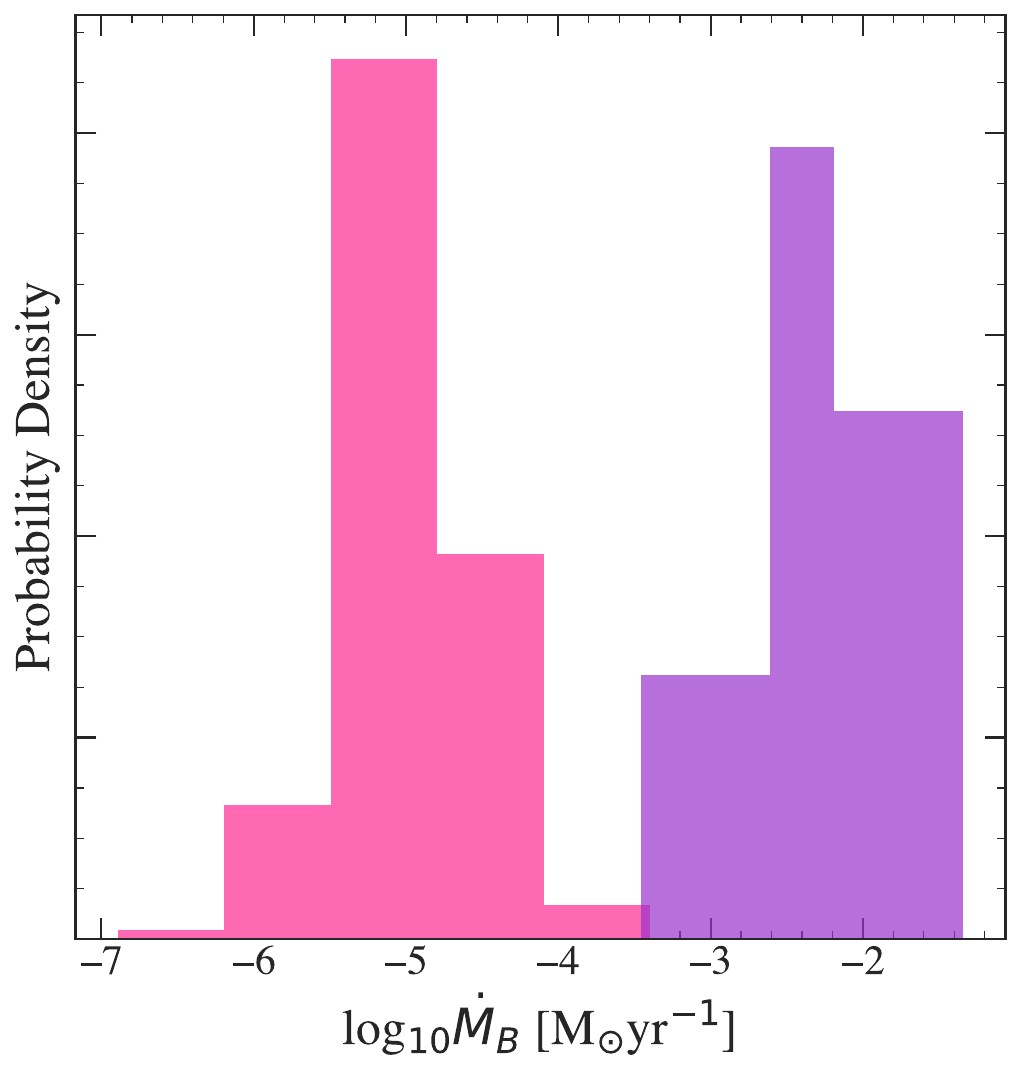}
    \caption{The inferred Bondi accretion rate for the 11 TDE host galaxies in our sample (pink) compared to the inferred Bondi accretion rates for 9 AGN in elliptical galaxies studied by \citep{Allen2006}. In the upper panel we show the accretion rate in Eddington-normalised units, in the lower panel in absolute units. The accretion rates inferred for the TDE sample are statistically significantly lower than those inferred for the AGN, suggesting a that the gas distribution in the nuclei of AGN is more dense than (quiescent) TDE hosts, an unsurprising result. The TDE sample has $\log_{10}f_{\rm{Edd}}=-3.96\pm0.34$ with $\log_{10}M_{\rm{BH,avg}}=6.64\pm0.39$, whereas the AGN sample has $\log_{10}f_{\rm{Edd}}=-2.79\pm0.53$ with $\log_{10}M_{\rm{BH,avg}}=8.2\pm0.3$.}
    \label{fig:bondi_mdots}
\end{figure}
The absolute value of the Bondi accretion rate is important in determining whether Bondi accretion alone can power jets that are ubiquitously observed in AGN, and particularly weaker jets observed in low-luminosity AGN.  
Observational constraints of the Bondi accretion rate (or Bondi accretion power) from high resolution X-ray observations can be compared with observed radio jet powers to determine whether radio-emitting outflows from AGN can be powered by Bondi accretion flows alone. 
\citet{Allen2006} found for a sample of 9 X-ray luminous elliptical galaxies (all with $d_L<135$\,Mpc and $M_{\rm{BH}}\sim10^8\,M_{\odot}$) Bondi accretion rates in the range $-3.46 <$log$_{10}\dot{M}_{\rm{Bondi}}[M_{\odot}\rm{yr}^{-1}]<-1.34$ (in other words $\log_{10}f_{\rm{Edd}}\approx-2.8$), where \cite{Allen2006} assumed an accretion efficiency of $\eta=0.1$, concluding that a strong correlation between jet power and Bondi power exists for large elliptical galaxies. A further sample of radiatively inefficient AGN in nearby galaxies studied by \citet{Ho2009} showed accretion rates (inferred from nuclear luminosities) that could be supplied by Bondi accretion of hot gas and local mass loss from evolved stars alone. In contrast, \citet{DiMatteo2001} showed in a sample of 7 nearby galaxies that their radio luminosities (tracing jet power) were significantly lower than expected for Bondi accretion rate estimates, concluding Bondi accretion alone is unlikely to power the observed radio jets.

In Figure \ref{fig:bondi_mdots} we plot the inferred Bondi accretion rate distributions for the TDE hosts (pink) and the AGN reported in \citet{Allen2006}. 
We constrain Bondi accretion rates in the range $f_{\rm{Edd}} \sim 10^{-5}-10^{-3.5}$ for our sample, consistent with the expectation that quiescent galaxies should have Bondi accretion rates lower than an AGN sample.

Our work presents strong evidence for Bondi accretion having influenced the circumnuclear density distribution in the 11 TDE host galaxies we examined. The majority of our sample (notably excluding ASASSN-14li) were in a quiescent galaxy phase (by which we mean there was no archival radio emission) prior to the TDE. These non-detections may well simply be explained by the lower accretion rates we infer (if low luminosity AGN radio jets are powered by Bondi accretion),  although observational evidence exists that Bondi accretion alone is not sufficient to power the full spectrum of nuclear activity observed in all galaxies \citep[e.g.][]{Ho2009}. This question warrants further scrutiny as the TDE population grows.

\subsection{A note on background X-ray luminosities}

We note that while we have inferred background Eddington ratios at the level of $f_{\rm Edd} \sim 10^{-4}$, we remind the reader that this implies background X-ray luminosities 
\begin{equation}
    L = f_{\rm Edd} \times {\eta_{\rm bondi}\over \eta} \times L_{\rm Edd},
\end{equation}
where $\eta = 0.1$ is the conventional efficiency we used when defining the Eddington ratio, and $\eta_{\rm bondi}$ is the actual radiative efficiency of the Bondi accretion process, which is much lower. Indeed, as $\eta_{\rm bondi}\ll 0.1$, we would expect Eddington luminosity ratios no higher than $\sim 10^{-7}$ from these host galaxies, far below typical detectability thresholds (as our TDEs are detected at large distances $d_L \gtrsim 100$ Mpc). 

\section{Summary}\label{sec:summary}
In this paper we have performed a detailed analysis of the circumnuclear density profiles of quiescent (TDE host) galaxies on sub-parsec scales. By developing a technique to simultaneously constrain the ambient density and swept up mass from broad-band radio spectral observations we have highlighted how shock microphysics parameters (namely the fraction of energy which has accelerated electrons $\epsilon_e$) can be constrained, leading to the removal of a key systematic uncertainty in previous analyses. 

We then applied this technique to a sample of 11 well-observed TDE hosts, where the radio emission was observed in the ``prompt'' phase. It is believed that this prompt phase of radio emission is produced by the interaction of a super-Eddington wind launched from the early phases of the TDE accretion disks evolution with the circumnuclear medium. In the case of a ballistically expanding outflow (like that of an accretion disk wind), the evolution of the radio emission is only dependent on the total energy in the initial outflow and the properties of the medium through which it propagates. 

We found that the density profiles in our sample were well approximated, in a relatively model-agnostic  power-law $n_e \propto r^{k}$ fashion, by profiles in reasonable agreement with the classic Bondi result $k \approx -3/2$. Explicitly assuming a classic Bondi solution, we are able to constrain the background mass accretion rate (the Bondi-Eddington fraction) of these host galaxies. 

We conclude that the classical Bondi solution for spherical black hole accretion provides a good description of the observed sub-parsec gas density distribution in our sample of quiescent (TDE host) galaxies. Our conclusions are summarised as follows: 

\begin{itemize}
    \item Radio observations of TDEs provide powerful constraints on the ambient density distribution at sub-parsec scales around quiescent SMBHs, well inside the Bondi sphere of the host black holes, where constraints from other approaches are lacking. \\
    \item The majority of the SMBHs in our sample follow power-law relationships which are reasonably close to the canonical $n_e\propto R^{-3/2}$, as expected for simple Bondi accretion.
    \item Under the assumption of a Bondi accretion profile, for each of the individual events in our sample, we can constrain the Bondi accretion rate in Eddington units. We find a sample average value of \,$\log_{10}f_{\rm{Edd}} = -3.96^{+0.30}_{-0.38}$.
    \item Future radio observations of larger samples will allow more precise measurements of these parameters, motivating densely sampled radio spectral follow-up observations of prompt radio-emitting TDEs. In particular, low-frequency radio observations which resolve the evolution of the radio spectral peak below 1\,GHz may provide constraints on $T_{\infty}$, the background gas temperature, which current radio observations of TDEs are unable to constrain in the majority of sources.  
\end{itemize}

\begin{acknowledgements}
AJG is grateful for support from the Forrest Research Foundation.  A.M. acknowledges support from the Ambrose Monell Foundation, the W.M. Keck Foundation and the John N. Bahcall Fellowship Fund at the Institute for Advanced Study. This research benefited from discussions at the Kavli Institute for Theoretical Physics (KITP) program on tidal disruption events in April 2024 that was supported in part by the National Science Foundation under PHY- 1748958.
\end{acknowledgements}

\bibliography{andy}{}

\begin{thebibliography}{}
\expandafter\ifx\csname natexlab\endcsname\relax\def\natexlab#1{#1}\fi
\providecommand{\url}[1]{\href{#1}{#1}}
\providecommand{\dodoi}[1]{doi:~\href{http://doi.org/#1}{\nolinkurl{#1}}}
\providecommand{\doeprint}[1]{\href{http://ascl.net/#1}{\nolinkurl{http://ascl.net/#1}}}
\providecommand{\doarXiv}[1]{\href{https://arxiv.org/abs/#1}{\nolinkurl{https://arxiv.org/abs/#1}}}

\bibitem[{{Alexander} {et~al.}(2016){Alexander}, {Berger}, {Guillochon}, {Zauderer}, \& {Williams}}]{Alexander16}
{Alexander}, K.~D., {Berger}, E., {Guillochon}, J., {Zauderer}, B.~A., \& {Williams}, P.~K.~G. 2016, \apjl, 819, L25, \dodoi{10.3847/2041-8205/819/2/L25}

\bibitem[{{Alexander} {et~al.}(2020){Alexander}, {van Velzen}, {Horesh}, \& {Zauderer}}]{Alexander2020}
{Alexander}, K.~D., {van Velzen}, S., {Horesh}, A., \& {Zauderer}, B.~A. 2020, \ssr, 216, 81, \dodoi{10.1007/s11214-020-00702-w}

\bibitem[{{Alexander} {et~al.}(2025){Alexander}, {Margutti}, {Gomez}, {Stroh}, {Chornock}, {Laskar}, {Cendes}, {Berger}, {Eftekhari}, {Franz}, {Hajela}, {Metzger}, {Terreran}, {Bietenholz}, {Christy}, {de Colle}, {Komossa}, {Nicholl}, {Ramirez-Ruiz}, {Saxton}, {Schroeder}, {Williams}, \& {Wu}}]{Alexander2025}
{Alexander}, K.~D., {Margutti}, R., {Gomez}, S., {et~al.} 2025, arXiv e-prints, arXiv:2506.12729, \dodoi{10.48550/arXiv.2506.12729}

\bibitem[{{Allen} {et~al.}(2006){Allen}, {Dunn}, {Fabian}, {Taylor}, \& {Reynolds}}]{Allen2006}
{Allen}, S.~W., {Dunn}, R.~J.~H., {Fabian}, A.~C., {Taylor}, G.~B., \& {Reynolds}, C.~S. 2006, \mnras, 372, 21, \dodoi{10.1111/j.1365-2966.2006.10778.x}

\bibitem[{{Barniol Duran} {et~al.}(2013){Barniol Duran}, {Nakar}, \& {Piran}}]{BarniolDuran2013}
{Barniol Duran}, R., {Nakar}, E., \& {Piran}, T. 2013, \apj, 772, 78, \dodoi{10.1088/0004-637X/772/1/78}

\bibitem[{{Bondi}(1952)}]{Bondi52}
{Bondi}, H. 1952, \mnras, 112, 195, \dodoi{10.1093/mnras/112.2.195}

\bibitem[{{Burn} {et~al.}(2025){Burn}, {Goodwin}, {Anderson}, {Miller-Jones}, {Cendes}, {Christy}, {Lu}, \& {van Velzen}}]{Burn2025}
{Burn}, M., {Goodwin}, A.~J., {Anderson}, G.~E., {et~al.} 2025, \apj, 993, 207, \dodoi{10.3847/1538-4357/ae0a16}

\bibitem[{{Cendes} {et~al.}(2021){Cendes}, {Alexander}, {Berger}, {Eftekhari}, {Williams}, \& {Chornock}}]{Cendes_2021_dsg}
{Cendes}, Y., {Alexander}, K.~D., {Berger}, E., {et~al.} 2021, \apj, 919, 127, \dodoi{10.3847/1538-4357/ac110a}

\bibitem[{{Chevalier}(1998)}]{Chevalier1998}
{Chevalier}, R.~A. 1998, \apj, 499, 810, \dodoi{10.1086/305676}

\bibitem[{{Christy} {et~al.}(2024){Christy}, {Alexander}, {Margutti}, {Wieringa}, {Cendes}, {Chornock}, {Laskar}, {Berger}, {Bietenholz}, {Coppejans}, {De Colle}, {Eftekhari}, {Holoien}, {Matsumoto}, {Miller-Jones}, {Ramirez-Ruiz}, {Saxton}, \& {van Velzen}}]{Christy2024}
{Christy}, C.~T., {Alexander}, K.~D., {Margutti}, R., {et~al.} 2024, \apj, 974, 18, \dodoi{10.3847/1538-4357/ad675b}

\bibitem[{{Christy} {et~al.}(2025){Christy}, {Alexander}, {Laskar}, {Franz}, {Goodwin}, {Pearson}, {Berger}, {Cendes}, {Chornock}, {Coppejans}, {Eftekhari}, {Margutti}, {Miller-Jones}, {Krips}, {Ramirez-Ruiz}, {Sand}, {Saxton}, {Shrestha}, \& {van Velzen}}]{Christy2025}
{Christy}, C.~T., {Alexander}, K.~D., {Laskar}, T., {et~al.} 2025, arXiv e-prints, arXiv:2509.14317, \dodoi{10.48550/arXiv.2509.14317}

\bibitem[{{Clark} {et~al.}(2025){Clark}, {Callow}, {Graur}, {Greenwell}, {Hu}, {Aguilar}, {Ahlen}, {Bianchi}, {Brooks}, {Claybaugh}, {Dawson}, {de la Macorra}, {Doel}, {Gontcho}, {Gutierrez}, {Honscheid}, {Juneau}, {Kehoe}, {Kisner}, {Kremin}, {Landriau}, {Le Guillou}, {Meisner}, {Miquel}, {Moustakas}, {P{\'e}rez-R{\`a}fols}, {Sanchez}, {Schubnell}, {Sprayberry}, {Tarl{\'e}}, {Weaver}, \& {Zou}}]{Clark2025}
{Clark}, P., {Callow}, J., {Graur}, O., {et~al.} 2025, arXiv e-prints, arXiv:2502.04080, \dodoi{10.48550/arXiv.2502.04080}

\bibitem[{{Croton} {et~al.}(2006){Croton}, {Springel}, {White}, {De Lucia}, {Frenk}, {Gao}, {Jenkins}, {Kauffmann}, {Navarro}, \& {Yoshida}}]{Croton2006}
{Croton}, D.~J., {Springel}, V., {White}, S. D.~M., {et~al.} 2006, \mnras, 365, 11, \dodoi{10.1111/j.1365-2966.2005.09675.x}

\bibitem[{{Di Matteo} {et~al.}(2001){Di Matteo}, {Carilli}, \& {Fabian}}]{DiMatteo2001}
{Di Matteo}, T., {Carilli}, C.~L., \& {Fabian}, A.~C. 2001, \apj, 547, 731, \dodoi{10.1086/318405}

\bibitem[{{Foreman-Mackey} {et~al.}(2013){Foreman-Mackey}, {Hogg}, {Lang}, \& {Goodman}}]{EMCEE}
{Foreman-Mackey}, D., {Hogg}, D.~W., {Lang}, D., \& {Goodman}, J. 2013, \pasp, 125, 306, \dodoi{10.1086/670067}

\bibitem[{{French} {et~al.}(2016){French}, {Arcavi}, \& {Zabludoff}}]{French2016}
{French}, K.~D., {Arcavi}, I., \& {Zabludoff}, A. 2016, \apjl, 818, L21, \dodoi{10.3847/2041-8205/818/1/L21}

\bibitem[{{Gillessen} {et~al.}(2019){Gillessen}, {Plewa}, {Widmann}, {von Fellenberg}, {Schartmann}, {Habibi}, {Jimenez Rosales}, {Baub{\"o}ck}, {Dexter}, {Gao}, {Waisberg}, {Eisenhauer}, {Pfuhl}, {Ott}, {Burkert}, {de Zeeuw}, \& {Genzel}}]{Gillessen2019}
{Gillessen}, S., {Plewa}, P.~M., {Widmann}, F., {et~al.} 2019, \apj, 871, 126, \dodoi{10.3847/1538-4357/aaf4f8}

\bibitem[{{Goodwin} \& {Mummery}(2026)}]{GoodwinMummery2026}
{Goodwin}, A.~J., \& {Mummery}, A. 2026, arXiv e-prints, arXiv:2602.14838, \dodoi{10.48550/arXiv.2602.14838}

\bibitem[{{Goodwin} {et~al.}(2022){Goodwin}, {van Velzen}, {Miller-Jones}, {Mummery}, {Bietenholz}, {Wederfoort}, {Hammerstein}, {Bonnerot}, {Hoffmann}, \& {Yan}}]{Goodwin22}
{Goodwin}, A.~J., {van Velzen}, S., {Miller-Jones}, J.~C.~A., {et~al.} 2022, \mnras, 511, 5328, \dodoi{10.1093/mnras/stac333}

\bibitem[{{Goodwin} {et~al.}(2023{\natexlab{a}}){Goodwin}, {Alexander}, {Miller-Jones}, {Bietenholz}, {van Velzen}, {Anderson}, {Berger}, {Cendes}, {Chornock}, {Coppejans}, {Eftekhari}, {Gezari}, {Laskar}, {Ramirez-Ruiz}, \& {Saxton}}]{Goodwin23}
{Goodwin}, A.~J., {Alexander}, K.~D., {Miller-Jones}, J.~C.~A., {et~al.} 2023{\natexlab{a}}, \mnras, 522, 5084, \dodoi{10.1093/mnras/stad1258}

\bibitem[{{Goodwin} {et~al.}(2023{\natexlab{b}}){Goodwin}, {Miller-Jones}, {van Velzen}, {Bietenholz}, {Greenland}, {Cenko}, {Gezari}, {Horesh}, {Sivakoff}, {Yan}, {Yu}, \& {Zhang}}]{Goodwin23b}
{Goodwin}, A.~J., {Miller-Jones}, J.~C.~A., {van Velzen}, S., {et~al.} 2023{\natexlab{b}}, \mnras, 518, 847, \dodoi{10.1093/mnras/stac3127}

\bibitem[{{Goodwin} {et~al.}(2024){Goodwin}, {Anderson}, {Miller-Jones}, {Malyali}, {Grotova}, {Homan}, {Kawka}, {Krumpe}, {Liu}, \& {Rau}}]{Goodwin24}
{Goodwin}, A.~J., {Anderson}, G.~E., {Miller-Jones}, J.~C.~A., {et~al.} 2024, \mnras, 528, 7123, \dodoi{10.1093/mnras/stae362}

\bibitem[{{Goodwin} {et~al.}(2025){Goodwin}, {Mummery}, {Laskar}, {Alexander}, {Anderson}, {Bietenholz}, {Bonnerot}, {Christy}, {Golay}, {Lu}, {Margutti}, {Miller-Jones}, {Ramirez-Ruiz}, {Saxton}, \& {van Velzen}}]{Goodwin25}
{Goodwin}, A.~J., {Mummery}, A., {Laskar}, T., {et~al.} 2025, \apj, 981, 122, \dodoi{10.3847/1538-4357/adb0b1}

\bibitem[{{Guo} {et~al.}(2025){Guo}, {Stone}, {Quataert}, \& {Springel}}]{Guo25}
{Guo}, M., {Stone}, J.~M., {Quataert}, E., \& {Springel}, V. 2025, \apj, 987, 202, \dodoi{10.3847/1538-4357/add1da}

\bibitem[{{Hajela} {et~al.}(2025){Hajela}, {Alexander}, {Margutti}, {Chornock}, {Bietenholz}, {Christy}, {Stroh}, {Terreran}, {Saxton}, {Komossa}, {Bright}, {Ramirez-Ruiz}, {Coppejans}, {Leung}, {Cendes}, {Wiston}, {Laskar}, {Horesh}, {Schroeder}, {A.~J.}, {Wieringa}, {Velez}, {Berger}, {Blanchard}, {Eftekhari}, {Gomez}, {Nicholl}, {Sears}, \& {Zauderer}}]{Hajela2025}
{Hajela}, A., {Alexander}, K.~D., {Margutti}, R., {et~al.} 2025, \apj, 983, 29, \dodoi{10.3847/1538-4357/adb620}

\bibitem[{{Ho}(2009)}]{Ho2009}
{Ho}, L.~C. 2009, \apj, 699, 626, \dodoi{10.1088/0004-637X/699/1/626}

\bibitem[{{Horesh} {et~al.}(2021){Horesh}, {Cenko}, \& {Arcavi}}]{Horesh21}
{Horesh}, A., {Cenko}, S.~B., \& {Arcavi}, I. 2021, Nature Astronomy, 5, 491, \dodoi{10.1038/s41550-021-01300-8}

\bibitem[{{Mummery} {et~al.}(2025){Mummery}, {Guolo}, {Matthews}, {Newsome}, {Lintott}, \& {Keel}}]{Mummery_et_al25}
{Mummery}, A., {Guolo}, M., {Matthews}, J., {et~al.} 2025, arXiv e-prints, arXiv:2503.14163, \dodoi{10.48550/arXiv.2503.14163}

\bibitem[{{Narayan} \& {Fabian}(2011)}]{Narayan2011}
{Narayan}, R., \& {Fabian}, A.~C. 2011, \mnras, 415, 3721, \dodoi{10.1111/j.1365-2966.2011.18987.x}

\bibitem[{{Newsome} {et~al.}(2024){Newsome}, {Arcavi}, {Howell}, {McCully}, {Terreran}, {Hosseinzadeh}, {Bostroem}, {Dgany}, {Farah}, {Faris}, {Padilla-Gonzalez}, {Pellegrino}, \& {Andrews}}]{Newsome2024_22upj}
{Newsome}, M., {Arcavi}, I., {Howell}, D.~A., {et~al.} 2024, \apj, 977, 258, \dodoi{10.3847/1538-4357/ad8a69}

\bibitem[{{Park} {et~al.}(2015){Park}, {Caprioli}, \& {Spitkovsky}}]{Park2015}
{Park}, J., {Caprioli}, D., \& {Spitkovsky}, A. 2015, \prl, 114, 085003, \dodoi{10.1103/PhysRevLett.114.085003}

\bibitem[{Proga(2003)}]{proga_numerical_2003}
Proga, D. 2003, \apj, 585, 406, \dodoi{10.1086/345897}

\bibitem[{{Russell} {et~al.}(2015){Russell}, {Fabian}, {McNamara}, \& {Broderick}}]{Russell2015}
{Russell}, H.~R., {Fabian}, A.~C., {McNamara}, B.~R., \& {Broderick}, A.~E. 2015, \mnras, 451, 588, \dodoi{10.1093/mnras/stv954}

\bibitem[{{Rybicki} \& {Lightman}(1979)}]{Rybicki79}
{Rybicki}, G.~B., \& {Lightman}, A.~P. 1979, {Radiative processes in astrophysics}

\bibitem[{{Somalwar} {et~al.}(2023){Somalwar}, {Ravi}, \& {Lu}}]{Somalwar2023}
{Somalwar}, J.~J., {Ravi}, V., \& {Lu}, W. 2023, arXiv e-prints, arXiv:2310.03795, \dodoi{10.48550/arXiv.2310.03795}

\bibitem[{{Stein} {et~al.}(2021){Stein}, {Velzen}, {Kowalski}, {Franckowiak}, {Gezari}, {Miller-Jones}, {Frederick}, {Sfaradi}, {Bietenholz}, {Horesh}, {Fender}, {Garrappa}, {Ahumada}, {Andreoni}, {Belicki}, {Bellm}, {B{\"o}ttcher}, {Brinnel}, {Burruss}, {Cenko}, {Coughlin}, {Cunningham}, {Drake}, {Farrar}, {Feeney}, {Foley}, {Gal-Yam}, {Golkhou}, {Goobar}, {Graham}, {Hammerstein}, {Helou}, {Hung}, {Kasliwal}, {Kilpatrick}, {Kong}, {Kupfer}, {Laher}, {Mahabal}, {Masci}, {Necker}, {Nordin}, {Perley}, {Rigault}, {Reusch}, {Rodriguez}, {Rojas-Bravo}, {Rusholme}, {Shupe}, {Singer}, {Sollerman}, {Soumagnac}, {Stern}, {Taggart}, {van Santen}, {Ward}, {Woudt}, \& {Yao}}]{Stein21}
{Stein}, R., {Velzen}, S.~v., {Kowalski}, M., {et~al.} 2021, Nature Astronomy, 5, 510, \dodoi{10.1038/s41550-020-01295-8}

\bibitem[{{Yao} {et~al.}(2023){Yao}, {Ravi}, {Gezari}, {van Velzen}, {Lu}, {Schulze}, {Somalwar}, {Kulkarni}, {Hammerstein}, {Nicholl}, {Graham}, {Perley}, {Cenko}, {Stein}, {Ricarte}, {Chadayammuri}, {Quataert}, {Bellm}, {Bloom}, {Dekany}, {Drake}, {Groom}, {Mahabal}, {Prince}, {Riddle}, {Rusholme}, {Sharma}, {Sollerman}, \& {Yan}}]{Yao2023}
{Yao}, Y., {Ravi}, V., {Gezari}, S., {et~al.} 2023, \apjl, 955, L6, \dodoi{10.3847/2041-8213/acf216}

\end{thebibliography}
\bibliographystyle{aasjournal}

\appendix

\section{Individual TDE fit results}

\begin{figure*}
    \centering
    \includegraphics[width=0.32\linewidth]{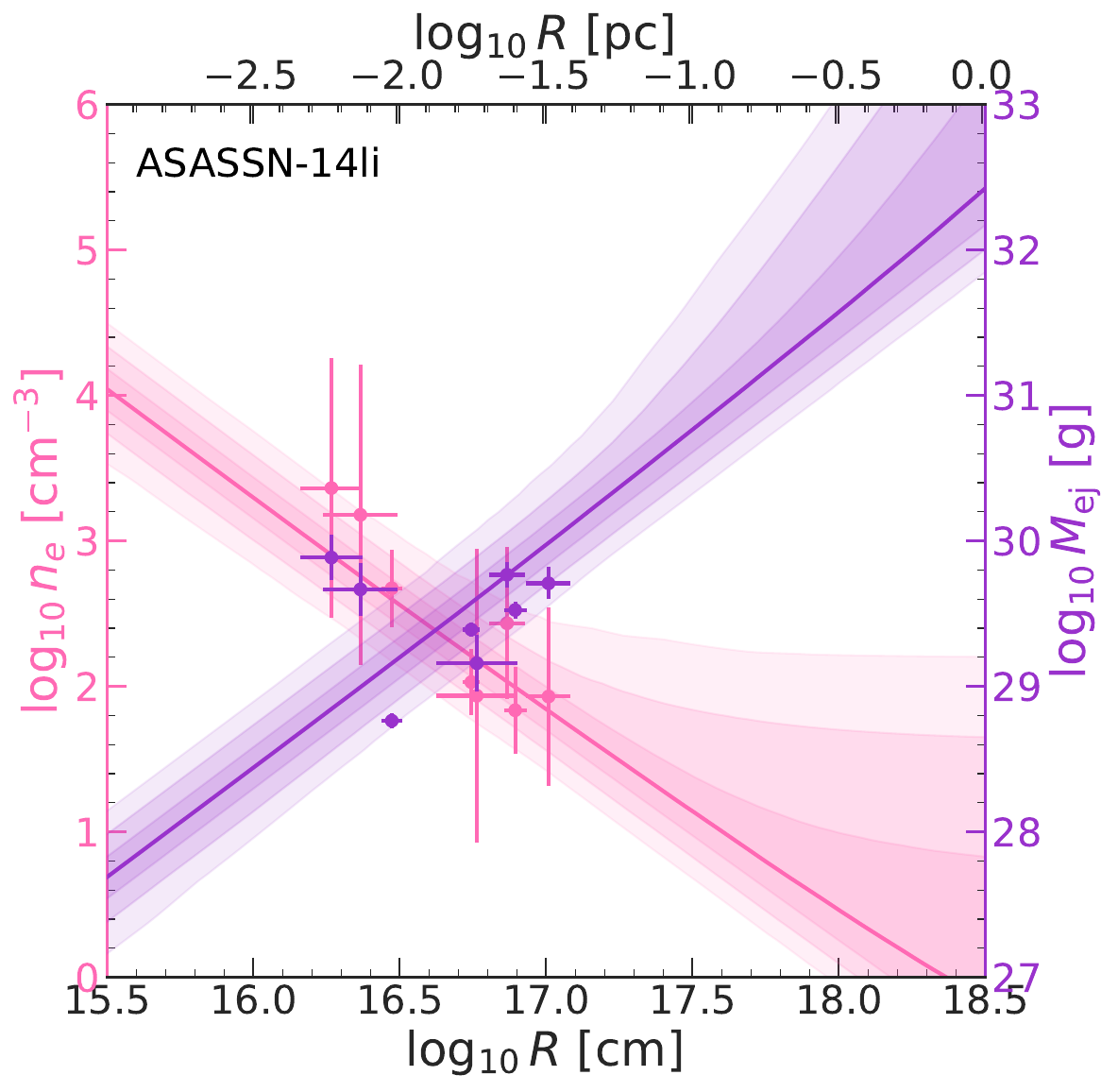}
    \includegraphics[width=0.32\linewidth]{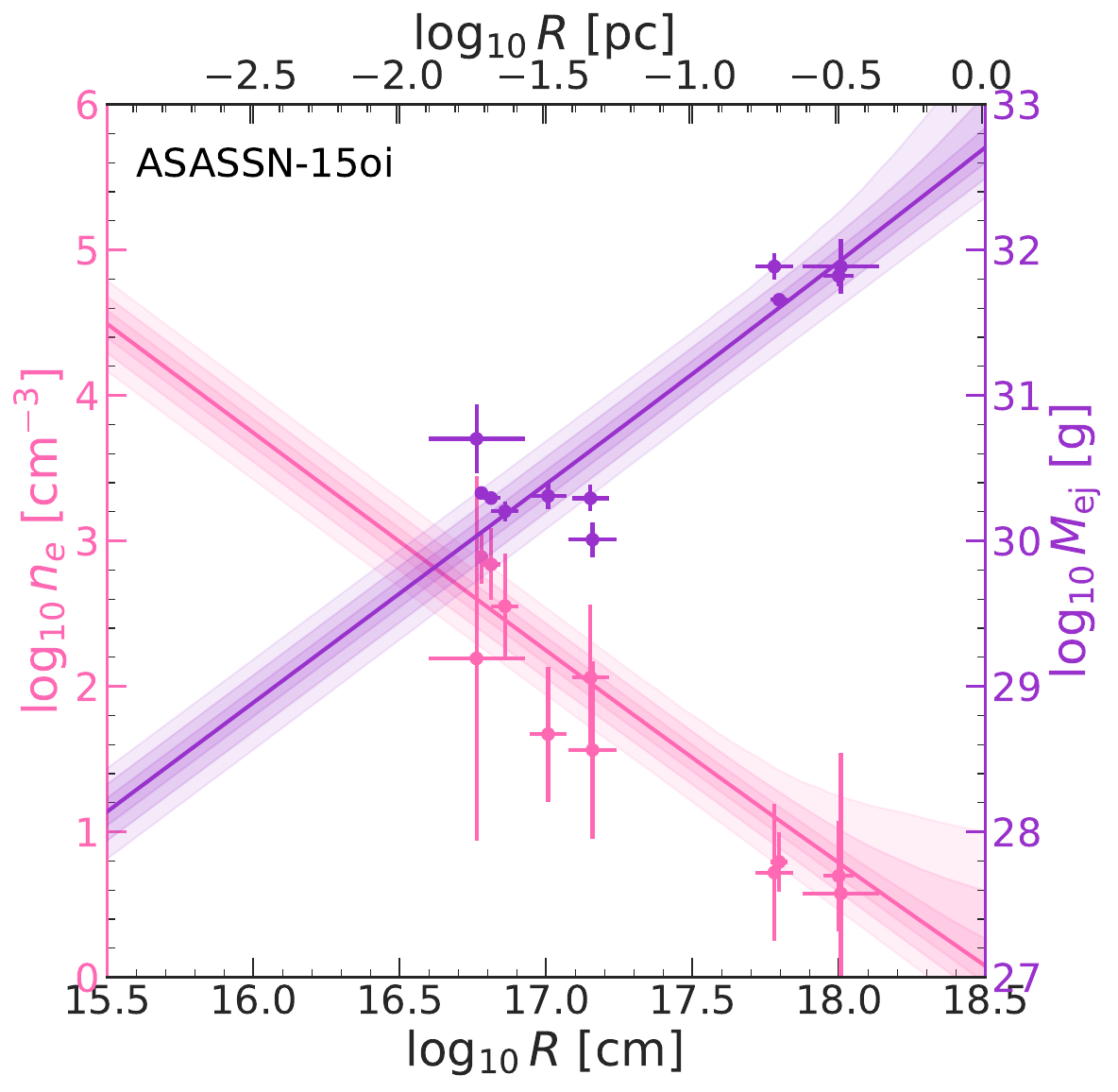}
    \includegraphics[width=0.32\linewidth]{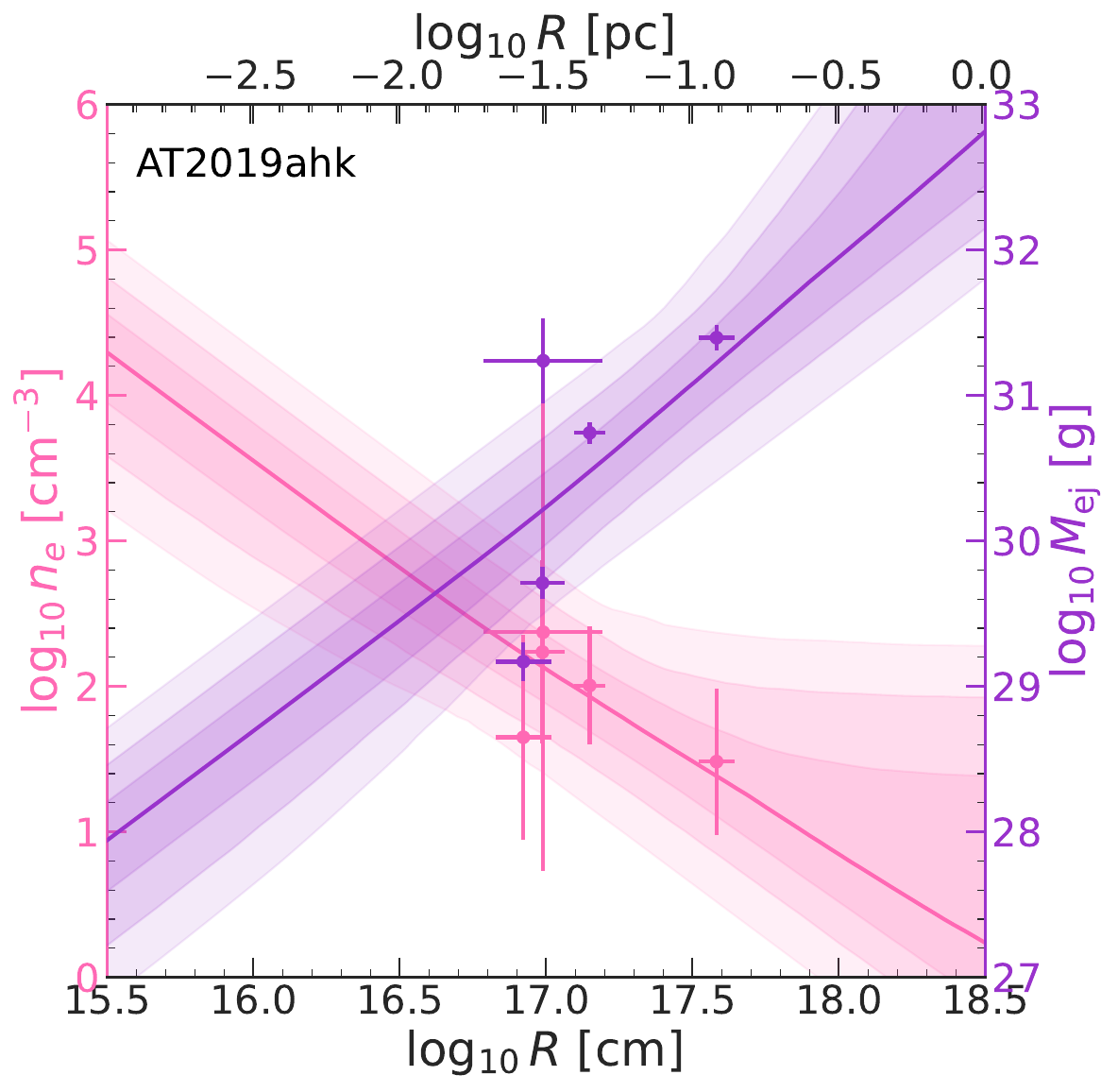}
    \includegraphics[width=0.32\linewidth]{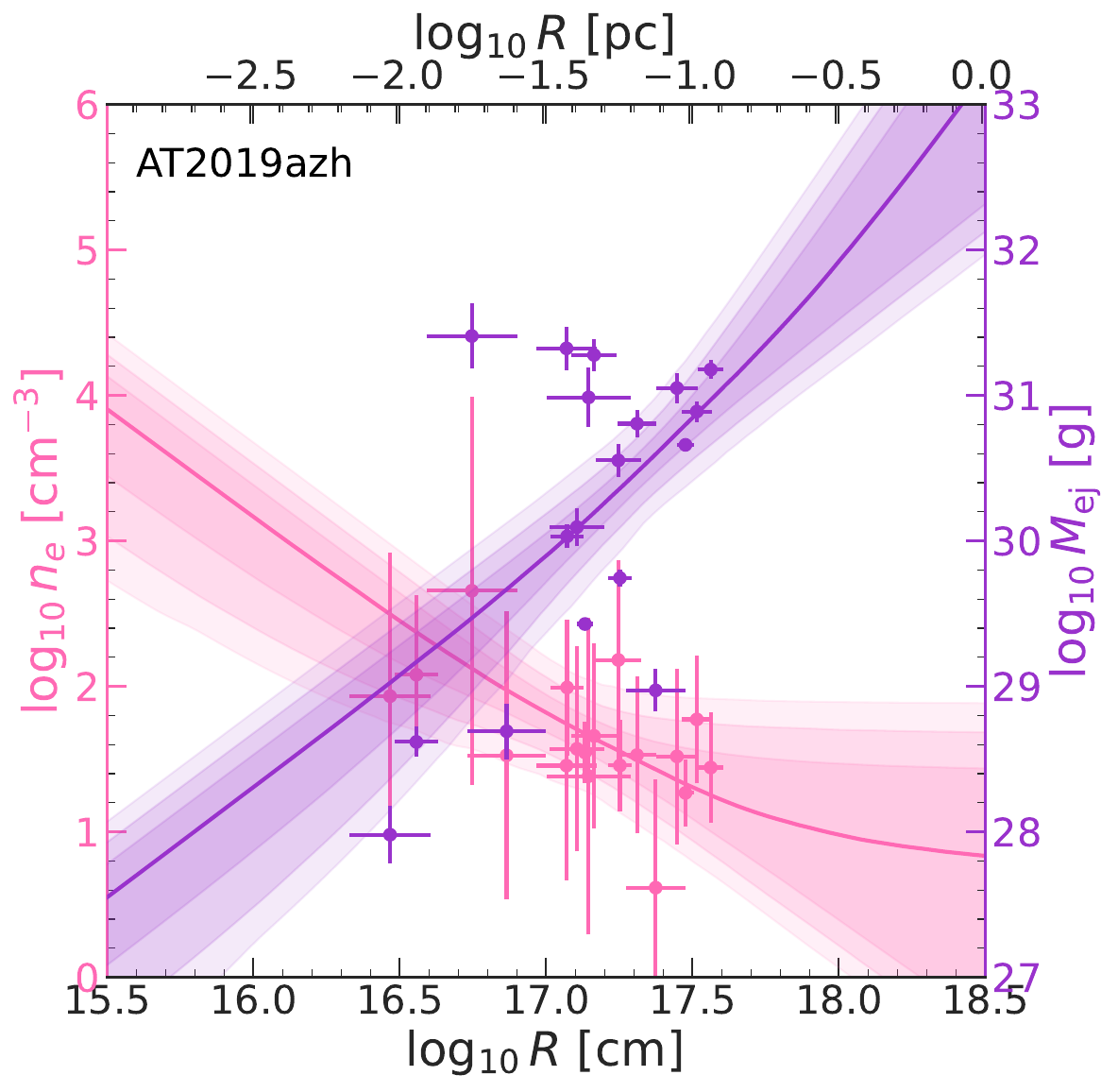}
    \includegraphics[width=0.32\linewidth]{ne_ind_fits/ne_m_AT2019dsg.pdf}
    \includegraphics[width=0.32\linewidth]{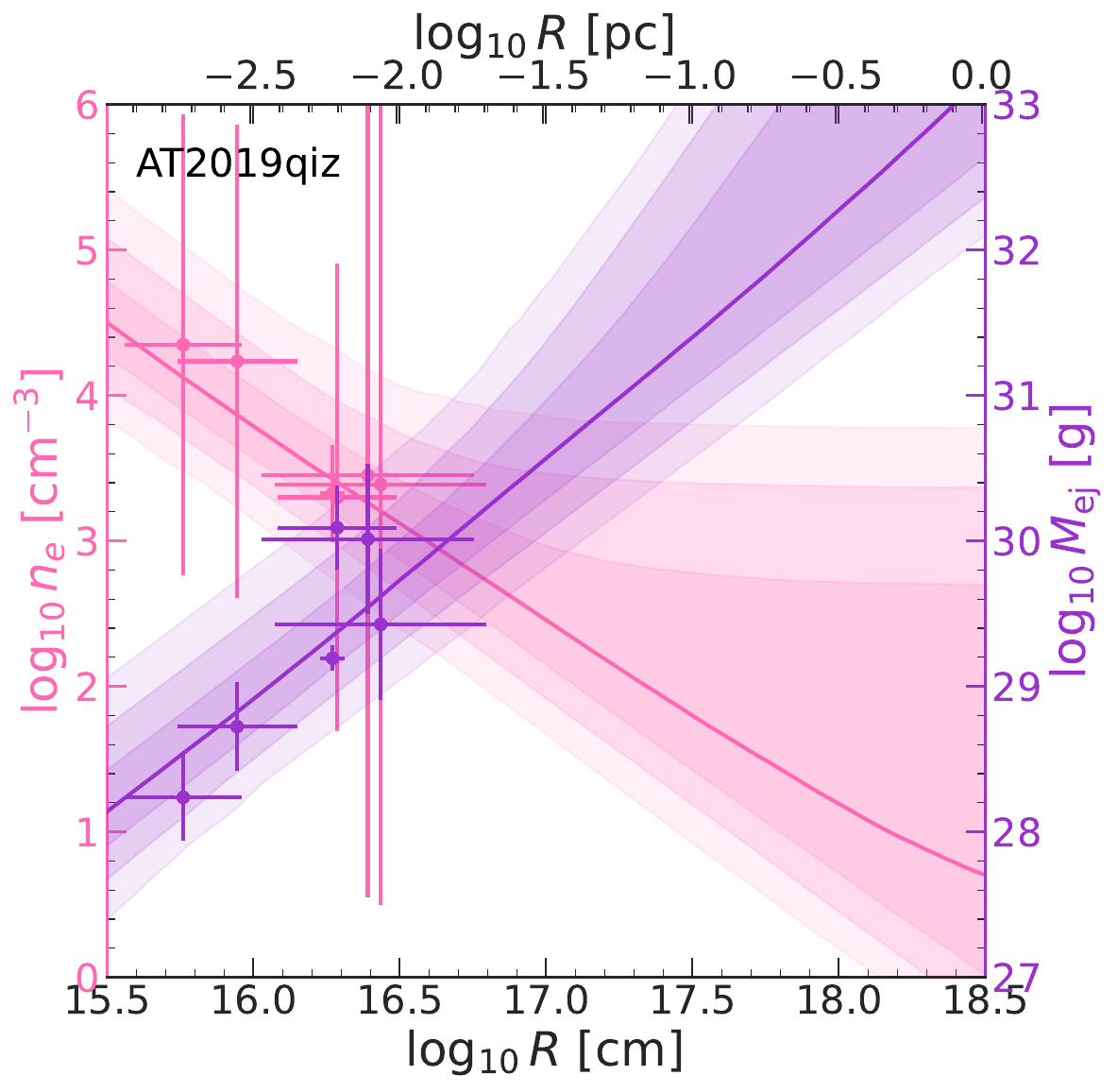}
    \includegraphics[width=0.32\linewidth]{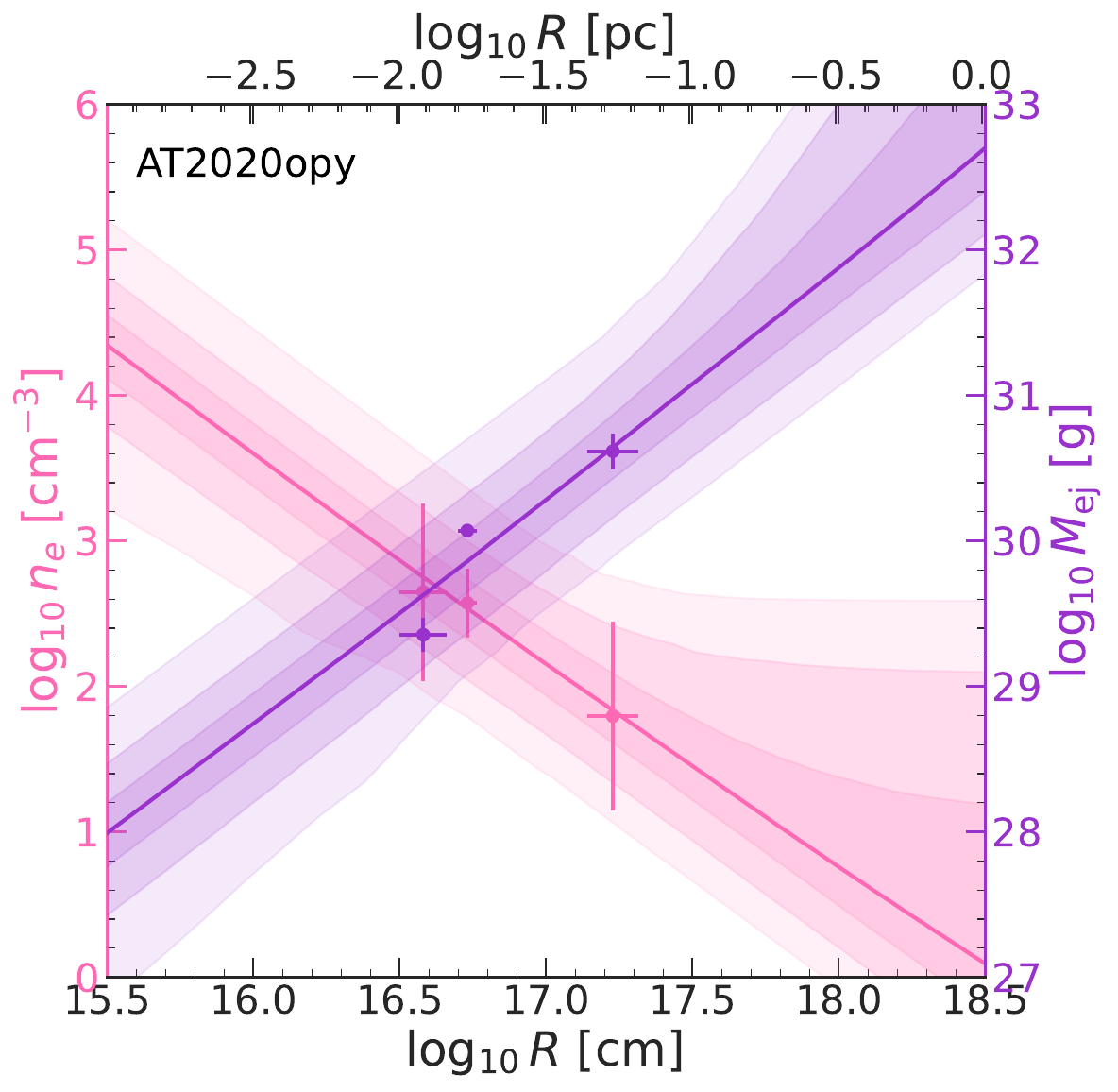}
    \includegraphics[width=0.32\linewidth]{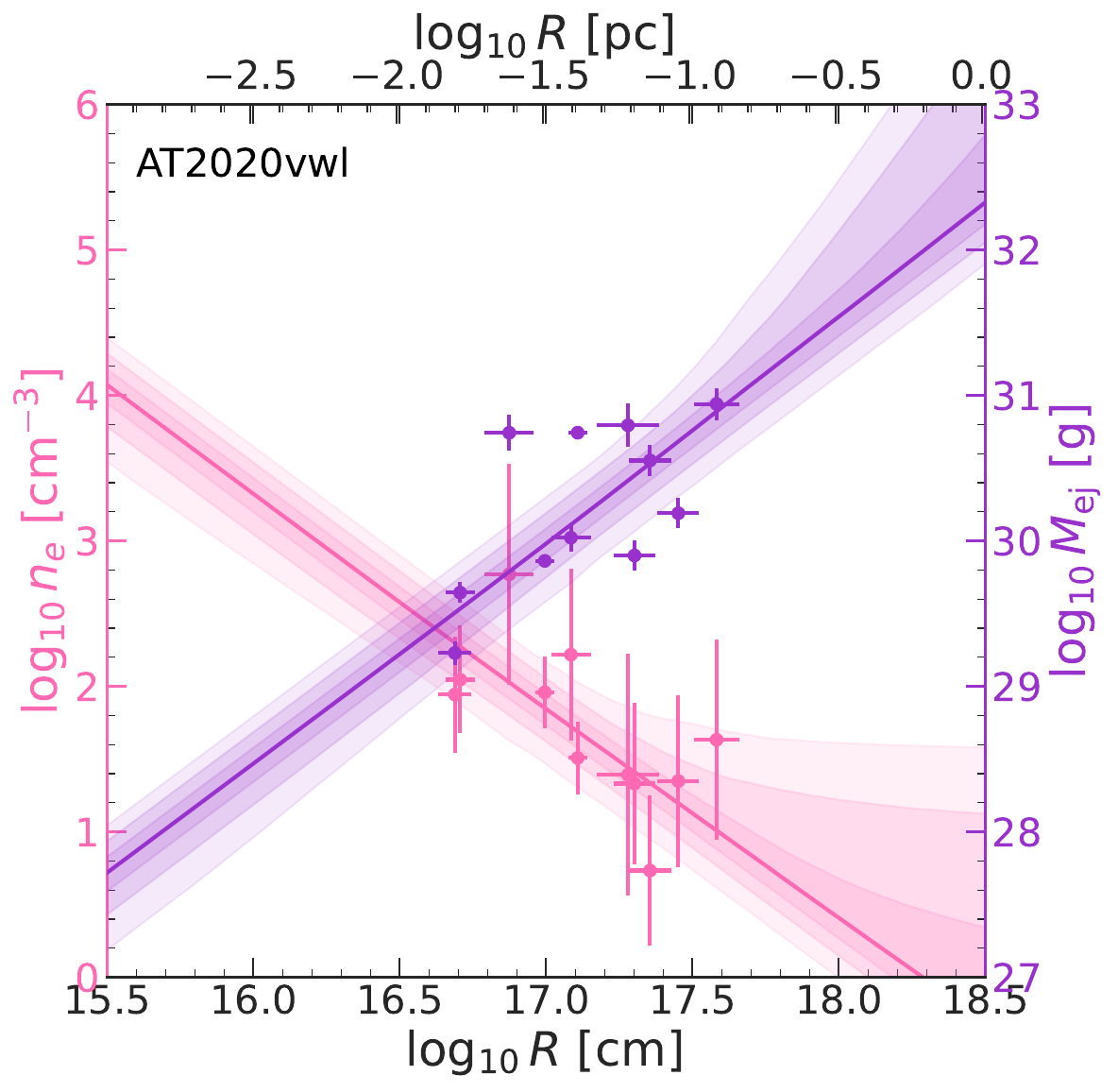}
    \includegraphics[width=0.32\linewidth]{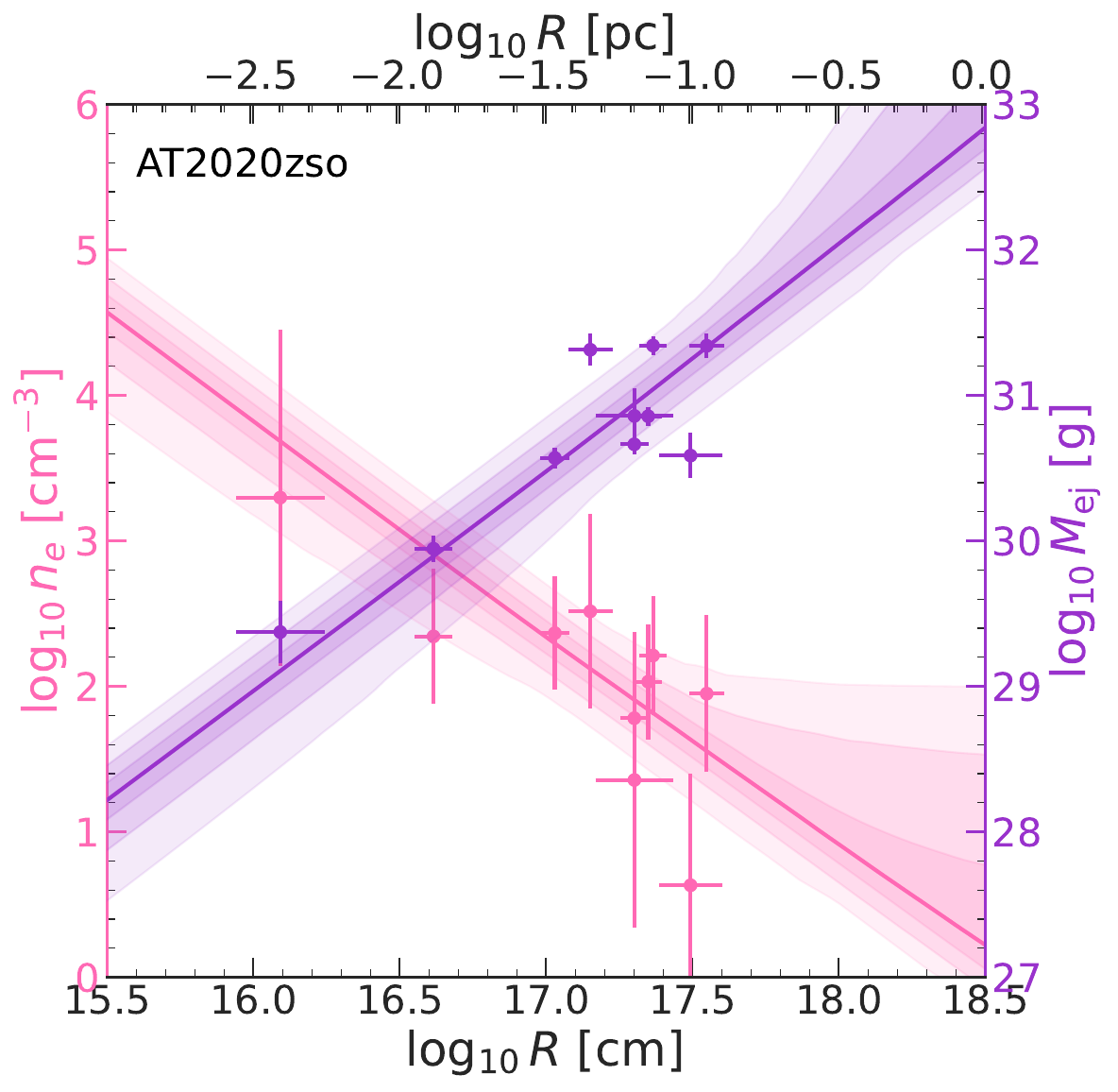}
    \includegraphics[width=0.32\linewidth]{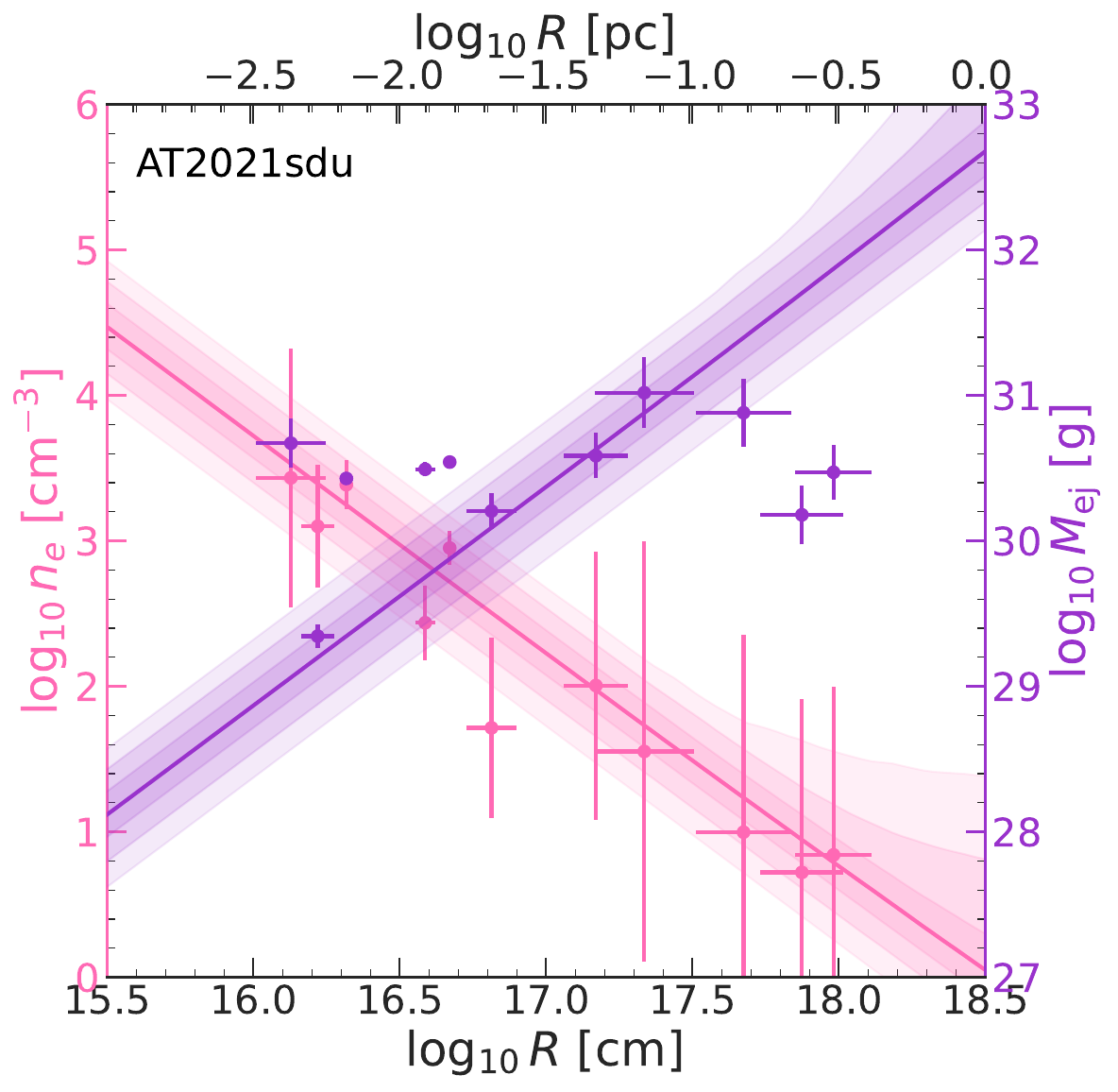}
    \includegraphics[width=0.32\linewidth]{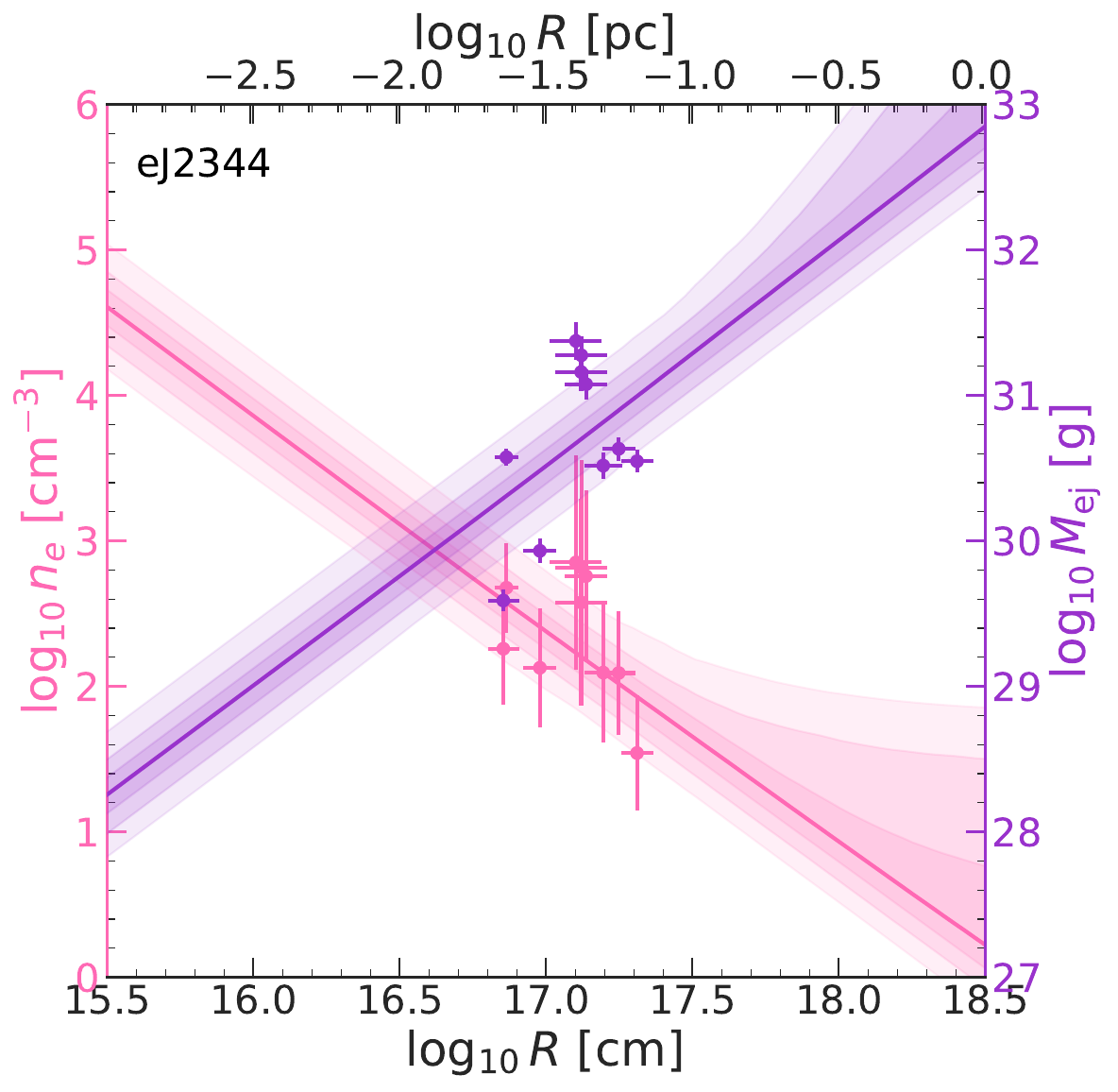}
    \caption{The individual fits for the joint mass and density fitting procedure for each source. In each panel, the ambient density with radius is plotted on the left axis (pink) and the outflow mass is plotted on the right axis (purple). The best-fit model is plotted in the dark line and the 1, 2, and 3$\sigma$ confidence-intervals are shaded. Note the vastly different scales of the two y-axes. }
    \label{fig:all_ne_fits}
\end{figure*}

The individual joint mass and density TDE fit results are listed in Table \ref{tab:ind_fit_results}, where $\log_{10} n_e = k \log_{10}R/R_0 + A$ was fit (i.e. free $k$) for the ``power-law" results. The joint mass-density fit results for the Bondi density profile described in Section \ref{sec:bondi_const} are also listed in Table \ref{tab:ind_fit_results}. The best-fitting radius, density, and mass values for each source for the Bondi profile fit are listed in Table \ref{tab:nes_all}. 
The individual density and mass fits for each TDE are plotted in Figure \ref{fig:all_ne_fits}.

\begin{table}[]
    \centering
    \begin{tabular}{l|lll|lll}
         TDE & Power law && & Bondi accretion  \\
         & $k$ & $A$ & $\log_{10}\epsilon_e$ & $\log_{10}f_{\rm{Edd}}$ & $\log_{10}T_\infty$ & $\log_{10}\epsilon_e$\\
         \hline

ASASSN-15oi & $-1.75^{+0.18}_{-0.19}$ & $4.18^{+0.22}_{-0.21}$ & $-1.67^{+0.15}_{-0.15}$ & $-3.58^{+0.14}_{-0.14}$ & $4.25^{+0.96}_{-0.87}$ & $-1.85^{+0.15}_{-0.14}$ \\
AT2020vwl & $-1.59^{+0.42}_{-0.41}$ & $3.59^{+0.38}_{-0.41}$ & $-1.63^{+0.16}_{-0.17}$ & $-4.17^{+0.21}_{-0.21}$ & $4.86^{+1.43}_{-1.29}$ & $-1.82^{+0.15}_{-0.15}$ \\
AT2019dsg & $-1.69^{+0.54}_{-0.55}$ & $3.87^{+0.32}_{-0.35}$ & $-1.99^{+0.27}_{-0.29}$ & $-3.98^{+0.19}_{-0.20}$ & $5.17^{+1.46}_{-1.40}$ & $-2.18^{+0.24}_{-0.24}$ \\
AT2019azh & $-0.82^{+0.40}_{-0.40}$ & $2.81^{+0.43}_{-0.43}$ & $-1.67^{+0.17}_{-0.17}$ & $-4.29^{+0.24}_{-0.45}$ & $6.59^{+0.85}_{-2.07}$ & $-1.93^{+0.17}_{-0.18}$ \\
ASASSN-14li & $-2.33^{+0.47}_{-0.40}$ & $3.83^{+0.17}_{-0.23}$ & $-1.80^{+0.28}_{-0.34}$ & $-4.03^{+0.15}_{-0.16}$ & $5.08^{+1.24}_{-1.41}$ & $-1.73^{+0.21}_{-0.21}$ \\
AT2019ahk & $-0.63^{+0.91}_{-0.90}$ & $2.90^{+0.93}_{-0.95}$ & $-1.63^{+0.36}_{-0.36}$ & $-4.04^{+0.28}_{-0.36}$ & $5.74^{+1.56}_{-1.88}$ & $-1.92^{+0.31}_{-0.32}$ \\
AT2019qiz & $-0.75^{+0.79}_{-0.78}$ & $3.87^{+0.26}_{-0.25}$ & $-0.64^{+0.58}_{-0.49}$ & $-3.56^{+0.28}_{-0.25}$ & $5.69^{+1.67}_{-1.83}$ & $-1.17^{+0.28}_{-0.38}$ \\
AT2020opy & $-1.33^{+0.78}_{-0.72}$ & $3.66^{+0.50}_{-0.58}$ & $-1.89^{+0.39}_{-0.39}$ & $-3.98^{+0.30}_{-0.31}$ & $5.23^{+1.66}_{-1.53}$ & $-2.10^{+0.36}_{-0.35}$ \\
eJ2344 & $-1.31^{+0.73}_{-0.72}$ & $3.85^{+0.67}_{-0.69}$ & $-1.47^{+0.23}_{-0.26}$ & $-4.14^{+0.15}_{-0.15}$ & $5.59^{+1.63}_{-1.75}$ & $-1.69^{+0.19}_{-0.20}$ \\
AT2021sdu & $-2.57^{+0.28}_{-0.22}$ & $4.43^{+0.16}_{-0.17}$ & $-1.91^{+0.21}_{-0.24}$ & $-3.87^{+0.26}_{-0.24}$ & $4.50^{+1.23}_{-1.02}$ & $-2.01^{+0.23}_{-0.22}$ \\
AT2020zso & $-1.63^{+0.28}_{-0.26}$ & $4.14^{+0.27}_{-0.29}$ & $-1.73^{+0.15}_{-0.15}$ & $-3.60^{+0.21}_{-0.21}$ & $4.79^{+1.29}_{-1.23}$ & $-1.88^{+0.15}_{-0.15}$ \\
\hline
    \end{tabular}
    \caption{MCMC fit results  for individual TDE joint mass and density fits. All fits assume spherical geometry, a weak assumption that does not impact the parameter inference.}
    \label{tab:ind_fit_results}
\end{table}

\clearpage
\begin{longtable}{cccc}
TDE & $\log_{10}R$ [cm] & $\log_{10}n_e$\,[cm$^{-3}$] & $\log_{10}m_{\rm{ej}}$ [g] \\
\hline
\hline
ASASSN-15oi & 16.76$\pm$0.16 & 2.19$\pm$1.25 & 30.70$\pm$0.24 \\
ASASSN-15oi & 17.80$\pm$0.03 & 0.79$\pm$0.20 & 31.66$\pm$0.04 \\
ASASSN-15oi & 18.00$\pm$0.05 & 0.70$\pm$0.38 & 31.82$\pm$0.07 \\
ASASSN-15oi & 18.01$\pm$0.13 & 0.58$\pm$0.97 & 31.89$\pm$0.19 \\
ASASSN-15oi & 17.78$\pm$0.06 & 0.72$\pm$0.47 & 31.89$\pm$0.09 \\
\hline
AT2020vwl & 16.87$\pm$0.08 & 2.77$\pm$0.76 & 30.74$\pm$0.12 \\
AT2020vwl & 17.11$\pm$0.03 & 1.51$\pm$0.25 & 30.74$\pm$0.05 \\
AT2020vwl & 17.35$\pm$0.07 & 0.73$\pm$0.52 & 30.55$\pm$0.11 \\
AT2020vwl & 17.28$\pm$0.11 & 1.39$\pm$0.83 & 30.79$\pm$0.15 \\
AT2020vwl & 17.58$\pm$0.08 & 1.63$\pm$0.69 & 30.94$\pm$0.11 \\
\hline
AT2019dsg & 16.25$\pm$0.02 & 3.07$\pm$0.13 & 29.05$\pm$0.03 \\
AT2019dsg & 16.85$\pm$0.05 & 1.80$\pm$0.38 & 28.97$\pm$0.08 \\
AT2019dsg & 16.56$\pm$0.03 & 3.30$\pm$0.25 & 29.79$\pm$0.04 \\
AT2019dsg & 16.75$\pm$0.03 & 2.68$\pm$0.24 & 30.33$\pm$0.04 \\
AT2019dsg & 17.09$\pm$0.07 & 2.00$\pm$0.56 & 30.58$\pm$0.10 \\
AT2019dsg & 17.34$\pm$0.16 & 1.03$\pm$1.21 & 30.54$\pm$0.22 \\
AT2019dsg & 16.87$\pm$0.07 & 1.88$\pm$0.53 & 31.14$\pm$0.10 \\
\hline
AT2019azh & 16.47$\pm$0.14 & 1.93$\pm$0.98 & 27.98$\pm$0.20 \\
AT2019azh & 16.56$\pm$0.07 & 2.08$\pm$0.55 & 28.62$\pm$0.10 \\
AT2019azh & 16.87$\pm$0.13 & 1.53$\pm$0.99 & 28.69$\pm$0.19 \\
AT2019azh & 17.37$\pm$0.10 & 0.62$\pm$0.75 & 28.97$\pm$0.15 \\
AT2019azh & 17.13$\pm$0.03 & 1.54$\pm$0.21 & 29.43$\pm$0.04 \\
AT2019azh & 17.25$\pm$0.04 & 1.46$\pm$0.32 & 29.74$\pm$0.06 \\
AT2019azh & 17.07$\pm$0.06 & 1.99$\pm$0.47 & 30.03$\pm$0.08 \\
AT2019azh & 17.11$\pm$0.09 & 1.57$\pm$0.70 & 30.09$\pm$0.13 \\
AT2019azh & 17.25$\pm$0.08 & 2.18$\pm$0.68 & 30.55$\pm$0.11 \\
AT2019azh & 17.48$\pm$0.03 & 1.27$\pm$0.23 & 30.66$\pm$0.05 \\
AT2019azh & 17.51$\pm$0.05 & 1.77$\pm$0.44 & 30.89$\pm$0.07 \\
AT2019azh & 17.31$\pm$0.07 & 1.53$\pm$0.54 & 30.81$\pm$0.10 \\
AT2019azh & 17.15$\pm$0.14 & 1.38$\pm$1.09 & 30.99$\pm$0.20 \\
AT2019azh & 17.45$\pm$0.07 & 1.52$\pm$0.60 & 31.05$\pm$0.10 \\
AT2019azh & 16.75$\pm$0.16 & 2.66$\pm$1.33 & 31.41$\pm$0.23 \\
AT2019azh & 17.56$\pm$0.04 & 1.44$\pm$0.38 & 31.18$\pm$0.06 \\
AT2019azh & 17.16$\pm$0.08 & 1.66$\pm$0.64 & 31.28$\pm$0.11 \\
AT2019azh & 17.07$\pm$0.10 & 1.46$\pm$0.79 & 31.32$\pm$0.15 \\
\hline
ASASSN-14li & 16.47$\pm$0.04 & 2.67$\pm$0.27 & 28.76$\pm$0.05 \\
ASASSN-14li & 16.76$\pm$0.14 & 1.94$\pm$1.01 & 29.16$\pm$0.20 \\
ASASSN-14li & 16.74$\pm$0.03 & 2.03$\pm$0.22 & 29.39$\pm$0.05 \\
ASASSN-14li & 16.37$\pm$0.13 & 3.18$\pm$1.03 & 29.67$\pm$0.18 \\
ASASSN-14li & 16.89$\pm$0.04 & 1.84$\pm$0.30 & 29.52$\pm$0.06 \\
ASASSN-14li & 16.27$\pm$0.11 & 3.36$\pm$0.89 & 29.89$\pm$0.15 \\
ASASSN-14li & 16.87$\pm$0.06 & 2.43$\pm$0.52 & 29.77$\pm$0.09 \\
ASASSN-14li & 17.01$\pm$0.08 & 1.93$\pm$0.61 & 29.71$\pm$0.11 \\
\hline
AT2019ahk & 16.92$\pm$0.09 & 1.65$\pm$0.70 & 29.17$\pm$0.13 \\
AT2019ahk & 16.99$\pm$0.08 & 2.24$\pm$0.63 & 29.71$\pm$0.11 \\
AT2019ahk & 17.15$\pm$0.05 & 2.01$\pm$0.41 & 30.74$\pm$0.08 \\
AT2019ahk & 16.99$\pm$0.20 & 2.37$\pm$1.64 & 31.24$\pm$0.29 \\
AT2019ahk & 17.58$\pm$0.06 & 1.48$\pm$0.50 & 31.40$\pm$0.09 \\
\hline
AT2019qiz & 15.76$\pm$0.20 & 4.35$\pm$1.59 & 28.24$\pm$0.30 \\
AT2019qiz & 15.94$\pm$0.20 & 4.23$\pm$1.63 & 28.72$\pm$0.30 \\
AT2019qiz & 16.27$\pm$0.04 & 3.33$\pm$0.33 & 29.19$\pm$0.09 \\
AT2019qiz & 16.44$\pm$0.36 & 3.39$\pm$2.89 & 29.43$\pm$0.52 \\
AT2019qiz & 16.39$\pm$0.36 & 3.45$\pm$2.90 & 30.01$\pm$0.52 \\
AT2019qiz & 16.29$\pm$0.20 & 3.30$\pm$1.61 & 30.09$\pm$0.29 \\
\hline
AT2020opy & 16.58$\pm$0.08 & 2.65$\pm$0.61 & 29.35$\pm$0.12 \\
AT2020opy & 16.73$\pm$0.03 & 2.57$\pm$0.24 & 30.07$\pm$0.05 \\
AT2020opy & 17.23$\pm$0.09 & 1.80$\pm$0.65 & 30.62$\pm$0.12 \\
\hline
eJ2344 & 16.85$\pm$0.05 & 2.26$\pm$0.39 & 29.59$\pm$0.08 \\
eJ2344 & 16.98$\pm$0.06 & 2.13$\pm$0.41 & 29.93$\pm$0.08 \\
eJ2344 & 16.86$\pm$0.04 & 2.68$\pm$0.31 & 30.57$\pm$0.06 \\
eJ2344 & 17.20$\pm$0.06 & 2.09$\pm$0.48 & 30.52$\pm$0.09 \\
eJ2344 & 17.25$\pm$0.06 & 2.09$\pm$0.43 & 30.63$\pm$0.08 \\
eJ2344 & 17.31$\pm$0.05 & 1.54$\pm$0.40 & 30.55$\pm$0.08 \\
eJ2344 & 17.14$\pm$0.07 & 2.76$\pm$0.59 & 31.08$\pm$0.10 \\
eJ2344 & 17.12$\pm$0.09 & 2.58$\pm$0.71 & 31.16$\pm$0.13 \\
eJ2344 & 17.12$\pm$0.09 & 2.81$\pm$0.74 & 31.28$\pm$0.13 \\
eJ2344 & 17.10$\pm$0.09 & 2.85$\pm$0.74 & 31.37$\pm$0.13 \\
\hline
AT2021sdu & 16.22$\pm$0.06 & 3.10$\pm$0.42 & 29.34$\pm$0.08 \\
AT2021sdu & 16.32$\pm$0.02 & 3.39$\pm$0.17 & 30.43$\pm$0.03 \\
AT2021sdu & 16.67$\pm$0.01 & 2.95$\pm$0.12 & 30.54$\pm$0.02 \\
AT2021sdu & 16.13$\pm$0.12 & 3.43$\pm$0.89 & 30.67$\pm$0.17 \\
AT2021sdu & 17.87$\pm$0.14 & 0.72$\pm$1.19 & 30.18$\pm$0.20 \\
AT2021sdu & 16.59$\pm$0.03 & 2.44$\pm$0.26 & 30.49$\pm$0.05 \\
AT2021sdu & 16.81$\pm$0.09 & 1.71$\pm$0.62 & 30.20$\pm$0.12 \\
AT2021sdu & 17.98$\pm$0.13 & 0.84$\pm$1.15 & 30.47$\pm$0.19 \\
AT2021sdu & 17.17$\pm$0.11 & 2.01$\pm$0.92 & 30.59$\pm$0.16 \\
AT2021sdu & 17.67$\pm$0.16 & 1.00$\pm$1.36 & 30.88$\pm$0.23 \\
AT2021sdu & 17.34$\pm$0.17 & 1.55$\pm$1.45 & 31.02$\pm$0.24 \\
\hline
AT2020zso & 16.09$\pm$0.15 & 3.30$\pm$1.15 & 29.37$\pm$0.22 \\
AT2020zso & 16.62$\pm$0.06 & 2.34$\pm$0.46 & 29.95$\pm$0.09 \\
AT2020zso & 17.03$\pm$0.05 & 2.37$\pm$0.39 & 30.57$\pm$0.07 \\
AT2020zso & 17.30$\pm$0.05 & 1.78$\pm$0.39 & 30.66$\pm$0.07 \\
AT2020zso & 17.35$\pm$0.05 & 2.03$\pm$0.40 & 30.86$\pm$0.07 \\
AT2020zso & 17.49$\pm$0.11 & 0.63$\pm$0.77 & 30.59$\pm$0.15 \\
AT2020zso & 17.30$\pm$0.13 & 1.36$\pm$1.02 & 30.86$\pm$0.19 \\
AT2020zso & 17.15$\pm$0.08 & 2.52$\pm$0.67 & 31.31$\pm$0.11 \\
AT2020zso & 17.37$\pm$0.05 & 2.21$\pm$0.41 & 31.34$\pm$0.07 \\
AT2020zso & 17.55$\pm$0.06 & 1.95$\pm$0.54 & 31.34$\pm$0.09 \\
\hline
    \caption{The best-fit ambient density density ($n_e$) and mass swept up by the outflow ($m_{\rm{ej}}$) at radius from the SMBH, $R$, for each of the TDE hosts studied in this work.}
    \label{tab:nes_all}
\end{longtable}

\end{document}